\documentclass[11pt]{article}
\usepackage[margin=1in]{geometry}
\usepackage{amsmath,amsfonts,amssymb,amsthm}
\usepackage{mathtools}
\usepackage{mathrsfs}
\usepackage{upgreek}
\usepackage{graphicx}
\usepackage{enumerate}
\usepackage{bm}
\usepackage{bbm}
\usepackage{verbatim}
\usepackage{hyperref,color}
\usepackage[capitalize,nameinlink]{cleveref}
\usepackage[dvipsnames]{xcolor}
\hypersetup{
	colorlinks=true,
	pdfpagemode=UseNone,
    citecolor=OliveGreen,
    linkcolor=NavyBlue,
    urlcolor=Magenta,
	pdfstartview=FitW
}
\usepackage{appendix}
\crefname{appsec}{Appendix}{Appendices}
\usepackage{tikz}
\usepackage[ruled,linesnumbered]{algorithm2e}
\SetKwComment{Comment}{/* }{ */}
\usepackage{tabularx}

\usepackage[colorinlistoftodos]{todonotes}

\theoremstyle{plain}
\newtheorem{thm}{Theorem}[section]
\newtheorem{prop}[thm]{Proposition}
\newtheorem{lem}[thm]{Lemma}
\newtheorem{lemma}[thm]{Lemma}

\newtheorem{cor}[thm]{Corollary}
\newtheorem{conj}[thm]{Conjecture}
\newtheorem{cond}[thm]{Condition}

\newtheorem{claim}[thm]{Claim}
\newtheorem{fact}[thm]{Fact}
\theoremstyle{definition}
\newtheorem{defn}[thm]{Definition}
\newtheorem{definition}[thm]{Definition}

\newtheorem{obs}[thm]{Observation}

\newtheorem*{ass*}{Assumption}

\theoremstyle{remark}

\newtheorem{rem}[thm]{Remark}
\newtheorem{ex}[thm]{Example}

\crefname{lem}{Lemma}{Lemmas}
\crefname{thm}{Theorem}{Theorems}
\crefname{defn}{Definition}{Definitions}
\crefname{fact}{Fact}{Facts}
\crefname{clm}{Claim}{Claims}
\crefname{prop}{Proposition}{Propositions}
\crefname{algocf}{Algorithm}{Algorithms}

\newcommand{\E}{\mathbb{E}}

\newcommand{\allone}{\mathbf{1}}

\DeclareMathOperator{\arctanh}{arctanh}

\DeclareMathOperator{\TV}{\mathsf{TV}}

\newcommand{\norm}[1]{\left\lVert #1 \right\rVert}
\newcommand*{\abs}[1]{\left|{#1}\right|}
\newcommand{\ceil}[1]{\left\lceil #1 \right\rceil}
\newcommand{\floor}[1]{\left\lfloor #1 \right\rfloor}

\newcommand{\diag}{\mathrm{diag}}
\newcommand{\poly}{\mathrm{poly}}
\newcommand{\dist}{\mathrm{dist}}

\newcommand{\eps}{\varepsilon}

\newcommand{\defeq}{\coloneqq} 
\newcommand{\p}{\bm{p}}
\newcommand{\y}{\bm{m}}
\newcommand{\x}{\bm{x}}
\newcommand{\w}{\bm{w}}
\newcommand{\N}{\mathbb{N}}

\newcommand{\R}{\mathbb{R}}
\newcommand{\RR}{\mathbb{R}}

\newcommand{\mcC}{\mathcal{C}}
\newcommand{\mcD}{\mathcal{D}}

\newcommand{\mcL}{\mathcal{L}}

\newcommand{\mfa}{\mathfrak{a}}
\newcommand{\mfb}{\mathfrak{b}}
\newcommand{\mfc}{\mathfrak{c}}

\newcommand{\ind}{\mathbf{1}}

\newcommand{\infl}{\Psi} 

\renewcommand{\Pr}{\mathbb{P}}

\newcommand*{\wrapp}[1]{\left({#1}\right)}

\newcommand*{\wrapc}[1]{\left\{{#1}\right\}}
\newcommand{\SM}{\mathsf{SM}} 
\newcommand{\INFL}{\mathsf{INFL}}
\newcommand{\SI}{\mathsf{SI}}
\newcommand{\lb}{\mathscr{L}}
\newcommand{\ub}{\mathscr{B}}

\newcommand{\dH}{d_{\mathrm{H}}}
\newcommand{\zcQ}{\mathcal{L}}
\newcommand{\zcS}{S}
\newcommand{\zcT}{\mathcal{T}}
\newcommand{\girth}{\mathsf{g}}
\newcommand{\zcDeg}{\mathrm{deg}}

\DeclareSymbolFont{fancygreek}{OML}{ztmcm}{m}{it}
\DeclareMathSymbol{\lambdafancy}{\mathord}{fancygreek}{"15}
\newcommand{\eigval}{\lambdafancy} 

\begin{document}
	
\title{Strong Spatial Mixing for Colorings on Trees\\ and its Algorithmic Applications}
\author{Zongchen Chen\thanks{email: \texttt{zchen83@buffalo.edu}, University at Buffalo} \and Kuikui Liu\thanks{email: \texttt{liukui@mit.edu}, Massachusetts Institute of Technology} \and Nitya Mani\thanks{email: \texttt{nmani@mit.edu}, Massachusetts Institute of Technology} \and Ankur Moitra\thanks{email: \texttt{moitra@mit.edu}, Massachusetts Institute of Technology}}
\date{\today}

\maketitle

\begin{abstract}
\emph{Strong spatial mixing (SSM)} is an important quantitative notion of correlation decay for Gibbs distributions arising in statistical physics, probability theory, and theoretical computer science.
A longstanding conjecture is that the uniform distribution on proper $q$-colorings on a $\Delta$-regular tree exhibits SSM whenever $q \ge \Delta+1$. 
Moreover, it is widely believed that as long as SSM holds on bounded-degree trees with $q$ colors, one would obtain an efficient sampler for $q$-colorings on all bounded-degree graphs via simple Markov chain algorithms. 
It is surprising that such a basic question is still open, even on trees, but then again it also highlights how much we still have to learn about random colorings.

In this paper, we show the following:
\begin{itemize}
    \item[(1)] For any $\Delta \ge 3$, SSM holds for random $q$-colorings on trees of maximum degree $\Delta$ whenever $q \ge \Delta + 3$. Thus we almost fully resolve the aforementioned conjecture. Our result substantially improves upon the previously best bound which requires $q \ge 1.59\Delta+\gamma^*$ for an absolute constant $\gamma^* > 0$.

    \item[(2)] For any $\Delta\ge 3$ and $\girth=\Omega_\Delta(1)$, we establish optimal mixing of the Glauber dynamics for $q$-colorings on graphs of maximum degree $\Delta$ and girth $\girth$ whenever $q \ge \Delta+3$. Our approach is based on a new general reduction from \emph{spectral independence} on large-girth graphs to SSM on trees that is of independent interest.
\end{itemize}
Using the same techniques, we also prove near-optimal bounds on weak spatial mixing (WSM), a closely-related notion to SSM, for the antiferromagnetic Potts model on trees. 
\end{abstract}
\thispagestyle{empty}

\clearpage 
\setcounter{tocdepth}{2} 
\tableofcontents
\thispagestyle{empty}
\clearpage 

\setcounter{page}{1}
\section{Introduction}

A notorious open problem in the field of approximate counting and sampling is to give an efficient algorithm for approximately counting and sampling proper $q$-colorings of the vertices of graphs of maximum degree $\Delta$ that works whenever $q \geq \Delta + 1$. Despite its long and storied history, we are still far away from proving this conjecture; the best known algorithms require substantially more colors, namely $q \ge (11/6-\eps_0)\Delta$ for some small constant $\eps_0 \approx 10^{-5}$ \cite{VIG99, CDM19}. 

It is understood that progress towards this problem is intertwined with a better understanding of the \emph{structural properties} of the uniform distribution on proper $q$-colorings.
In particular, one important such structural property, which is widely believed to hold down to $q \geq \Delta + 1$ and has been exploited to great effect in algorithms for approximate counting and sampling, is \emph{strong spatial mixing (SSM)} or \emph{decay of correlations}. Intuitively, SSM says that the strength of the correlation between color assignments of far away pairs of vertices should be (exponentially) small in their graph distance, even under the conditional distribution that results from pinning the colors of some other vertices. More precisely:

\begin{defn}[Strong Spatial Mixing]
The uniform distribution $\mu$ on proper $q$-colorings of a graph $G=(V,E)$ satisfies \emph{strong spatial mixing (SSM) with exponential decay rate $1-\delta$ and constant $C$} if for every vertex $v \in V$, every subset $\Lambda \subseteq V \setminus\{v\}$, and every pair of pinnings $\sigma,\tau : \Lambda \to [q]$ which differ on $\partial_{\sigma,\tau} = \{u \in \Lambda : \sigma(u) \neq \tau(u)\}$, we have the inequality
\begin{align*}
    \norm{\mu_{v}^{\sigma} - \mu_{v}^{\tau}}_{\TV} \leq C (1-\delta)^{K}.
\end{align*}
where $\mu^\sigma_v$ (resp., $\mu^\tau_v$) denotes the marginal distribution on the color of vertex $v$ conditional on $\sigma$ (resp., $\tau$), $\|\cdot\|_{\TV}$ denotes the total variation distance, and $K = \dist_G(v,\partial_{\sigma,\tau})$ denotes the graph distance between $v$ and $\partial_{\sigma,\tau}$.
\end{defn}

There is a vast literature studying SSM and other related notions for proper $q$-colorings~\cite{MOS94,MO94,JER95,VIG99,WEI04,GMP05,GKM15,EGHSV19}. Weitz~\cite{WEI04} made a striking conjecture that simple and local Markov chain sampling algorithms like the Glauber dynamics ought to mix rapidly if and only if the distribution exhibits SSM. Thus, we believe that algorithms for sampling from distributions go hand-in-hand with showing that the distribution in question does not exhibit long range correlations. 
Some results along this direction are known, including for spin systems on the integer grid $\mathbb{Z}^d$ (including for example, \cite{DS87,SZ92,MO94,MO94ii,Cesi01,DSVW04,CP20,BCPSV22}) and more generally the class of \emph{neighborhood-amenable graphs} \cite{GMP05,FGY22}. 
However these results leave out important families of graphs, like random regular graphs, where the neighborhood around each vertex is locally tree-like and grows superpolynomially, which causes these techniques to break down. 
In general, black box implications between SSM and rapid mixing of the Glauber dynamics in \textit{either direction} are not known. 
Nevertheless many of the methods for establishing SSM turn out to have algorithmic implications.

In this paper, our main focus is on the following longstanding conjecture \cite{JON02,MAR99,WEI04} and its algorithmic implications:

\begin{conj}[Folklore] \label{c:coloring-folklore}
For $q \ge \Delta + 1$, the uniform distribution on $q$-colorings on trees of maximum degree $\Delta$ exhibits \textit{strong spatial mixing} with exponential decay rate.
\end{conj}

In 2002, Jonasson~\cite{JON02} showed that a related but weaker notion, called {\em weak spatial mixing (WSM)}, holds for $q$-colorings on $\Delta$-regular trees provided that $q \ge \Delta + 1$, a tight bound. However, the proof required computer verification for $q \leq 1000$, making it difficult to generalize to strong spatial mixing. 
Ge and {\v{S}}tefankovi\v{c}~\cite{GS11} proved that SSM holds for $4$-colorings on binary trees. More generally,
Goldberg, Martin, and Paterson~\cite{GMP05} showed that SSM holds for \textit{neighborhood-amenable, triangle-free graphs} of maximum degree $\Delta$ when $q \ge 1.76 \Delta$. Subsequently this was extended to cover (list) colorings on all triangle-free graphs, in particular including trees, by Gamarnik, Katz, and Misra~\cite{GKM15}. 
In an important recent work, Efthymiou et al. \cite{EGHSV19} showed SSM holds on $\Delta$-regular trees provided that $q \ge 1.59\Delta+\gamma^*$ where $\gamma^* > 0$ is an absolute constant. 
This remains the best bound for SSM on trees, but is still quite far from the conjectured bounds. It is surprising that such a basic question is still open, even on trees, but then again it also highlights how much we still have to learn about random colorings.

Our first main result is to almost fully resolve~\cref{c:coloring-folklore}.

\begin{thm}\label{t:strong-ssm}
For all integers $\Delta \ge 3$ and $q \ge \Delta + 3$, the uniform distribution on $q$-colorings exhibits strong spatial mixing with exponential decay rate on any tree $T$ with maximum degree $\Delta$.
\end{thm}
Our proof also establishes SSM for list colorings on trees where at each vertex $v$ the number of available colors is at least $\zcDeg(v) + 3$; see \cref{thm:SSM+SSI} for more details. The techniques that go into our proof are discussed in more detail in~\cref{sec:outline}. A major advantage of our methods is that, with the right potential function in hand, everything is conceptually quite simple; in fact in~\cref{sec:colorings-simple}, we give a short proof of SSM on trees with maximum degree $\Delta$ whenever $q \ge \Delta + 3\sqrt{\Delta}$, which is already a substantial improvement over what was previously known and also within striking distance of optimal.

\subsection{Algorithmic applications for sampling colorings}
In this section we discuss the algorithmic implications of our new bounds for SSM. 
As we already discussed, sampling random proper $q$-colorings is a major open problem.
In 1994, Jerrum~\cite{JER95} proved that a natural and widely studied Markov chain called the \textit{Glauber dynamics} \cite{JER95,BD97} succeeds whenever $q \ge 2\Delta + 1$. 
Glauber dynamics (also known as Gibbs sampling) is a simple and popular Markov chain algorithm for generating random objects from an exponentially large state space. 
In each step of Glauber dynamics, we pick a vertex $v$ uniformly at random and update its color with a uniformly random available color that does not already appear in $v$'s neighborhood. 
In 1999, Vigoda~\cite{VIG99} introduced an alternate Markov chain called the \textit{flip dynamics}, which he showed succeeds whenever $q > \frac{11}{6} \Delta$. This was recently slightly improved to $q > \wrapp{\frac{11}{6} - \eps_{0}} \Delta$ for $\eps_0 \approx 10^{-5}$ \cite{CDM19}, and under the same condition Glauber dynamics was shown to mix as well (see also \cite{BCCPSV22,Liu21}). 
The main conjecture \cite{JER95} that concerns sampling proper $q$-colorings is the following, posed by Jerrum in 1995.

\begin{conj}[\cite{JER95}]\label{c:color-mix}
The Glauber dynamics for sampling $q$-colorings from a graph of maximum degree $\Delta$ mixes in polynomial time whenever $q \ge \Delta + 2$.
\end{conj}

We remark that $\Delta+2$ is the best possible for the number of colors since the Glauber dynamics may not be ergodic with $q = \Delta+1$ colors for some graph (e.g., a clique on $\Delta+1$ vertices). The conjecture can be thought of as an algorithmic counterpart to~\cref{c:coloring-folklore}.
In one direction, it has been shown~\cite{GJS76, GSV15} that for even $q < \Delta$ there is no randomized polynomial time approximation scheme for sampling $q$-colorings unless $\textsf{NP = RP}$. However, as noted above, positive results in the other direction are only known for substantially larger $q$.
For completeness, we summarize known results on WSM, SSM and approximate counting and sampling for proper $q$-colorings in \cref{table}.

\begin{table}[t]
\centering
\begin{tabular}{|l|l|p{9em}|p{12em}|}
\hline
 & Trees & Triangle-free & General graphs \\
 \hline
WSM     & $q \ge \Delta + 1$~\cite{JON02} & $q \ge 1.76\Delta$~\cite{GKM15} & $q \ge 2\Delta$~\cite{GKM15}  \\
SSM     & $q \ge 1.59\Delta$~\cite{EGHSV19} & $q \ge 1.76\Delta$~\cite{GKM15} & $q \ge 2\Delta$~\cite{GKM15} \\
Sampling & Trivial & $q \ge 1.76\Delta$~\cite{LSS19,CGSV21,FGYZ22} & $q \ge \left(11/6 - \eps_0 \right)\Delta$~\cite{VIG99,CDM19} \\
\hline 
\end{tabular}
\caption{Known results for colorings}\label{table}
\end{table}
\normalsize

There has also been considerable progress in obtaining better bounds under additional assumptions, namely assuming a lower bound on the girth or working in the large degree setting. For triangle-free graphs, \cite{CGSV21,FGYZ22} recently showed that the Glauber dynamics mixes rapidly provided that $q \ge 1.76\Delta$ using the recently introduced \emph{spectral independence} approach. For graphs of girth at least $7$, Dyer et al. \cite{DFHV13} showed that $q \ge 1.489\Delta$ colors suffice. When the graph has girth at least $11$ and $\Delta = \Omega_{\eps}(\log n)$, Hayes and Vigoda~\cite{HV03} showed that $q \ge (1 + \eps)\Delta$ colors suffice. These results all point in the direction that as the girth becomes larger, it becomes easier to approach the conjectured bounds. 
Intuitively, this is because the distribution on colorings in a neighborhood around $v$ becomes easier to understand. When the maximum degree of a graph is logarithmic, one can even exploit concentration bounds, say, on the number of missing colors in the neighborhood of each vertex. We will be particularly interested in taking these arguments to their natural limit, motivated by the following question that arises from these previous works. {\em For what families of graphs can we get nearly optimal bounds, like $q \geq \Delta + O(1)$? } To the best of our knowledge, such bounds were only previously known for trees \cite{MSW07,SZ17}, which is an extremely special case, as one can count the number of colorings exactly via simple dynamic programming. 

Thus, our second main result is a general reduction from rapid mixing of Glauber dynamics to strong spatial mixing on local structures.
In particular we establish spectral independence on general graphs of maximum degree $\Delta$ and sufficiently large girth $\Omega_\Delta(1)$ whenever $q \ge \Delta + 3$, matching our strong spatial mixing result \cref{t:strong-ssm} on trees.
The spectral independence approach is a powerful framework for proving optimal (near-linear) mixing time of Glauber dynamics for many families of models; we refer to \cref{subsec:pre-SI} for further background on spectral independence.   
In turn, this implies $O_{\Delta}(n \log n)$ mixing time bounds for the associated Glauber dynamics. Thus we essentially resolve~\cref{c:color-mix} for graphs of large girth.

\begin{thm}\label{t:SI-girth}
For every $\Delta \ge 3$ there exists $\girth>0$ such that, for any graph $G$ of maximum degree at most $\Delta$ and girth at least $\girth$, spectral independence holds for random $q$-colorings of $G$ with constant $O_{\Delta}(1)$ if $q \ge \Delta+3$. 
Furthermore, under the same assumptions the Glauber dynamics for sampling random $q$-colorings mixes in $O_{\Delta}(n \log n)$ time. 
\end{thm}
\begin{rem}
How large does the girth need to be? If we relax our goal, we can get by with relatively mild lower bounds in terms of the dependence on $\Delta$. In particular when $q \ge (1 + \eps)\Delta$ we only need the girth to be at least $C(\eps) \log^2 \Delta$ for some constant $C(\eps)$ that depends on $\eps$; see \cref{rem:(1+eps)Delta} for details. Thus, our bounds are meaningful even in relatively high-degree settings. 
\end{rem}

To put our reduction in context, in the setting of $2$-spin systems (e.g.\ Ising model or hardcore model) there is a generic reduction from sampling on general graphs to SSM on trees. In particular Weitz~\cite{Wei06} and Scott--Sokal \cite{SS05} showed how the marginal distribution at some vertex can be exactly computed by solving the same problem on a tree of self-avoiding walks with a carefully constructed boundary condition. 
For multi-spin systems like colorings, where $q > 2$, unfortunately the picture is significantly more complicated. There are approaches based on the {\em computation tree} \cite{GK12}. However the reduction from general graphs results in trees with branching factor $(\Delta - 1)q$ instead of $\Delta-1$. 
Na\"ive attempts to reduce the branching factor at best can reduce it to $(\Delta-1)(q-1)$, which is still much too large.
Because of the increase in the degree, even tight bounds for trees do not translate into tight bounds for bounded-degree graphs. Even some of the results we expect to carry over from trees either require many new ideas, or are still unknown. For example, \cite{LSS19,CGSV21,FGYZ22} showed only recently that there are efficient algorithms for sampling colorings on triangle-free graphs when $q \ge 1.76\Delta$. 
And for the previous record for SSM on trees which required $q \ge 1.59\Delta$ \cite{EGHSV19}, it seems challenging to translate it into mixing time bounds for the Glauber dynamics on bounded-degree graphs. Nevertheless~\cref{t:SI-girth} demonstrates how approximately tight guarantees for high-girth graphs can be deduced from SSM on trees.

\subsection{Weak spatial mixing for Potts model on trees}
A similar but weaker notion of correlation decay is known as weak spatial mixing. 
It is defined as follows.

\begin{defn}[Weak Spatial Mixing]
The uniform distribution $\mu$ on proper $q$-colorings of a graph $G=(V,E)$ satisfies \emph{weak spatial mixing (WSM) with exponential decay rate $1-\delta$ and constant $C$} if for every $v \in V$, every subset $\Lambda \subseteq V$, and every pair of pinnings $\sigma,\tau: \Lambda \to [q]$, we have the inequality
\begin{align*}
    \norm{\mu_{v}^{\sigma} - \mu_{v}^{\tau}}_{\TV} \leq C (1-\delta)^{K},
\end{align*}
where $\mu^\sigma_v$ (resp., $\mu^\tau_v$) denotes the marginal distribution on the color of vertex $v$ conditional on $\sigma$ (resp., $\tau$) and $K = \dist_G(v,\Lambda)$ denotes the graph distance between $v$ and $\Lambda$.
\end{defn}
Informally, the above says that the combined influence of vertices far away from a given central vertex $v$ is exponentially small in their distance. This influence should be small \emph{regardless} of the size of the ``boundary'' $\Lambda$.  Notice that for WSM, if it holds, the marginal distribution at $v$ converges to uniform as the boundary becomes farther away. 
From a physics perspective, WSM is interesting because it implies \emph{uniqueness} of the Gibbs measure when the size of the graph $|V|$ goes to infinity. Of particular relevance is the uniqueness of the Gibbs measure on \emph{infinite $\Delta$-regular trees} (or \emph{Bethe lattices}).\footnote{While we will not rigorously define what a Gibbs measure on an infinite graph means, this is not important for this paper. Roughly, it means that the conditional marginal distribution induced on any finite subset of vertices under any boundary condition is the Gibbs measure on that finite induced subgraph.} 

Using similar methods to those used to obtain~\cref{t:strong-ssm}, we are also able to obtain improved and nearly optimal bounds for WSM in the more permissive setting of the \textit{antiferromagnetic Potts model}, a Gibbs distribution on $q$-spin systems where adjacent vertices are allowed to have the same color at the cost of some penalty, controlled by inverse temperature $\beta$ (see~\cref{sec:antiferro-wsm} for a precise definition). It is widely believed that there is a sharp threshold, that is the natural analogue of \cref{c:coloring-folklore}, for WSM and SSM in the antiferromagnetic Potts model~\cite{PdLM83,PdLM87,JON02,SST14,GGY18,dBBR23,BBR22}.

\begin{conj}
\label{conjecture:antiferro-Potts-wsm}
Let $q \geq 2$ be an integer and consider the $q$-state antiferromagnetic Potts model with inverse temperature $\beta \in [0,1]$. 
Then both weak and strong spatial mixing hold for all graphs of maximum degree $\Delta$ if and only if $\beta \geq \beta_{c} \defeq \max\left\{0, 1 - \frac{q}{\Delta}\right\}$ (with strict inequality if $q = \Delta$). In particular, there is a unique Gibbs measure on the infinite $\Delta$-regular tree if and only if $\beta \geq \beta_{c}$ (again, with strict inequality if $q = \Delta$).
\end{conj}

In one direction~\cite{GSV15} showed that approximately sampling from the antiferromagnetic Potts model is hard for even $q < (1- \beta)\Delta$ and that $q = (1 -\beta)\Delta$ is the threshold for the uniqueness of semi-translation invariant Gibbs measures on an infinite $\Delta$-regular tree. Towards understanding larger $q$, Yin and Zhang~\cite{YZ15} established SSM for graphs of maximum degree $\Delta$ whenever $q \ge 3(1-\beta)\Delta + O(1)$. Substantial work has gone into studying the antiferromagnetic Potts models, including several works studying specific graphs such as infinite lattices.

In a recent breakthrough Bencs, de Boer, Buys, and Regts~\cite{BBR22} established WSM and uniqueness of the Gibbs measure for the antiferromagnetic Potts model on the infinite $\Delta$-regular tree down to the conjectured bounds. Unfortunately they require a sufficiently large lower bound on the degree $\Delta$ which depends nontrivially on $q$. To the best of our knowledge, their bounds are ineffective for small $\Delta$ because their argument relies on studying the tree recursion in the $\Delta \to \infty$ limit, and then pulling back to finite $\Delta$ via a compactness argument.

Our third result is to prove WSM and uniqueness of the Gibbs measure almost down to the conjectured threshold for all $q$ and $\Delta$. Thus we bypass the assumption that the degree is sufficiently large. Moreover our result also strengthens earlier works by removing the restriction that the tree must be regular. This result further highlights the flexibility of the techniques used to show~\cref{t:strong-ssm}.

\begin{thm}\label{thm:wsm-potts}
Let $T$ be an arbitrary tree of maximum degree $\Delta$. The Gibbs distribution $\mu$ of the $q$-state antiferromagnetic Potts model on $T$ with inverse temperature $\beta \in [0,1]$ exhibits weak spatial mixing with exponential decay rate whenever $q \geq (1 - \beta)(\Delta + 5) + 1$.
\end{thm}
We hopefully expect further refinements and optimizations to our techniques can establish strong spatial mixing on trees roughly or even exactly down to the precise threshold $q \geq (1-\beta)\Delta$.

\subsection{Organization of the paper}

We begin in~\cref{sec:outline} by giving a technical overview of the arguments that go into proving our main results: \cref{t:strong-ssm,t:SI-girth}. We then proceed with some preliminaries on colorings in~\cref{sec:prelim} that we use immediately in~\cref{sec:colorings-simple} to show a weaker version of~\cref{t:strong-ssm}. Specifically, we prove in~\cref{t:weak-ssm} that strong spatial mixing holds for colorings on trees when $q \ge \Delta + 3\sqrt{\Delta}$, giving a short proof that contains the key ideas of our main result but has minimal calculations. We defer the proof of~\cref{t:strong-ssm} to~\cref{sec:colorings-weighted}.

In \cref{s:girth}, we prove our second main result,~\cref{t:SI-girth}, by giving a generic reduction from sampling on large-girth graphs to strong spatial mixing on trees, introducing spectral independence formally and a related notion of influence decay along the way. We then turn our attention to the antiferromagnetic Potts model, establishing weak spatial mixing on trees in~\cref{sec:antiferro-wsm}. Throughout, we defer verification of a few technical claims to the appendices.

Rather than directly proving the main theorems,~\cref{sec:colorings-simple,sec:colorings-weighted,sec:antiferro-wsm} all instead prove intermediate results about \textit{contraction} of the Jacobian of a matrix associated to the distribution, the primary novelty in this work. Deriving strong (resp.\ weak) spatial mixing, spectral independence, and influence decay from such contraction-type results is relatively standard, and thus we leave the formal verification of the statements of~\cref{t:weak-ssm,t:strong-ssm,thm:wsm-potts} to~\cref{sec:condition}, articulating a black-box property called the \textit{Jacobian norm condition} that we show implies the aforementioned desired structural properties of an associated Gibbs distribution. Subsequently, we briefly conclude with some open questions in~\cref{sec:conclusion}.
\newline \newline 
\textbf{Acknowledgements.} 
We thank Ferenc Bencs, Khallil Berrekkal, Guus Regts and Xiaoyu Chen, Xinyuan Zhang for pointing out errors in earlier versions of this paper. 
K.L. was supported by NSF-2022448. N.M. was supported by an Hertz Graduate Fellowship and the NSF GRFP. A.M. was supported by a Microsoft Trustworthy AI Grant, an ONR Grant and a David and Lucile Packard Fellowship.

\section{Proof outline}\label{sec:outline}

In this section, we give an overview of our techniques. There are two main steps:
\begin{enumerate}
    \item\label{outline:item:trees} \textbf{Contraction on trees:} It is well known that the marginal distributions of vertices in trees can be computed recursively up from the leaves via the \emph{tree recursion} (see \cref{d:tree-recurse-coloring}), and that a contractive property of this recursion implies exponential decay of correlation, i.e.\ weak/strong spatial mixing. A major challenge in the context of colorings is to find a suitable metric with which to measure the contraction. Many natural approaches immediately break down for $q$ below $2\Delta$, which is the threshold at which simple coupling arguments (e.g.\ Dobrushin) stop working.
    
    Our approach is based on a combination of a good choice of a univariate \textit{potential function} $\phi$, which we apply entry-wise, and a carefully designed \emph{weighted $L^{2}$-metric}. Although some of these ingredients have appeared separately in previous works studying SSM and related problems, what turns out to be crucial is the combination of a metric and potential that work in concert to amplify the natural ``anti-correlation'' between a vertex's marginal distribution and those of its children.
    \item\label{outline:item:reduction} \textbf{From trees to large-girth graphs}: To establish spectral independence beyond trees, we give a \emph{generic reduction} from graphs with sufficiently large girth to trees. We emphasize this reduction extends beyond the setting of proper $q$-colorings, and is of independent interest. We are motivated by the intuition that if a graph has girth $\girth$, then the depth-$(\girth/2)$ neighborhood of any vertex is a tree, where we can hope to leverage our SSM results on trees. 
    
    However, na\"{i}ve attempts to implement this idea require $\girth = \Omega(\log n)$, which is extremely restrictive. Instead, we introduce a new ``local coupling'' procedure which not only allows us to break the $\Omega(\log n)$ barrier, but completely removes the dependence on $n$ altogether. In the end, our girth lower bound only has a mild dependence on the maximum degree.
\end{enumerate}

Establishing spectral independence via contraction of the tree recursion had been previously carried out for many $2$-spin models \cite{ALO20,CLV20,CLV21,CFYZ22}. However, as we discussed, there are serious complications that arise in the setting of multi-spin systems. 

\subsection{Contraction on trees via the potential method}
To show WSM or SSM on trees, a natural strategy is to establish a \emph{contractive property} for the tree recursion. We will use $g$ to denote the function that maps from the marginals of the children to the induced marginal on the parent (see \cref{d:tree-recurse-coloring}). Thus we want to show that for every vertex $r \in V$ with children $u_{1},\dots,u_{d}$ on a (rooted) tree $T=(V,E)$, we have
\begin{align*}
    d(\bm{p}_{r}, \bm{p}_{r}') = d(g(\bm{p}_{1},\dots,\bm{p}_{d}), g(\bm{p}_{1}',\dots,\bm{p}_{d}')) \leq (1-\delta) \cdot \max_{i \in [d]} d(\bm{p}_{i}, \bm{p}_{i}'),
\end{align*}
where $d(\cdot,\cdot)$ is some metric on probability measures on $[q]$. Here, for each $i\in[d]$, $\bm{p}_{i}$ denotes the marginal distribution $\mu_{u_{i}}^{\tau}$ in the subtree $T_{u_{i}}$ rooted at $u_i$ conditioned on some boundary pinning $\tau : \Lambda \to [q]$, and $\bm{p}_{i}'$ denotes the analogous conditional marginal distribution but with respect to some other boundary pinning $\tau' : \Lambda \to [q]$. By iterating this inequality inductively, we get that $d(\bm{p}_{r},\bm{p}_{r}') \leq (1-\delta)^{\ell}$ where $\ell$ is the distance from $r$ to $\Lambda$. We would be immediately done if $d(\cdot,\cdot)$ could be taken to be the total variation distance. Unfortunately, in general, one-step contraction for total variation distance fails.

A natural strategy is to use an alternative distance metric $d(\cdot,\cdot)$, paying a multiplicative constant when converting to and from total variation. In general, designing such metrics is ad hoc, at least as of this writing. A natural recipe \cite{LY13,GKM15,YZ15} for constructing such metrics is to use a univariate monotone \emph{potential function} $\phi : \R \to \R$ and take $d(\cdot,\cdot)$ to be an appropriate $L^{p}$-like distance after mapping each marginal distribution $\bm{p}_{i}$ to $\bm{m}_{i} = \phi(\bm{p}_{i})$; here, $\phi$ is applied entrywise. Each ``message'' $\bm{m}_{i}$ is then passed up the tree via the \emph{modified tree recursion}
\begin{align*}
    \bm{m}_{r} = g^{\phi}(\bm{m}_{1},\dots,\bm{m}_{d}) \defeq \phi(g(\phi^{-1}(\bm{m}_{1}),\dots,\phi^{-1}(\bm{m}_{d}))).
\end{align*}
Establishing contraction then becomes a matter of showing that an appropriate matrix norm of the \emph{Jacobian} of $g^{\phi}$ is strictly less than one. This is formalized in~\cref{cond:jacobian-norm}, which we call the \textit{strong (resp.~weak) Jacobian norm condition}. It is not hard to show via standard techniques that such contraction, if it holds, implies both SSM (\cref{thm:ssm-from-contraction}) and spectral independence (\cref{thm:si-from-contraction}).

We will make use of a potential function  $\phi(p)$ considered by Lu and Yin~\cite{LY13} which is defined implicitly by its derivative $$\phi'(p) = \Phi(p) \defeq \frac{1}{\sqrt{p}(1-p)}.$$
Lu and Yin~\cite{LY13} used it to give a deterministic algorithm for approximately counting proper $q$-colorings that works whenever $q \geq 2.58\Delta$. The potential function was applied in conjunction with an $L^{\infty}$ distance, which seems necessary in terms of deriving bounds for general graphs via the computation tree approach. Our first key observation is that we can get much better mileage on trees by using a (weighted) $L^{2}$ distance instead. 

At this juncture, the potential function may seem mysterious. By using the chain rule we can compute the Jacobian of the modified  tree recursion $g^{\phi} = \phi \circ g \circ \phi^{-1}$ (see \cref{o:jacob-phi}). The key contributions to this Jacobian turn out to be diagonal matrices of the form
$$\diag \left(\sqrt{\p_i \odot \p_r}\right),$$
where $\bm{a} \odot \bm{b}$ denotes the entrywise product of two vectors, $\p_r$ denotes the marginal distribution of a vertex $r$, and $\p_i$ denotes the marginal distribution of the $i$th child of $r$.
The key point is that such a matrix accommodates the \emph{repulsive} nature of proper colorings. For example, in an extreme case, if $\p_i(\mfc) = 1$, meaning that child $i$ is always colored $\mfc$, then $\p_r(\mfc) = 0$, meaning that the parent is never colored $\mfc$. Thus we have good control the norm of a such a matrix in the extremes, and actually everywhere in between. In the regime where $q$ is close to $\Delta$ and we are allowed to pin the colors of some vertices, we expect large oscillations in the marginal distributions as one computes up the tree via the tree recursion. Thus, it is crucial to have bounds that control the matrices' norms, even when associated vertices have marginal distribution far from uniform.

We formalize this intuition, beginning by giving a short and simple proof of SSM for colorings on trees when $q \geq \Delta + 3\sqrt{\Delta}$ in \cref{sec:colorings-simple} using an unweighted $L^{2}$ distance and the above choice of potential. A more refined, weighted $L^{2}$ distance is used to get all the way down to $q \geq \Delta + 3$. This analysis appears in \cref{sec:colorings-weighted}. Our choice of weights falls out of studying why the simple $L^{2}$ analysis in \cref{sec:colorings-simple} fails to go below $\Delta + \Omega(\sqrt{\Delta})$.

Finally, in~\cref{sec:antiferro-wsm}, we prove~\cref{thm:wsm-potts} for the antiferromagnetic Potts model in an analogous manner. Here, we again consider a weighted $L^{2}$ norm, where the potential $\phi$ now has derivative $\phi'(p) = \Phi(p) \defeq \frac{1}{\sqrt{p}(1 - (1-\beta)p)}$. For technical reasons, we are only able to prove \emph{weak} spatial mixing, but we hopefully expect our strategy to also give nearly optimal bounds for SSM.

\subsection{Reduction from large-girth graphs to trees: a new local coupling}

Our approach to proving \cref{t:SI-girth} leverages \textit{spectral independence}, which is a powerful framework that can be used to show optimal (near-linear) convergence of Glauber dynamics and has been successfully applied to many classes of spin systems. 
Moreover, it is closely related to SSM in the sense that many approaches for showing SSM can be used to show spectral independence too. 
We will follow a similar gameplan here.

For simplicity, in this overview, we will describe our argument in the setting of the Gibbs distribution $\mu$ without any pinned vertices.
An informal version of spectral independence can be described as follows: 
For every vertex $u \in V$ and every pair of distinct colors $\mfb,\mfc \in [q]$, one can construct a coupling $\mathcal{C}$ between the conditional Gibbs measures $\mu^{u \gets \mfb}$ and $\mu^{u \gets \mfc}$ such that 
\begin{equation}\label{eq:outline-goal}
    W_1\left( \mu^{u \gets \mfb},\mu^{u \gets \mfc} \right) 
    \le \E_{(X,Y)\sim \mathcal{C}} \left[ \dH\left( X,Y \right) \right] = O(1),
\end{equation}
where $\dH$ denotes the Hamming distance between two colorings, and $W_1$ denotes the 1-Wasserstein distance between two distributions. 
In words, the expected number of disagreements between $X \sim \mu^{u \gets \mfb}$ and $Y \sim \mu^{u \gets \mfc}$ under the coupling is at most some constant independent of the size of the graph.
We remark that similar ``local coupling'' procedures have been used to great effect in other settings such as for solutions of constraint satisfaction problems \cite{Moi19,CMM23,GGGH22}, the ferromagnetic Ising model \cite{CZ23}, and colorings of triangle-free graphs \cite{GMP05,CGSV21,FGYZ22} and general graphs \cite{BCCPSV22,Liu21}. Our local coupling procedure also bears some resemblance to the local sampling algorithm of Anand--Jerrum \cite{AJ22,AJ23}.

We give a new iterative construction of such a coupling. In each step, we carefully select a new vertex $v \neq u$ which has not already been colored, and optimally couple the marginal distribution of $v$ conditioned on all previously determined color assignments. The major challenge one faces is ensuring that the disagreements \emph{do not proliferate out of control}. In particular, one must ensure that the conditional marginal distributions remain very close. Obtaining such refined control is where we use a girth lower bound and SSM on trees.

Let $R > 0$ be some constant to be specified later, and let $S(u,R)$ denote the set of all vertices at distance \textit{exactly} $R$ away from $u$. 
We iterate through all vertices in $S(u,R)$ one by one in a random ordering, and in each step reveal the colors of one vertex from $S(u,R)$ in both $X$ and $Y$, under an optimal coupling conditioned on all previously revealed colors.

\begin{itemize}
    \item Good case: If we manage to successfully couple all vertices in $S(u,R)$ at distance $R$, then all disagreements are necessarily contained within the ball of radius $R$ centered at $u$, of which there can be at most $\Delta^R = O(1)$ many disagreements.
    \item Bad case: If we fail to couple at some vertex $v \in S(u,R)$, say $X(v) = \mfb' \neq \mfc' = Y(v)$, then we stop the coupling procedure immediately. We then ``restart'' two coupling procedures at both $u$ and the new vertex $v$ of disagreement to construct a coupling $\mathcal{C}$ between $\mu^{u \gets \mfb, v\gets\mfb'}$ and $\mu^{u \gets \mfc, v\gets\mfc'}$.
    \begin{itemize}
        \item Couple $\mu^{u \gets \mfb, v\gets\mfb'}$ and $\mu^{u \gets \mfc,v\gets\mfb'}$ recursively, where the disagreement is at $u$;
        \item Couple $\mu^{u\gets\mfc,v\gets\mfb'}$ and $\mu^{u\gets\mfc,v\gets\mfc'}$ recursively, where the disagreement is at $v$;
        \item Glue the above two couplings together to get $\mathcal{C}$.
    \end{itemize}
\end{itemize}
Note that the way we deal with the ``bad case'' is essentially by applying the triangle inequality for the Wasserstein distance:
\[
W_1\left( \mu^{u \gets \mfb,v\gets\mfb'},\mu^{u \gets \mfc,v\gets\mfc'} \right)
\le 
W_1\left( \mu^{u \gets \mfb,v\gets\mfb'},\mu^{u \gets \mfc,v\gets\mfb'} \right)
+
W_1\left( \mu^{u \gets \mfc,v\gets\mfb'},\mu^{u \gets \mfc,v\gets\mfc'} \right).
\]
We succeed because we encounter the ``good case'' with much higher probability, and therefore the number of disagreements will decay in expectation.
In particular, if $D$ is some upper bound we want to prove in \cref{eq:outline-goal} by induction, then the coupling procedure above informally gives:
\[
D \le \Pr\left(\text{good case}\right) \cdot \Delta^R 
+ \Pr\left(\text{bad case}\right) \cdot 2D. 
\]
In particular, we would be done if $\Pr\left(\text{bad case}\right) < 1/2$. 

Suppose the graph $G$ has girth $\girth$ which is much larger than $R$.
The depth-$(\girth/2-1)$ neighborhood of $u$ is a tree rooted at $u$, and contraction for the tree recursion allows us to say that the \textit{total influence} of $u$ on $S(u, R)$ is small assuming $R$ is sufficiently large. 
This means that regardless of how the colors $\mfb,\mfc \in [q]$ for $u$ are chosen, the conditional marginal distributions $\mu_{v}^{u \gets \mfb}, \mu_{v}^{u \gets \mfc}$ are close, relative to the size of the boundary of the depth-$R$ neighborhood.
From this fact we are able to prove $\Pr\left(\text{bad case}\right) < 1/2$.
We make this intuition precise in \cref{s:girth}.

\section{Preliminaries}\label{sec:prelim}
For a positive integer $n \geq 1$, we write $[n] = \{1,\dots,n\}$. In this paper, we only work with finite undirected simple graphs $G=(V,E)$. If $u,v \in V$ are two vertices in a graph $G=(V,E)$, we write $\dist_G(u,v)$ for the length of the shortest path between them in $G$ (i.e.\ the graph distance). Similarly, if $S \subseteq V$ and $v \in V$, we write $\dist_G(v,S) = \min\{\dist_G(u,v) : u \in S\}$ for the length of the shortest path from $v$ to any vertex in $S$. For every vertex $v \in V$, we write $\zcDeg(v) = \#\{u \in V : u \sim v\}$ for the \emph{degree} of $v$ in $G$. We write $\Delta \defeq \max_{v \in V} \zcDeg(v)$ for the \emph{maximum degree} of $G$.

If $T=(V,E)$ is a tree, we will often write distinguish a \textit{root vertex }$r \in V$. We call the pair $(T,r)$ a \emph{rooted tree}. If $(T,r)$ is a rooted tree, then every vertex $v \in V \setminus\{r\}$ has exactly one \emph{parent}, and exactly $\Delta_{v} \defeq \zcDeg(v) - 1 \leq \Delta - 1$ \emph{children}. In particular, if $T$ has maximum degree $\Delta$, then it has \emph{branching factor} at most $\Delta - 1$. For every $v \in T$, we write $T_{v} \subseteq T$ for the unique rooted \emph{subtree} of $T$ rooted at $v$; more precisely, $T_{v}$ is the induced subgraph of $T$ with respect to all descendants of $v$ (including $v$ itself).

For two probability measures $\mu,\nu$ on a common (for convenience, finite) state space $\Omega$, we write $\norm{\mu - \nu}_{\TV} = \frac{1}{2} \sum_{\omega \in \Omega} \abs{\mu(\omega) - \nu(\omega)} = \sup_{A \subseteq \Omega} \abs{\mu(A) - \nu(A)}$ for the \emph{total variation distance} between $\mu,\nu$. If $\mu,\nu$ are two probability distributions on (possibly different) finite state spaces $\Omega,\Lambda$ respectively, then a \emph{coupling} of $\mu,\nu$ is a joint distribution $\gamma$ on $\Omega \times \Lambda$ with $\mu,\nu$ as its marginals, i.e.\ $\mu(x) = \sum_{y \in \Lambda} \gamma(x,y)$ and $\nu(y) = \sum_{x \in \Omega} \gamma(x,y)$ for every $x \in \Omega, y \in \Lambda$.

If $\bm{u},\bm{v} \in \R_{\geq0}^{q}$ are two vectors of the same dimension, then we write $\bm{u} \odot \bm{v} \in \R_{\geq0}^{q}$ for their \emph{Hadamard product} (or \emph{entrywise product}), where $(\bm{u} \odot \bm{v})_{i} = \bm{u}_{i} \cdot \bm{v}_{i}$ for all $i \in [q]$. If $\bm{u} \in \R^{q}$ is a vector, we write $\diag(\bm{u}) \in \R^{q \times q}$ for the diagonal matrix with the entries of $\bm{u}$ on its diagonal. Linear algebraic and analytic lemmata are stated and proved in~\cref{sec:technical2}.

\subsection{List coloring}
We will need to handle sampling from the distribution of uniform $q$-colorings (or list colorings) conditioned on an assignment for a subset of vertices. Let $\Lambda \subseteq V$ be a subset of vertices. We say a partial assignment $\tau : \Lambda \to [q]$ is a \emph{(valid) pinning} if there exists $\sigma : V \to [q]$ with $\mu(\sigma) > 0$ such that $\sigma$ agrees with $\tau$ on $\Lambda$, i.e.\ $\sigma\mid_{\Lambda} = \tau$; in other words, $\tau$ is a valid pinning if and only if it can be extended to full configuration in the support of $\mu$.

Given a pinning $\tau : \Lambda \to [q]$ on a subset of vertices $\Lambda \subseteq V$, we may then write $\mu^{\tau}$ for the conditional Gibbs measure on $[q]^{V}$ given by $\mu^{\tau}(\sigma) \propto \mu(\sigma)$ if $\sigma\mid_{\Lambda} = \tau$ and $\mu^{\tau}(\sigma) = 0$ otherwise. In other words, we simply restrict attention to full configurations agreeing with $\tau$ on all of $\Lambda$. If $\Lambda$ is a singleton $\{v\}$, we often simply write $v \gets \mfc$ for the pinning $\tau : \Lambda \to [q]$ satisfying $\tau(v) = \mfc \in [q]$; similarly, we write $\mu^{v \gets \mfc}$. Our notation for the conditional marginal distributions extend naturally: If $\Lambda,S \subseteq V$ are disjoint subsets of vertices and $\tau : \Lambda \to [q]$ is a pinning on $\Lambda$, then we write $\mu_{S}^{\tau}$ for the marginal distribution over configurations on $S$ conditioned on $\tau$.

In the special case of graph colorings, where $A = J - I$, the conditional Gibbs measures may be described more combinatorially as follows: For each vertex $v \in V$, let $\mcL_{v} \subseteq [q]$ denote \emph{list} of colors contained in the global color palette $[q]$, and denote $q_{v} = |\mcL_{v}|$ for convenience. We write $\mcL = \{\mcL_{v}\}_{v \in V}$, and call $(G,\mcL)$ a \emph{list-coloring instance}. This gives rise a conditional Gibbs measure $\mu$, where $\mu$ is uniform over assignments $\sigma:V \to [q]$ simultaneously satisfying the coloring constraint $\sigma(u) \neq \sigma(v)$ for $u \sim v$, as well as the constraint $\sigma(v) \in \mcL_{v}$ for all $v \in V$. Intuitively, one should imagine each $\mcL_{v}$ as being obtained by pinning some neighbors of $v$ to each of the colors in $[q] \setminus \mcL_{v}$, thereby disallowing $v$ itself to be assigned those colors.

In our arguments, we will need some control over the marginal distributions of vertices. This motivates the following convenient piece of notation.
\begin{defn}[Bounded Subdistributions]\label{def:bdd-subdistr}
Given a finite set $\Omega$ and a real number $\ub \in [0,1]$, let
\begin{align*}
    \mcD(\Omega,\ub) \defeq \wrapc{\bm{p} \in [0,\ub]^{\Omega} : \sum_{x \in \Omega} \bm{p}(x) \leq 1}
\end{align*}
denote the space of all subdistributions $\bm{p}$ with entries upper bounded by $\ub$. We will write $\mcD(\Omega)$ for the space of all subdistributions $\mcD(\Omega,1)$. Similarly, if $q \in \N$, we write $\mcD(q,\ub)$ for $\mcD([q], \ub)$.
\end{defn}
It turns out that in our proofs, we must handle subdistributions as opposed to merely distributions, due to reasons related to our use of potential functions. 

\subsection{Tree recursion and marginal bounds for list coloring}
In this subsection, we specialize to the uniform distribution on proper \textit{list colorings} on trees. Let $(T,\mcL)$ be a list coloring instance,
and recall the associated Gibbs distribution is given by 
$$\mu(\sigma) = \mu_{T, \mcL}(\sigma) \propto \prod_{v \in V} \ind\{\sigma(v) \in \mcL_{v}\} \prod_{uv \in E} \ind\{\sigma(u) \neq \sigma(v)\},$$
i.e.\ the uniform distribution on all valid proper list colorings of $T$ given vertex color palettes $\mcL.$
We let $\p_v(\mfc)$ be the probability under this Gibbs distribution that vertex $v \in V$ is assigned color $\mfc$ (which is zero if $\mfc \not \in \mcL_v$).

The following tree recursion is well-known \cite{GK12,GKM15,LY13}.
\begin{obs}[Tree recursion]\label{d:tree-recurse-coloring}
Consider list coloring instance $(T, \mcL)$ where $T$ is rooted at $r$; suppose $r$ has children $v_1, \ldots, v_d$ (for $d = d_{r}$) and associated palette $\mcL_{r}$. Given distributions $\p_{v_i} \in \mcD(\mcL_{v_i})$ on the colors of each $v_i$, we can compute the \textit{marginal distribution} of $r$ as follows.
$$\p_{r} = (g_{\mfc}(\p_1, \ldots, \p_d))_{\mfc \in \mcL_r} \in \mcD(\mcL_r),$$
where 
$$g_{\mfc}(\p_1, \ldots, \p_d) = \frac{\prod_{i \in [d]} (1 - \p_i(\mfc))}{\sum_{\mfc' \in \mcL_r} \prod_{i \in [d]} (1 - \p_i(\mfc'))},$$
where $\p_i \in \mcD(\mcL_r)$ is obtained from $\p_{v_i} \in \mcD(\mcL_{v_i})$ by only including entries from $\p_{v_i}$ corresponding to colors in $\mcL_r \cap \mcL_{v_i}$ and adding a zero for all colors in $\mcL_r \backslash \mcL_{v_i}$.
\end{obs}

Note that given a list coloring instance $(T, \mcL)$ and some $v \in V(T)$, the induced subtree rooted at $v$ yields an associated list coloring instance. We make the following simple observation about $\p_{v}(\mfc)$ for any $\mfc \in \mcL_v$.

\begin{fact}\label{l:marginal}
Consider a list coloring instance $(T, \mcL)$ where $T$ is rooted at $r$ and suppose that $r$ has children $v_1, \ldots, v_d$ for $d = d_r$ and associated color palette $\mcL_r$ of size $q_r \ge d_r + \gamma$. Then for all $\mfc \in \mcL_r$, whenever $\p_i \in \mcD(\mcL_r)$, we have that 
$$\p_r(\mfc) = g_{\mfc}(\p_1, \ldots, \p_d) \le \frac{1}{\gamma}.$$ 
\end{fact}

By iterating the tree recursion of~\cref{d:tree-recurse-coloring}, we obtain the following stronger bound.
\begin{lemma}\label{l:marginal-strong}
Consider a list coloring instance $(T, \mcL)$ where $T$ is rooted at $r$ and suppose that $r$ has children $v_1, \ldots, v_d$ for $d = d_r$. Suppose that for all $v \in V$, vertex $v$ has associated color palette $\mcL_v$ of size $q_v \ge d_v + \gamma$. Then for all $\mfc \in \mcL_r$, whenever $\p_i \in \mcD(\mcL_r, 1/\gamma)$, we have that 
$$\frac{\p_{r}(\mfc)}{1 - \p_{r}(\mfc)} = \frac{g_{\mfc}(\p_1, \ldots, \p_d)}{1 - g_{\mfc}(\p_1, \ldots, \p_d)}\le \frac{1}{q_r-1} \left(1 + \frac{1}{\gamma - 1} \right)^{\frac{\gamma d_r}{q_r-1}}.$$
\end{lemma}
We provide a proof in \cref{sec:pf-marg-bounds}. We will also find the following \textit{lower bound} on the marginal probabilities of colors helpful.
\begin{lemma}[Lemma 3~\cite{GKM15}] \label{lem:marginal-lb}
Consider a list coloring instance $(G, \mcL)$ on a graph $G$ where $q_v \ge d_v + \gamma$. Then for all $\mfc \in \mcL_{v}$, we have that 
$$\p_{v}(\mfc) \ge \frac{1}{q} \left(1 - \frac{1}{\gamma} \right)^{d_v}.$$
\end{lemma}

\subsection{The Jacobian of the tree recursion}
Let $d,q \geq 1$ be positive integers, and let $g:(\R_{\geq0}^{q})^{d} \to \R_{\geq0}^{q}$ be a multivariate map taking as inputs $\bm{x} \in (\R_{\geq0}^{q})^{d}$ (for example, we will be interested in $g = (g_{\mfc}(\p))_{c \in [q]}$). 
Motivated by this choice of $g$, we will index the entries of $\bm{x}$ by pairs $(i,\mfb)$, where $i \in [d]$ and $\mfb \in [q]$. Recall the \emph{Jacobian} of $g$ is defined as the matrix $(Jg)(\bm{x}) \in \R^{q \times dq}$ of partial derivatives, with entries
\begin{align*}
    (Jg)_{\mfc, (i,\mfb)}(\bm{x}) \defeq (\partial_{\bm{x}_{i,\mfb}} g_{\mfc})(\bm{x}), \qquad \forall i \in [d], \mfb,\mfc \in [q].
\end{align*}
It will also be convenient to stratify $Jg$ by $i \in [d]$. More specifically, for every $i \in [d]$, define $(J_{i}g)(\bm{x})$ to be the $q \times q$ matrix with entries
\begin{align*}
    (J_{i}g)_{\mfb, \mfc}(\bm{x}) \defeq (\partial_{\bm{x}_{i,\mfb}} g_{\mfc})(\bm{x}), \qquad \forall \mfb,\mfc \in [q].
\end{align*}
One may view $(Jg)(\bm{x})$ as the concatenation of the $(J_{i}g)(\bm{x})$ over all $i \in [d]$. 

As noted above, we will be interested in the cases where $g = (g_{\mfc}(\p))_{\mfc \in [q]}$ is the tree recursion of a spin system, or $g = g^{\phi}$, where for smooth univariate $\phi: [0, 1] \to \RR_{\ge 0}$ with smooth inverse, we define $g^{\phi}$ entrywise via
\begin{equation}\label{eq:gphi}
    g_{\mfc}^{\phi}(\bm{x}) \defeq \phi(g_{\mfc}(\phi^{-1}(\bm{x}_{1}),\dots,\phi^{-1}(\bm{x}_{d}))), \qquad \forall \mfc \in [q].
\end{equation}

We make a few observations about the Jacobian of the above class of functions $g$. We formally verify these in~\cref{sec:technical2}.
\begin{obs}\label{o:jacob-plain}
Let $g =(g_{\mfc}(\p))_{\mfc \in [q]}$, for $g_{\mfc}(\p)$ defined as in~\cref{d:tree-recurse-coloring}. Then,
$$(J_i g)(\p) = (g(\p) g(\p)^{\top} - \diag(g(\p))) \diag(1 - \p_i)^{-1} \in \RR^{q \times q}.$$
\end{obs}

By applying the Chain Rule and the Inverse Function Theorem, we have the following consequence for the Jacobian of $g^{\phi}$.
\begin{obs}\label{o:jacob-phi}
Let $\phi: [0, 1] \rightarrow \RR_{\ge 0}$ be a smooth univariate function with smooth inverse and derivative $\Phi$, and take $g = (g_{\mfc}(\p))_{\mfc \in [q]}$, for $g_{\mfc}(\p)$ defined as in~\cref{d:tree-recurse-coloring}. If $\p = \phi^{-1}(\y)$, then 
$$(J_i g^{\phi})(\y) = \diag(\Phi(g(\p))) \left(g(\p) g(\p)^{\top} - \diag(g(\p))\right) \diag(1 - \p_i)^{-1} \diag(\Phi(\p_i))^{-1}.$$
\end{obs}

\section{Strong spatial mixing for colorings on trees when $q \ge \Delta + 3\sqrt{\Delta}$}\label{sec:colorings-simple}
As an illustration of our methods, we begin by proving a slightly weaker result:~we show that strong spatial mixing holds for the uniform distribution on proper $q$-colorings on trees of maximum degree $\Delta$ provided $q \ge \Delta + 3\sqrt{\Delta}$. The argument below contains the same essential ingredients as our stronger main result,~\cref{t:strong-ssm}, but avoids the technical complications that arise in proving the tighter bound.

\begin{prop}\label{t:weak-ssm}
For all positive integers $\Delta \ge 3$ and $q \ge \Delta + 3\sqrt{\Delta}$, there exists $\delta, C > 0$ such that the uniform distribution on $q$-colorings exhibits strong spatial mixing with exponential decay rate $1 - \delta$ and constant $C$ on any tree $T$ with maximum degree $\Delta$. 
\end{prop}

Consider an arbitrary rooted tree $T = (V,E)$ of maximum degree $\Delta$. 
Let $\Lambda \subseteq V$ be a subset of vertices and $\tau$ be an arbitrary pinning on $\Lambda$. 
This induces the following instances of list coloring: for a free vertex $v \in V \setminus \Lambda$, we consider the tree recursion at $v$ in the subtree $T_v$ rooted at $v$.
We use $\Delta_v$ to denote the total number of children of $v$ and $d_v$ to denote the number of free (unpinned) children of $v$. 
Also let $\zcQ_v \subseteq [q]$ be the list of available colors for $v$ under the pinning $\tau$, and let $q_v = |\zcQ_v|$. 

As described in~\cref{sec:outline}, our strategy will be to define a potential function and matrix norm under which our message recursion contracts.
We first define a suitable matrix norm for the recursion at $v$.

\begin{defn} \label{d:norm-star}
Define vector norm $\| \cdot \|_\star$ on $\RR^{\Delta_v q}$ as
$$\|\x\|_\star := \max_{i \in [\Delta_v]} \|\x_i\|_2,$$
where each $\x_i \in \RR^q$. 
For $M \in \RR^{q \times \Delta_v q}$ define matrix norm 
$$\| M \|_{\star\star} := \sup_{\x \in \RR^{\Delta_v q}} \frac{\|M\x\|_2}{\|\x\|_\star}.$$
\end{defn}

The above instantiations of $\|\cdot\|_\star$ and $\|\cdot\|_{\star\star}$ are well defined. 
We will bound the Jacobian of message recursion $g^{\phi}$ (see~\cref{eq:gphi}), where we choose $\phi = 2 \arctanh(\sqrt{x})$ so that 
\begin{equation}\label{eq:phi-color}
    \Phi(x) := \phi'(x) = \frac{1}{\sqrt{x}(1-x)},
\end{equation}
and
$\phi^{-1}(y) = \left( \frac{e^y-1}{e^y+1} \right)^2$. Note that $\phi$ is smooth with smooth inverse and derivative.
This is the same potential function used in the previous work of Lu and Yin~\cite{LY13}; they used a potential analysis of correlation decay on a computation tree to give an FPTAS for counting $q$-colorings in maximum degree $\Delta$ graphs for $q \ge 2.58 \Delta + 1$. We give some further intuition for this choice of potential in our analysis in the following proof.

We prove that under such choices of matrix norm and potential function, the Jacobian of our message recursion is strictly less than one when $q \ge \Delta + 3\sqrt{\Delta}$. 
\begin{prop}[Jacobian bound]\label{p:weak}
Consider $\phi$ and matrix norm $\|\cdot\|_{\star\star}$ as above. 
If $q \ge \Delta + 3\sqrt{\Delta}$, then there exists $\delta > 0$ such that for any vertex $v$, for any $\p = (\p_1, \ldots, \p_{\Delta_v})$ with $\p_i \in \mcD(q,\frac{1}{q-\Delta})$ (free children) or $\p_i = \ind_{\mfc}$ for some $\mfc \in [q]$ (pinned children), and for $\y = \phi(\p)$, we have that $g(\p) \in \mcD(q,\frac{1}{q-\Delta})$ and
$$\left\| \left( (J_i g^{\phi})(\y) \right)_{i \in [\Delta_v]} \right\|_{\star\star} \le 1 - \delta.$$
\end{prop}

\begin{rem}
The choice of $\frac{1}{q-\Delta}$ for the upper bound on entries of $\p_i$ for each free child is given by \cref{l:marginal}. 
In particular, $g(\p) \in \mcD(q,\frac{1}{q-\Delta})$ also follows from \cref{l:marginal}. 
\end{rem}

Once we have a bound on the Jacobian less than one, it then becomes relatively easy to deduce strong spatial mixing from it.
We prove \cref{t:weak-ssm} from \cref{p:weak} in \cref{sec:condition}, where we give general reductions from notions of correlation decay to the Jacobian bound, following the same path as in many previous works \cite{LLY13,LY13,CLV20,CFYZ22}. 
Our bound on the Jacobian yields some $\delta > 0$ for which the \textit{strong Jacobian condition} of~\cref{cond:jacobian-norm} holds; we will then be able to conclude strong spatial mixing via~\cref{thm:ssm-from-contraction} and deduce two closely related notions, \emph{spectral independence} and \emph{total influence decay}, via~\cref{thm:si-from-contraction}.
The notion of total influence decay is especially important for us to prove our main algorithmic result \cref{t:SI-girth} in \cref{s:girth}.

\begin{proof}[Proof of \cref{p:weak}]
Note that for $\p = (\p_1, \ldots ,\p_{\Delta_v}) = \phi^{-1}(\y) \in \RR^{\Delta_v q}$, fixed $i \in [\Delta_v]$, and $\Phi$ as in~\cref{eq:phi-color}, we have by~\cref{o:jacob-phi} that
\begin{align*}
(J_i g^{\phi}&)(\y) = \diag(\Phi(g(\p))) (J_i g)(\p) \diag(\Phi(\p_i))^{-1} \\
&= \diag \left( \frac{1}{\sqrt{g(\p)}(1 - g(\p))} \right) \left(g(\p) g(\p)^{\top} - \diag(g(\p))\right) \diag(1 - \p_i)^{-1} \diag \left(\frac{1}{\sqrt{\p_i}(1 - \p_i)}\right)^{-1} \\
&= \diag \underbrace{\left(1 - g(\p) \right)^{-1}}_{(1)} \underbrace{\left(\sqrt{g(\p)} \sqrt{g(\p)}^{\top} - I\right)}_{(2)} \underbrace{\diag \left(\sqrt{\p_i \odot g(\p)}\right)}_{(3)}.
\end{align*}
Notice that already, this choice of potential seems promising, as it simplifies the Jacobian into the product of three terms that we might hope to have control on:
\begin{enumerate}[(1)]
    \item We can use upper bounds on the marginal probabilities as in~\cref{l:marginal} to show that the matrix norm is not too large.
    \item This matrix has $L^2$ norm $1$ since $I-u u^{\top}$ is a projection matrix for any unit vector $u$.
    \item This diagonal matrix has entries of the form $\p_{i}(\mfc) g_{\mfc}(\p)$; note that these two quantities are negatively correlated (in the extreme setting, if e.g.\ $\p_i(\mfc) = 1$, then $g_{\mfc}(\p) = 0$), and thus we can hope that the norm is not too large.
\end{enumerate}
Thus, the choice of $\Phi$ allows us to group up pairs of negatively correlated terms with minimal additional error factors.
To simplify notation, we also make the following observations:
\begin{enumerate}[(i)]
    \item If $\p_i = \ind_\mfc$ for some $\mfc \in [q]$, then $(J_i g^{\phi})(\y) = 0$. Namely, if the child $i$ is pinned, then the Jacobian matrix with respect to the child $i$ is zero.
    \item If $\mfc \notin \zcQ_v$, then $\frac{\partial g^{\phi}_\mfc}{\partial \y_{i,\mfb}} (\y) = \frac{\partial g^{\phi}_\mfb}{\partial \y_{i,\mfc}} (\y) = 0$ for all $i \in [\Delta_v]$ and $\mfb \in [q]$. Namely, if the color $\mfc$ is not available for $v$ (because it is used by a pinned child of $v$), then all rows and columns associated with $\mfc$ are zero.
\end{enumerate}
By the definition of our matrix norm $\|\cdot\|_{\star\star}$, it suffices to consider only the non-zero block of the Jacobian matrix. 
Thus, with slight abuse of notation, we consider the submatrix denoted by $\left( (J_i g^{\phi})(\y) \right)_{i \in [d]}$ where $d = d_v$ is the number of free children, and each $(J_i g^{\phi})(\y) \in \RR^{q \times q}$ with $q = q_v = |\zcQ_v|$ being the number of available colors. Note that $q \ge d + 3\sqrt{d}$ by our assumption.

We thus proceed with computing $\left\| \left( (J_i g^{\phi})(\y) \right)_{i \in [d]} \right\|_{\star\star},$ letting $\gamma := q - d \ge 3\sqrt{d}.$
For any $\x_1, \ldots, \x_d \in \RR^q$, 
\begin{align*}
\left\| \sum_{i \in [d]} (J_i g^\phi)(\y) \x_i \right\|_2 
&= \left\|\diag \left(1 - g(\p) \right)^{-1} \left(\sqrt{g(\p)} \sqrt{g(\p)}^{\top} - I\right)  \sum_{i \in [d]}  \diag \left(\sqrt{\p_i \odot g(\p)}\right) \x_i \right\|_2 \\
& \le \left\|\diag \left(1 - g(\p) \right)^{-1}\right\|_2  \cdot \left\| \sqrt{g(\p)} \sqrt{g(\p)}^{\top} - I \right\|_2 \cdot  \left\| \sum_{i \in [d]}  \diag \left(\sqrt{\p_i \odot g(\p)}\right) \x_i \right\|_2 \\
& \le e^{\frac{1}{\gamma-1}} \left\| \sum_{i \in [d]}  \diag \left(\sqrt{\p_i \odot g(\p)}\right) \x_i \right\|_2,
\end{align*}
where the final inequality follows by noting that $g(\p) \le 1/\gamma$ by~\cref{l:marginal}, and thus we have that  $\left\|\diag \left(1 - g(\p) \right)^{-1}\right\|_2 \le \frac{\gamma}{\gamma-1} \le e^{\frac{1}{\gamma-1}}$, and by observing that $I - \sqrt{g(\p)} \sqrt{g(\p)}^{\top}$ is a projection matrix, which consequently satisfies $\left\| \sqrt{g(\p)} \sqrt{g(\p)}^{\top} - I \right\|_2  \le 1$. It thus remains to understand the last term that appears in the above bound. Using \cref{lem:norm-concat-diag}, which can be verified straightforwardly by the Cauchy--Schwarz inequality, we obtain
\begin{align*}
\left\| \sum_{i \in [d]}  \diag \left(\sqrt{\p_i \odot g(\p)}\right) \x_i \right\|_2^2 
&\le \max_{\mfc \in [q]} \left\{ g_\mfc(\p) \sum_{i \in [d]} \p_i(\mfc)\right\} \sum_{i \in [d]} \|\x_i\|_2^2 \\
&\le d \max_{\mfc \in [q]} \left\{ g_\mfc(\p) \sum_{i \in [d]} \p_i(\mfc)\right\} \|\x\|_\star^2.
\end{align*}
Therefore, we have
\begin{align}\label{eq:starstar}
\left\| \left( (J_i g^{\phi})(\y) \right)_{i \in [d]} \right\|_{\star\star}^2 = \sup_{\x_1, \ldots, \x_d \in \RR^q} \frac{\left\| \sum_{i \in [d]} (J_i g)(\y) \x_i \right\|_2^2}{\|\x\|_\star^2} 
\le d e^{\frac{2}{\gamma-1}} \max_{\mfc \in [q]} \left\{ g_\mfc(\p) \sum_{i \in [d]} \p_i(\mfc)\right\}.
\end{align}

We then apply the following bound.
\begin{claim}\label{l:tech1}
We have for all $\mfc \in [q]$ and $q = d + \gamma$ that
$$ g_\mfc(\p) \sum_{i \in [d]} \p_i(\mfc) \le  \frac{1}{q} \exp\left( -\frac{\gamma}{q} + \frac{1}{\gamma-1} \right).$$
\end{claim}
\begin{proof}
First we have that
\[
g_\mfc(\p) \sum_{i \in [d]} \p_i(\mfc) 
= \frac{\prod_{i \in [d]} (1 - \p_i(\mfc)) \sum_{i \in [d]} \p_i(\mfc)}{\sum_{\mfb \in [q]} \prod_{i \in [d]} (1 - \p_{i}(\mfb))}.
\]
By AM--GM, the numerator is at most
\[
\left( \frac{1}{d+1} \left( \sum_{i \in [d]} (1 - \p_i(\mfc)) + \sum_{i \in [d]} \p_i(\mfc) \right) \right)^{d+1}
= \left( 1-\frac{1}{d+1} \right)^{d+1} \le \frac{1}{e}.
\]
The denominator is bounded by the following sequence of inequalities:
\begin{align*}
 \frac{1}{q} \sum_{\mfb \in [q]} \prod_{i \in [d]} (1 - \p_{i}(\mfb)) 
 &\ge \left(\prod_{\mfb \in [q]} \prod_{i \in [d]} (1 - \p_{i}(\mfb)) \right)^{1/q} \tag{AM-GM} \\ 
 &\ge \exp\left( - \frac{1}{q} \sum_{i \in [d]}  \sum_{\mfb \in [q]} \frac{\p_{i}(\mfb)}{1 - \p_{i}(\mfb)}\right) \tag{$1 - x \ge e^{- \frac{x}{1-x}}$ for $x \in [0, 1]$} \\
 &\ge \exp\left( - \frac{1}{q} \left( 1+\frac{1}{\gamma-1} \right) \sum_{i \in [d]}  \sum_{\mfb \in [q]} \p_{i}(\mfb)\right) \tag{$\p_{i}(\mfb) \le \frac{1}{\gamma}$ by assumption}\\ 
 &\ge \exp\left( - \left( 1 - \frac{\gamma}{q} \right) \left( 1+\frac{1}{\gamma-1} \right) \right) \tag{$\sum_{\mfb \in [q]} \p_{i}(\mfb) \le 1$} \\
 &\ge \exp\left( - 1 + \frac{\gamma}{q} - \frac{1}{\gamma-1} \right).
\end{align*}
Combining the two bounds we obtain the claim.
\end{proof}
We can then combine our earlier inequality \cref{eq:starstar} with~\cref{l:tech1} to obtain:
\begin{align}\label{eq:3sqrtd}
\left\| \left( (J_i g^{\phi})(\y) \right)_{i \in [d]} \right\|_{\star\star}^2
\le \frac{d}{q} \exp\left(\frac{2}{\gamma-1}\right) \exp\left( -\frac{\gamma}{q} + \frac{1}{\gamma-1} \right)
\le \exp\left( -\frac{2\gamma}{q} + \frac{3}{\gamma-1} \right) < 1,
\end{align}
where the second inequality follows from $d/q = 1-\gamma/q \le e^{-\gamma/q}$ 
and the last inequality holds for $\gamma = q-d \ge 3\sqrt{d}$.
\end{proof}

In the course of the above proof, we also showed the following $L^2$ bound in more generality; we will use this as an ingredient in the proof of~\cref{t:strong-ssm} in~\cref{sec:colorings-weighted}.
\begin{cor}\label{c:l2-j-bound}
For any $\p = (\p_1, \ldots, \p_d)$ with $\p_i \in \mcD(q)$ and for $\y = \phi(\p)$, we have that
$$\left\| \left( (J_i g^{\phi})(\y) \right)_{i \in [d]} \right\|_2^2 
\le \max_{\mfc \in [q]} \left\{ \frac{1}{(1 - g_\mfc(\p))^2} \right\} \cdot
\max_{\mfc \in [q]} \left\{ g_\mfc(\p) \sum_{i \in [d]} \p_i(\mfc) \right\}.$$
\end{cor}

A consequence of our proof is that if $q = (1+\eps) \Delta$ for some small constant $\eps > 0$, we can take $\delta = \delta(\eps),$ i.e.\ we obtain strong spatial mixing with exponential decay rate $1 - \delta$ for $\delta$ that does not have a dependence on $\Delta$. 
See \cref{prop:(1+eps)Delta} for the full statement, which is based on the following corollary. 
\begin{cor}[Jacobian bound]
\label{coro:(1+eps)Delta}
If $q = (1 + \eps) \Delta$ for $\eps \in (0,1)$ and $\Delta \ge 7/\eps^2$, 
then for any vertex $v$, for any $\p = (\p_1, \ldots, \p_{\Delta_v})$ with $\p_i \in \mcD(q,\frac{1}{q-\Delta})$ (free children) or $\p_i = \ind_{\mfc}$ for some $\mfc \in [q]$ (pinned children), and for $\y = \phi(\p)$, we have that 
$$\left\| \left( (J_i g^{\phi})(\y) \right)_{i \in [\Delta_v]} \right\|_{\star\star} \le  \exp\left(-\frac{\eps}{4} \right) .$$
\end{cor}
\begin{proof}
Let $\gamma := q_v - d_v \ge q - \Delta_v \ge q - \Delta \ge \eps \Delta$. Then, we obtain from \cref{eq:3sqrtd} that
\begin{align*}
\left\| \left( (J_i g^{\phi})(\y) \right)_{i \in [d]} \right\|_{\star\star}^2 
\le \exp\left( -\frac{2\gamma}{q_v} + \frac{3}{\gamma-1} \right)
\le  \exp\left(-\frac{\eps}{2} \right),
\end{align*}
since $q_v \le q = (1+\eps)\Delta \le 2\Delta$ and $3/(\gamma-1) \le \eps /2$ when $\Delta \ge 7/\eps^2$.
\end{proof}

\begin{rem}
Our analysis in this section has a certain ``tightness'', exemplified by~\cref{ex:bad}. To get around this, we will modify our matrix norm to include vertex-neighbor specific weights that will allow us to achieve the full strength of~\cref{t:strong-ssm}. This argument follows in~\cref{sec:colorings-weighted}.
\end{rem}
 
\section{Sampling colorings on large-girth graphs}\label{s:girth}

In this section, we study the mixing time of the Glauber dynamics for sampling random colorings by establishing spectral independence, thereby proving~\cref{t:SI-girth}. 
We begin in \cref{subsec:pre-SI} some preliminaries about \textit{spectral independence} and its implications for fast mixing of the Glauber dynamics; we also define some related notions of mixing and influence decay. We proceed in \cref{subsec:proof-strategy-girth} to give an outline of our proof strategy for establishing spectral independence on large-girth graphs given sufficiently strong correlation decay properties on trees. 
In \cref{subsec:decay-infl,subsec:tree2girth} we prove the two main technical lemmas respectively and complete the proof. 

\subsection{Spectral independence and influence decay}
\label{subsec:pre-SI}

\subsubsection{General spin systems}
While our main result is stated in the language of sampling $q$-colorings, our reduction in this section is substantially more general; we actually prove that for a broad class of \textit{spin systems}, sufficiently strong correlation decay properties on trees imply spectral independence on large-girth graphs in a black-box manner. 

Thus, below we introduce general \emph{spin systems} and their associated \textit{Gibbs distributions.}
We consider spin systems on a graph $G = (V,E)$ where each vertex $v \in V$ receives a \emph{color} (or \emph{spin}) from a list $\zcQ_v$ associated with $v$. 
Colors are often denoted by lower case letters $\mfa,\mfb,\mfc$, etc.
Every edge $(u,v)$ is associated with an interaction function $A_{u,v}: \zcQ_u \times \zcQ_v \to \R_{\ge 0}$ which is symmetric. 
Every vertex $v$ is associated with an external field function $\lambda_v: \zcQ_v \to \R^+$. 
The Gibbs distribution $\mu$ is defined over the product space $\prod_{v \in V} \zcQ_v$, given by
\begin{align*}
    \mu(\sigma) \propto \prod_{(u,v) \in E} A_{u,v}(\sigma(u),\sigma(v)) \prod_{v \in V} \lambda_{v}(\sigma(v)), \qquad \forall \sigma : V \to \prod_{v \in V} \zcQ_v.
\end{align*}
If $A_{u,v}(\mfb,\mfc) = \ind\{\mfb \neq \mfc\}$ for every edge $(u,v)$ (forbidding monochromatic edges) and $\lambda_v(\mfc) = 1$ for every vertex $v$ (no external fields/no biases towards colors), then we get the uniform distribution over all proper list colorings of $G$. 
If further $\zcQ_v = [q]$ for all $v$ then we get the uniform distribution on all proper $q$-colorings of $G$. A permissive distribution on $q$-colorings, the \textit{antiferromagnetic Potts model}, is the spin system we study later in~\cref{sec:antiferro-wsm}.

\subsubsection{Spectral independence and optimal mixing of Glauber dynamics}
Here we give the formal definition of spectral independence and its implication for optimal mixing of Glauber dynamics. To do this, we formally define pinnings, influences, and influence matrices.

\begin{defn}
A pinning $(\Lambda,\tau)$ comprises a subset $\Lambda \subseteq V$ of vertices and a \textit{feasible} partial coloring $\tau: \Lambda \to \prod_{v \in \Lambda} \zcQ_v$ of vertices in $\Lambda$, where partial coloring $\tau$ is feasible if and only if it can be extended to a proper vertex coloring with positive probability. A vertex is called \emph{free} or \emph{unpinned} if it is not pinned. 

The collection of all pinnings is denoted by $\zcT$. Given a pinning $(\Lambda,\tau) \in \zcT$, we let $\zcQ^\tau_v$ denote the set of feasible colors for a vertex $v$ conditioned on $\tau$.
Also let $\mu^\tau_v$ denote the marginal distribution at $v$ conditioned on $\tau$ which is supported on $\zcQ^\tau_v$.
\end{defn}

When $\Lambda$ is clear from context, we often just use $\tau$ to denote a pinning of a collection of vertices.

The \emph{spectral independence} approach was introduced by Anari, Liu, and Oveis Gharan \cite{ALO20}, building upon ideas from high-dimensional expanders \cite{AL20}.
To formally define it, we first give the definitions of influences and influence matrices. 

\begin{defn}[Influence Matrices]
Let $\mu$ be the Gibbs distribution of a spin system on a finite graph $G=(V,E)$. Let $\Lambda \subseteq V$ be a subset of vertices, and let $\tau$ be a valid pinning on $\Lambda$. For vertices $u,v \in V \setminus \Lambda$ and colors $\mfb \in \zcQ_u, \mfc \in \zcQ_v$, we define the \emph{influence} of the vertex-color pair $(u,\mfb)$ on the vertex-color pair $(v,\mfc)$ by the quantity
\begin{align*}
    \infl_{\mu^{\tau}}((u,\mfb) \to (v,\mfc)) &\defeq \mu_{v}^{\tau, u \gets \mfb}(\mfc) - \mu_{v}^{\tau}(\mfc) \\
    &= \Pr_{\sigma \sim \mu}[\sigma(v) = \mfc \mid \sigma_{\Lambda} = \tau, \sigma(u) = \mfb] - \Pr_{\sigma \sim \mu}[\sigma(v) = \mfc \mid \sigma_{\Lambda} = \tau].
\end{align*}
If $u=v$, or if the assignments $u \gets \mfb$ or $v \gets \mfc$ are not feasible in $\mu^{\tau}$, then we define this quantity to be zero. 

We write $\infl_{\mu^{\tau}}$ for the corresponding \emph{influence matrix} with entries given by the influences defined above. In this paper, it will be convenient to also introduce the influence matrix $\infl_{\mu^{\tau}}^{u \to v} \in \R^{|\zcQ_u| \times |\zcQ_v|}$ of a vertex $u$ onto another vertex $v$, which is given by
\begin{align*}
    \infl_{\mu^{\tau}}^{u \to v}(\mfb, \mfc) \defeq \infl_{\mu^{\tau}}((u,\mfb) \to (v,\mfc)), \qquad \forall \mfb \in \zcQ_u, \mfc \in \zcQ_v.
\end{align*}
\end{defn}

We remark that $\infl_{\mu^{\tau}}$ is a square matrix with real eigenvalues; see \cite{AL20,ALO20,CGSV21,FGYZ22}.

\begin{defn}[Spectral Independence]
We say the Gibbs distribution $\mu$ of a spin system on a finite $n$-vertex graph $G=(V,E)$ is \emph{spectrally independent with constant $C_\SI$} if for every $\Lambda \subseteq V$ with $|\Lambda| \leq n - 2$ and every pinning $\tau$ on $\Lambda$, we have the inequality $\eigval_{\max}(\infl_{\mu^{\tau}}) \leq C_\SI$, where $\eigval_{\max}$ denotes the maximum eigenvalue of a square matrix.
\end{defn}

The \textit{Glauber dynamics}, also known as Gibbs sampling, is a heavily studied Markov chain for sampling from a \textit{Gibbs distribution} $\mu$. In each step of the chain, a vertex $v$ is selected uniformly at random; the color of $v$ is then updated by randomly selecting a color from $\zcQ_v$ according to the conditional distribution $\mu_v^\sigma$, where $\sigma$ represents the current coloring of all other vertices. 
In the setting of list colorings, this corresponds to select an available color in $\zcQ_v$ uniformly at random that is not currently being used by any of the neighbors of $v$. 
We assume that for any pinning $\tau$ the Glauber dynamics for the conditional distribution $\mu^\tau$ is ergodic. This is in general a mild and necessary condition; for example, it holds for list colorings if $|\zcQ_v| \ge \zcDeg(v)+2$ for every vertex $v$. 

It is known that spectral independence implies optimal mixing of Glauber dynamics for spin systems on bounded-degree graphs. 

\begin{thm}[\cite{ALO20,CLV21,BCCPSV22}]\label{thm:specind-opt-mix}
Let $\mu$ be the Gibbs distribution of a spin system on a finite $n$-vertex graph $G=(V,E)$ with maximum degree $\Delta\ge 3$. Suppose that $\mu$ is spectrally independent with constant $C_\SI>0$. 
Suppose that $\mu_v^\tau(\mfc) \ge \eps$ for any pinning $\tau$, any free vertex $v$, and any color $\mfc \in \zcQ^\tau_v$ for some constant $\eps > 0$.
Then the Glauber dynamics for sampling from $\mu$ mixes in $O_{\Delta,C_\SI,\eps}(n\log n)$ steps.
\end{thm}

\subsubsection{Spectral independence via Wasserstein distance}
One way to upper bound the spectral independence constant is via studying the Wasserstein distance between two conditional Gibbs distributions that disagree at a single vertex. This type of approach has been recently used to great effect in some recent works~\cite{BCCPSV22,Liu21,CZ23,CMM23,GGGH22}. We introduce this framework in the context of our problem below.

Consider a pinning $(\Lambda,\tau)\in \zcT$, a free vertex $u \in V \setminus \Lambda$, and two colors $\mfb,\mfc \in \zcQ^\tau_v$. Let $U = V \setminus \Lambda \setminus \{u\}$ be the set of all other free vertices.
For two colorings $X,Y$ on $U$, their \textit{Hamming distance} is defined as
\[
\dH\left( X,Y \right) := \sum_{v \in U} \ind\{X(v) \neq Y(v)\};
\]
namely, the number of vertices with disagreeing colors from $X$ and $Y$.
The \textit{$1$-Wasserstein distance} (with respect to the Hamming distance) between $\mu^{\tau,u\gets\mfb}_U$ and $\mu^{\tau,u\gets\mfc}_U$ is defined as
\begin{equation}\label{eq:W1-def}
W_1 \left( \mu^{\tau,u\gets\mfb}_U, \, \mu^{\tau,u\gets\mfc}_U \right)
= \min_{\mathcal{C}} \left\{ \E_{(X,Y) \sim \mathcal{C}}\left[ \dH\left( X,Y \right) \right] \right\},    
\end{equation}
where the minimum is over all couplings $\mcC$ of the two distributions $\mu^{\tau,u\gets\mfb}_U$ and $\mu^{\tau,u\gets\mfc}_U$. 

\begin{fact}\label{fact:W1}
$
W_1 \left( \mu^{\tau, u\gets\mfb}_U, \, \mu^{\tau, u\gets\mfc}_U \right)
\ge
\sum_{v \in U} \left\| \mu^{\tau, u\gets\mfb}_v - \mu^{\tau, u\gets\mfc}_v \right\|_{\TV}.
$
\end{fact}
\begin{proof}
Let $\mathcal{C}$ be an optimal coupling for \cref{eq:W1-def}. Then we have
\begin{align*}
W_1 \left( \mu^{\tau,u\gets\mfb}_U, \, \mu^{\tau,u\gets\mfc}_U \right)
&= \E_{(X,Y) \sim \mathcal{C}}\left[ \dH\left( X,Y \right) \right] 
= \E_{(X,Y) \sim \mathcal{C}}\left[ \sum_{v \in U} \ind\{X(v) \neq Y(v)\} \right] \\
&= \sum_{v \in U} \Pr_{(X,Y) \sim \mathcal{C}}\left( X(v) \neq Y(v) \right)
\ge \sum_{v \in U} \left\| \mu^{\tau,u\gets\mfb}_v - \mu^{\tau,u\gets\mfc}_v \right\|_{\TV},
\end{align*}
as claimed.
\end{proof}

The following lemma gives us a way to establish $O(1)$ spectral independence via sufficiently strong upper bounds on the $W_1$-distance of pairs of conditional distributions differing at one vertex in the conditioning.

\begin{lemma}[\cite{BCCPSV22,CZ23}]
\label{lem:W1-SI}
For any pinning $(\Lambda,\tau) \in \zcT$, we have 
\[
\eigval_{\max}(\infl_{\mu^{\tau}}) 
\le 
\norm{\infl_{\mu^{\tau}}}_\infty
\le
2 \max_{u \in V \setminus \Lambda} \max_{\mfb,\mfc \in \zcQ_u^\tau}
\left\{ W_1 \left( \mu^{\tau,u\gets\mfb}_{V \setminus \Lambda \setminus \{u\}}, \, \mu^{\tau,u\gets\mfc}_{V \setminus \Lambda \setminus \{u\}} \right) \right\}.
\]
\end{lemma}
\begin{proof}
The first inequality is a standard fact from linear algebra. 
For any vertex $u \in V \setminus \Lambda$ and color $\mfb \in \zcQ^\tau_u$, let $U = V \setminus \Lambda \setminus \{u\}$.
We have for all $v \in U$ that 
\begin{align*}
\sum_{\mfa \in \zcQ^\tau_v} 
\left| \infl_{\mu^{\tau}}^{u \to v}(\mfb, \mfa) \right| 
&= \sum_{\mfa \in \zcQ^\tau_v}
\left| \mu_{v}^{\tau, u \gets \mfb}(\mfa) - \mu_{v}^{\tau}(\mfa) \right| \\
&= 2\left\| \mu_{v}^{\tau, u \gets \mfb} - \mu_{v}^{\tau} \right\|_{\TV} 
\le 2 \sum_{\mfc \in \zcQ^\tau_u} \mu_u^\tau(\mfc) \left\| \mu_{v}^{\tau, u \gets \mfb} - \mu_{v}^{\tau, u\gets \mfc} \right\|_{\TV}.
\end{align*}
Summing over $v$, we deduce that
\begin{align*}
\sum_{v \in U} 
\sum_{\mfa \in \zcQ^\tau_v} 
\left| \infl_{\mu^{\tau}}^{u \to v}(\mfb, \mfa) \right| 
&\le 2 
\sum_{\mfc \in \zcQ^\tau_u} \mu_u^\tau(\mfc) 
\sum_{v \in U} \left\| \mu^{\tau, u \gets \mfb}_v - \mu^{\tau, u \gets \mfc}_v \right\|_{\TV} \\
&\le 2
\max_{\mfc \in \zcQ^\tau_u} 
\left\{
\sum_{v \in U} \left\| \mu^{\tau, u \gets \mfb}_v - \mu^{\tau, u \gets \mfc}_v \right\|_{\TV}
\right\} \\
&\le 2
\max_{\mfc \in \zcQ^\tau_u} \left\{ W_1 \left( \mu^{\tau, u \gets \mfb}_U, \, \mu^{\tau, u\gets \mfc}_U \right) \right\},
\end{align*}
where the last inequality follows from \cref{fact:W1}.
The lemma then follows by taking the maximum over $u \in V \setminus \Lambda$ and $\mfb \in \zcQ^\tau_u$.
\end{proof}

\subsubsection{Strong spatial mixing and total influence decay on trees}

Here, we give an alternate definition of \emph{strong spatial mixing (SSM)} and a related notion, called \textit{total influence decay (TID)} for a spin system on a graph $G=(V,E)$. Our definition here of strong spatial mixing is slightly different from the one in the introduction and has been used in other related works (e.g.~\cite{FGY22}); it is straightforward to see that the two definitions we give of SSM are equivalent up to constants assuming lower bounds on the marginal probabilities as given in \cref{lem:marginal-lb}.

\begin{definition}[Strong spatial mixing (SSM)]
\label{def:SSM}
We say the 
Gibbs distribution $\mu$ on $G = (V, E)$ exhibits \textit{strong spatial mixing }with exponential decay rate $1 - \delta$ and constant $C_{\SM} > 0$ if the following holds.
For any integer $K>0$, any pinning $(\Lambda,\tau) \in \zcT$, any vertex $u \in V \setminus \Lambda$, any color $\mfc\in \zcQ_v^\tau$, and any two feasible partial colorings $\sigma,\eta$ on $\zcS(u,K) \setminus \Lambda$, we have
\[
\left| \frac{\mu^{\tau,\sigma}_u (\mfc)}{\mu^{\tau,\eta}_u (\mfc)} - 1 \right|
\le C_{\SM}(1-\delta)^K,
\]
where $S(u, K) = \{w \in V \mid \dist_G(u, w) = K\}$ denotes the set of all vertices at graph distance \textit{exactly} $K$ from $u$.
\end{definition}

The notion of total influence decay is a sibling of strong spatial mixing. Indeed, we are able to prove both of them simultaneously using the potential function method (as we do in~\cref{thm:si-from-contraction})
\begin{definition}[Total influence decay (TID)]
\label{def:SSI}
We say the 
Gibbs distribution $\mu$ on $G = (V, E)$ exhibits \textit{total influence decay} with exponential decay rate $1 - \delta$ and constant $C_{\INFL}$ if the following holds. 
For any integer $K>0$, any pinning $(\Lambda,\tau) \in \zcT$, any vertex $u \in V \setminus \Lambda$, and any two colors $\mfb,\mfc \in \zcQ_v^\tau$, we have
\[
\sum_{v \in \zcS(u,K) \setminus \Lambda} \left\| \mu^{\tau,u\gets\mfb}_v - \mu^{\tau,u\gets\mfc}_v \right\|_{\TV} 
\le C_{\INFL}(1-\delta)^K,
\]
where $S(u, K) = \{w \in V \mid \dist_G(u, w) = K\}$ denotes the set of all vertices at graph distance \textit{exactly} $K$ from $u$.
\end{definition}

Using our results from \cref{sec:colorings-simple} and later in \cref{sec:colorings-weighted}, we have the following theorem that establishes SSM and TID for list colorings on trees; we formally verify this result in~\cref{sec:main-result-proofs}.

\begin{thm}
\label{thm:SSM+SSI}
Consider list colorings on a tree $T=(V,E)$ with maximum degree $\Delta = \max_v \zcDeg(v)$ and maximum list size $q = \max_v |\zcQ_v|$. 
If $|\zcQ_v| \ge \zcDeg(v) + 3$ for each vertex $v$, then there exist constants $\delta = \delta(\Delta,q) \in (0,1)$, $C_{\SM} = C_{\SM}(\Delta,q) > 0$, and $C_{\INFL} = C_{\INFL}(\Delta,q) > 0$ such that:
\begin{enumerate}
\item Strong spatial mixing holds with exponential decay rate $1-\delta$ and constant $C_{\SM}$;
\item Total influence decay holds with exponential decay rate $1-\delta$ and constant $C_{\INFL}$.
\end{enumerate}
\end{thm}

In general, implications between strong spatial mixing and total influence decay (on trees) are not known in either direction. However, as in this paper they are often proved together by the potential function method; see, e.g., for two-spin systems \cite{LLY13,CLV20,CFYZ22}. 

\subsection{Proof strategy for spectral independence}
\label{subsec:proof-strategy-girth}

In this section we prove \cref{t:SI-girth}.
In fact, we prove the following more general theorem about list colorings which implies \cref{t:SI-girth}.
\begin{thm}\label{thm:list-colorings}
For all $\Delta \ge 3$ and $q \ge \Delta+3$ there exists $\girth>0$ such that the following is true. 
Let $G = (V,E)$ be an $n$-vertex graph of maximum degree $\Delta = \max_{v \in V} \zcDeg(v)$ and girth $\girth$. Suppose that each vertex $v \in V$ is associated with a list $\zcQ_v$ of colors and the maximum list size is $q = \max_{v \in V} |\zcQ_v|$. 
If $|\zcQ_v| \ge \zcDeg(v) + 3$ for all $v \in V$, then spectral independence holds for the uniform distribution over all list colorings of $G$ with constant $O_{\Delta,q}(1)$. 
Furthermore, under the same assumptions the Glauber dynamics for sampling random list colorings mixes in $O_{\Delta,q}(n \log n)$ time. 
\end{thm}

By \cref{thm:specind-opt-mix,lem:W1-SI}, it suffices to establish spectral independence by upper bounding certain Wasserstein distances associated with the Gibbs distribution. 
In this work, we present a novel approach towards proving spectral independence on all bounded-degree graphs with sufficiently large (yet constant) girth, assuming knowledge of strong spatial mixing and total influence decay on trees.

\begin{rem}\label{rem:general-spin-rem}
As noted at the beginning of the section, our approach works not only for list colorings but in fact also for arbitrary spin systems with pairwise interactions between neighboring vertices, where the hard constraints of the spin system satisfy the following mild condition.
For any pinning $(\Lambda,\tau) \in \zcT$ and any two distinct vertices $u,v$ such that $\dist_G(u,v) \ge 2$, we have
\begin{equation}\label{eq:mild}
\zcQ^\tau_v = \zcQ^{\tau,u\gets\mfb}_v, \quad\forall \mfb \in \zcQ^\tau_u.
\end{equation}
In other words, fixing the color of a vertex $u$ does not forbid any color of $v$ if they are not adjacent. 
We also require that our spin system is closed under taking pinnings and subgraphs.

These conditions are very mild and hold naturally in our setting, as they hold for list colorings whenever $|\zcQ_v| \ge \zcDeg(v)+1$. 
\end{rem}

Our proof of~\cref{t:SI-girth} proceeds via the following two major steps. 

\paragraph{Step 1. A new condition implying spectral independence.} We present a new condition, \textit{$(R,\eps)$-influence decay} (\cref{cond:inf-decay} below), which implies spectral independence for any spin system defined on bounded-degree graphs. 
$(R,\eps)$-influence decay states that the sum of influences of a vertex $u$ on all vertices at distance \textit{exactly} $R$ from $u$ is upper bounded by some (tiny) $\eps$.

\begin{cond}[$(R,\eps)$-influence decay]
\label{cond:inf-decay}
For integer $R \ge 2$ and real $\eps > 0$, 
Gibbs distribution $\mu$ on graph $G = (V, E)$ has \textit{$(R,\eps)$-influence decay} if the following holds.
For any pinning $(\Lambda,\tau) \in \zcT$, any vertex $u \in V \setminus \Lambda$, and any two colors $\mfb,\mfc \in \zcQ_u^\tau$, we have
\[
\sum_{v \in \zcS(u,R) \setminus \Lambda} \left\| \mu^{\tau,u\gets\mfb}_v - \mu^{\tau,u\gets\mfc}_v \right\|_{\TV}
\le \eps,
\]
where  $S(u, R) = \{w \in V \mid \dist_G(u, w) = R\}$ denotes the set of all vertices at graph distance \textit{exactly} $R$ from $u$.
\end{cond}

We prove that if $(R,\eps)$-influence decay holds for \textit{one specific} choice of $R$ and $\eps = O_{\Delta}(1/R)$, we will be able to deduce spectral independence on graphs of constant degree.

\begin{lem}
\label{thm:cond-to-SI}
Let $\Delta \ge 3$ and $R \ge 2$ be integers. 
Consider a spin system on a graph $G=(V,E)$ of maximum degree $\Delta \ge 3$ satisfying \cref{eq:mild}, with associated Gibbs distribution $\mu$. Suppose that the distribution has $(R,\eps)$-influence decay for some constants $R$ and $\eps \le 1/(8R\ln\Delta)$. 
Then, for any pinning $(\Lambda,\tau) \in \zcT$, any vertex $u \in V \setminus \Lambda$, and any two colors $\mfb,\mfc \in \zcQ_u^\tau$, we have
\[
W_1 \left( \mu^{\tau,u\gets\mfb}_U, \, \mu^{\tau,u\gets\mfc}_U \right) \le 2\Delta^R
\]
where $U = V \setminus \Lambda \setminus \{u\}$ and $W_1(\cdot,\cdot)$ denotes the $1$-Wasserstein distance with respect to the Hamming distance. 
Consequently, spectral independence holds with constant $4\Delta^R$ by \cref{lem:W1-SI}. 
\end{lem}

The proof of \cref{thm:cond-to-SI} is based on a simple coupling procedure and is given in \cref{subsec:decay-infl}.
Note that for this result we do not require any lower bound on the girth.
We believe \cref{thm:cond-to-SI} is interesting on its own and may have other applications for showing spectral independence.

\paragraph{Step 2. SSM and TID on trees $\implies$ $(R,\eps)$-influence decay on graphs of large girth.} 
We show that if the model satisfies both \emph{strong spatial mixing} and \emph{total influence decay} on trees, then $(R,\eps)$-influence decay holds on graphs of sufficiently large girth. 

\begin{lem}
\label{thm:tree-to-cond}
Consider a family of spin systems that is closed under taking subgraphs and pinnings. 
Suppose that for all \emph{trees} of maximum degree $\Delta \ge 3$ in the family both of the following properties hold:
\begin{enumerate}
\item Strong spatial mixing with exponential decay rate $1-\delta$ and constant $C_{\SM}$;
\item Total influence decay with exponential decay rate $1-\delta$ and constant $C_{\INFL}$.
\end{enumerate}
Then there exist integers 
$$
R = \Omega\left( \delta^{-1}(\log\log\Delta + \log C_{\INFL} + \log(1/\delta)) \right) 
\quad\text{and}\quad
\girth = \Omega\left( \delta^{-1}(R\log\Delta + \log C_{\SM}) \right)
$$
such that $(R,\eps)$-influence decay (\cref{cond:inf-decay}) holds with $\eps \le 1/(8R\ln\Delta)$ for all graphs of degrees at most $\Delta$ and girth at least $\girth$ in the family.
\end{lem}

\begin{rem}
In fact, we can deduce from \cref{thm:SSM+SSI} that $(R,\eps)$-influence decay holds for all sufficiently large $R$ with exponential decay rate $\eps = \exp(-\Omega(R))$.
\end{rem}

Our proof of \cref{thm:tree-to-cond} utilizes both strong spatial mixing and total influence decay on trees. 
To prove $(R,\eps)$-influence decay (\cref{cond:inf-decay}) for graphs with large girth $\girth$, we fix an arbitrary pinning on all vertices at distance $\floor{\girth/2}-1$ away from the center $u$, and claim from strong spatial mixing that the total influence of $u$ on distance $R$ is not affected much if $\girth$ is much larger than $R$. Then the problem is reduced to trees and we can apply our knowledge of total influence decay on trees. 
The full proof of \cref{thm:tree-to-cond} is provided in \cref{subsec:tree2girth}.

\medskip
We give here the proof of \cref{thm:list-colorings} which implies \cref{t:SI-girth}. 

\begin{proof}[Proof of \cref{thm:list-colorings}]
Combining \cref{thm:cond-to-SI,thm:tree-to-cond} and also \cref{thm:SSM+SSI}, we establish spectral independence for list colorings on all bounded-degree large-girth graphs. 
Optimal mixing of Glauber dynamics then follows by \cref{thm:specind-opt-mix}.
\end{proof}

\begin{rem}\label{rem:(1+eps)Delta}
In particular, if $q = (1+\eps) \Delta$ for $\eps \in (0,1)$ and $\Delta \ge 7/\eps^2$, then by \cref{prop:(1+eps)Delta} we can take $R = \Omega_\eps(\log \Delta)$ and our girth requirement is $\girth = \Omega_\eps(\log^2\Delta)$.
\end{rem}

\subsection{Decay of influences implies spectral independence}
\label{subsec:decay-infl}

We construct a coupling between $\mu^{\tau,u\gets\mfb}_U$ and $\mu^{\tau,u\gets\mfc}_U$ where $U = V \setminus \Lambda \setminus \{u\}$. 
The coupling procedure is very simple and is described in \cref{alg:CP}. 
Let $B(u,R) = \{v\in V: \dist_G(u,v) \le R\}$ be the set of vertices at distance at most $R$ from $u$. In a nutshell,~\cref{alg:CP} picks a uniformly random unpinned vertex at distance $R$ from $u$ and tries to couple it optimally; then it optimally couples all the remaining unpinned vertices at distance $R$ from $u$.

\begin{algorithm}[t]
\caption{A simple coupling procedure}\label{alg:CP}
\KwIn{$(\Lambda,\tau) \in \zcT$, $u \in V \setminus \Lambda$, $U = V \setminus \Lambda \setminus \{u\}$, $\mfb,\mfc \in \zcQ_u^\tau$} 
\KwOut{$(X,Y)$ distributed as a coupling between $\mu^{\tau,u\gets\mfb}_U$ and $\mu^{\tau,u\gets\mfc}_{U}$}

\medskip 
\eIf{$\zcS(u,R) \setminus \Lambda = \emptyset$}{
	Sample $X_{U \cap B(u,R)}$ and $Y_{U \cap B(u,R)}$ independently\;
	Couple $X_{U \setminus B(u,R)} = Y_{U \setminus B(u,R)}$ identically\;
	\KwOut{$(X,Y) = ( X_{U \cap B(u,R)} \cup X_{U \setminus B(u,R)}, Y_{U \cap B(u,R)} \cup X_{U \setminus B(u,R)} )$}
}
(){
Pick $v \in \zcS(u,R) \setminus \Lambda$ uniformly at random\;\label{line:pick}
Sample $(X_v,Y_v) = (\mfb',\mfc')$ from a TV-optimal coupling between $\mu^{\tau,u\gets\mfb}_v$ and $\mu^{\tau,u\gets\mfc}_v$\;
Sample $(X',Y')$ from a $W_1$-optimal coupling between $\mu^{\tau, u\gets\mfb, v\gets\mfb'}_{U \setminus \{v\}}$ and $\mu^{\tau, u\gets\mfc, v\gets\mfc'}_{U \setminus \{v\}}$\;

\KwOut{$(X,Y) = (X' \cup \{v\gets\mfb'\}, Y' \cup \{v\gets\mfc'\})$}
}
\end{algorithm}

We refer to the pair of distributions $\mu^{\tau,u\gets\mfb}_U$ and $\mu^{\tau,u\gets\mfc}_U$ which we want to couple as an \emph{instance}.
We consider all such instances obtained by taking pinnings from the original spin system. Hence, each instance is specified by a pinning $(\Lambda,\tau) \in \zcT$, a vertex $u \in V \setminus \Lambda$, and two colors $\mfb,\mfc \in \zcQ_u^\tau$, and we use the tuple $(\tau,u,\mfb,\mfc)$ to represent an instance by a slight abuse of notation. 
We in particular keep track of two quantities which characterize how big an instance is.
Recall that a vertex is called free if its color is not pinned.

\begin{itemize}
\item The \emph{size} $k$ of an instance $(\tau,u,\mfb,\mfc)$ is the number of free vertices; i.e., 
$$ k = |U| = |V \setminus \Lambda \setminus \{u\}|. $$ 
\item The \emph{breadth} $\ell$ of an instance $(\tau,u,\mfb,\mfc)$ is the number of free vertices in $\zcS(u,R)$; i.e., 
$$ \ell = |\zcS(u,R) \setminus \Lambda|. $$ 
\end{itemize}
We then define for a given spin system on some graph $G = (V,E)$, 
\[
\mathbf{D}(k,\ell) = \max_{\substack{ \text{instances of size $k$} \\ \text{and breadth $\ell$} }} 
W_1 \left( \mu^{\tau,u\gets\mfb}_U, \, \mu^{\tau,u\gets\mfc}_U \right).
\]
We upper bound $\mathbf{D}(k,\ell)$ recursively via the coupling in \cref{alg:CP}. 

\begin{lemma}
Assume that $(R,\eps)$-influence decay (\cref{cond:inf-decay}) holds for constants $R \ge 2$ and $\eps > 0$.
Then we have
\begin{equation}\label{eq:recursion-f}
\mathbf{D}(k,\ell) \le
\begin{cases}
\Delta^R, &\ell=0; \\
\mathbf{D}(k-1,\ell-1) + \dfrac{\eps}{\ell} \left( \max\limits_{m \le \Delta^R} \{\mathbf{D}(k-1,m)\} + 1 \right), & \ell \ge 1.
\end{cases}
\end{equation}
\end{lemma}

\begin{proof}
Consider an arbitrary instance $(\tau,u,\mfb,\mfc)$ of size $k$ and breadth $\ell$. 
We upper bound the $W_1$-distance between $\mu^{\tau,u\gets\mfb}_U$ and $\mu^{\tau,u\gets\mfc}_U$ by the expected Hamming distance $\E[\dH(X,Y)]$ under the simple coupling procedure in \cref{alg:CP}.

If $\ell = |\zcS(u,R) \setminus \Lambda| = 0$, then we can couple such that all discrepancies appear inside the ball $B(u,R-1)$, and by \cref{alg:CP} we get 
\[
\E[\dH(X,Y)] 
\le
|B(u,R-1) \setminus \Lambda| 
\le \Delta^R. 
\]
Next, assume $\ell \ge 1$ and consider the coupling given in \cref{alg:CP}. 
For a free vertex $v \in \zcS(u,R) \setminus \Lambda$, we first consider the expected Hamming distance conditional on picking $v$ in Line~\ref{line:pick}.
Let $\mathcal{C}$ denote the optimal coupling between $\mu^{\tau,u\gets\mfb}_v$ and $\mu^{\tau,u\gets\mfc}_v$ for TV distance.
Note that $\zcQ^{\tau,u\gets\mfb}_v = \zcQ^{\tau,u\gets\mfc}_v = \zcQ^\tau_v$ by \cref{eq:mild} since $\dist_G(u,v) \ge 2$.
Then we have
\begin{align*}
\E \left[ \dH(X,Y) \mid \text{choose $v$} \right] 
&= \sum_{\mfb',\mfc' \in \zcQ^\tau_v} \Pr_{\mathcal{C}}(\mfb',\mfc') 
\left( W_1 \left( \mu^{\tau, u\gets\mfb, v\gets\mfb'}_{U \setminus \{v\}}, \, \mu^{\tau, u\gets\mfc, v\gets\mfc'}_{U \setminus \{v\}} \right) + \ind\{\mfb' \neq \mfc' \} \right) \\
\le{}& \left( 1-\left\| \mu^{\tau,u\gets\mfb}_v - \mu^{\tau,u\gets\mfc}_v \right\|_{\TV} \right) 
\cdot \max_{\mfb' \in \zcQ^\tau_v} \left\{ W_1 \left( \mu^{\tau, u\gets\mfb, v\gets\mfb'}_{U \setminus \{v\}}, \, \mu^{\tau, u\gets\mfc, v\gets\mfb'}_{U \setminus \{v\}} \right) \right\} \\
&+ \left\| \mu^{\tau,u\gets\mfb}_v - \mu^{\tau,u\gets\mfc}_v \right\|_{\TV}
\cdot \left( \max_{\mfb',\mfc' \in \zcQ^\tau_v} \left\{ W_1 \left( \mu^{\tau, u\gets\mfb, v\gets\mfb'}_{U \setminus \{v\}}, \, \mu^{\tau, u\gets\mfc, v\gets\mfc'}_{U \setminus \{v\}} \right) \right\} + 1 \right). 
\end{align*}
Now, observe that for any $\mfb' \in \zcQ^\tau_v$ we have
\[
W_1 \left( \mu^{\tau, u\gets\mfb, v\gets\mfb'}_{U \setminus \{v\}}, \, \mu^{\tau, u\gets\mfc, v\gets\mfb'}_{U \setminus \{v\}} \right) 
\le \mathbf{D}(k-1,\ell-1)
\]
since both the size and the breadth decrease when the color of $v$ is fixed. 
Meanwhile, for any $\mfb',\mfc' \in \zcQ^\tau_v$ we deduce from the triangle inequality that 
\begin{align*}
W_1 \left( \mu^{\tau, u\gets\mfb, v\gets\mfb'}_{U \setminus \{v\}}, \, \mu^{\tau, u\gets\mfc, v\gets\mfc'}_{U \setminus \{v\}} \right)
&\le 
W_1 \left( \mu^{\tau, u\gets\mfb, v\gets\mfb'}_{U \setminus \{v\}}, \, \mu^{\tau, u\gets\mfc, v\gets\mfb'}_{U \setminus \{v\}} \right)
+
W_1 \left( \mu^{\tau, u\gets\mfc, v\gets\mfb'}_{U \setminus \{v\}}, \, \mu^{\tau, u\gets\mfc, v\gets\mfc'}_{U \setminus \{v\}} \right) \\
&\le \mathbf{D}(k-1,\ell-1) + \max_{m \le |S(v, R) \backslash \Lambda|} \{\mathbf{D}(k-1,m)\},
\end{align*}
where the upper bound for the second $W_1$-distance is because the discrepancy is at $v$ and we have no control on the breadth of the instance. 
Hence, we obtain that
\[
\E \left[ \dH(X,Y) \mid \text{choose $v$} \right] 
\le \mathbf{D}(k-1,\ell-1) + \left\| \mu^{\tau,u\gets\mfb}_v - \mu^{\tau,u\gets\mfc}_v \right\|_{\TV} \cdot \left( \max_{m \le \Delta^R} \{\mathbf{D}(k-1,m)\} + 1 \right).
\]
Taking expectation over the random choice of $v \in \zcS(u,R) \setminus \Lambda$, we finally deduce by $(R,\eps)$-influence decay that
\begin{align*}
\E \left[ \dH(X,Y) \right] 
&\le \mathbf{D}(k-1,\ell-1) + \frac{1}{\ell} \left( \sum_{v \in \zcS(u,R) \setminus \Lambda} \left\| \mu^{\tau,u\gets\mfb}_v - \mu^{\tau,u\gets\mfc}_v \right\|_{\TV} \right)
\left( \max_{m \le \Delta^R} \{\mathbf{D}(k-1,m)\} + 1 \right) \\
&\le \mathbf{D}(k-1,\ell-1) + \frac{\eps}{\ell} \left( \max_{m \le \Delta^R} \{\mathbf{D}(k-1,m)\} + 1 \right),
\end{align*} 
as claimed.
\end{proof}

\begin{lemma}\label{lem:D-bound}
Assume \cref{cond:inf-decay} holds for constants $R \ge 2$ and $0<\eps \le 1/(8R\ln\Delta)$.
Then we have
\begin{equation}\label{eq:induction-f}
\mathbf{D}(k,\ell) \le \left( 1+2\eps H(\ell) \right) \Delta^R,
\end{equation}
where $H(\ell) = \sum_{i=1}^\ell 1/i \le \ln\ell+1$ for $\ell \ge 1$ and $H(0) = 0$.
\end{lemma}

\begin{proof}
We prove the lemma by induction on $k$, the size of the instance. 

For $k \le \Delta^R$, \cref{eq:induction-f} holds trivially. 
Suppose \cref{eq:induction-f} holds for some $k-1$ and all $\ell \ge 0$, and consider now the inequality for $k$ and all $\ell \ge 0$.
For $\ell = 0$, \cref{eq:induction-f} is already shown in \cref{eq:recursion-f}. 
Next we consider $\ell \ge 1$. 
We first get from the induction hypothesis that
\begin{align*}
\max_{m \le \Delta^R} \left\{\mathbf{D}(k-1,m)\right\}
&\le \max_{m \le \Delta^R} \left\{ \left( 1+2\eps H(m) \right) \Delta^R \right\}\\
&\le \left( 1+2\eps(R\ln \Delta + 1) \right) \Delta^R \\
&\le 2\Delta^R - 1,
\end{align*}
where the last inequality follows from the assumption that $\eps \le 1/(8R\ln\Delta)$. 
For $\ell \ge 1$, we deduce from \cref{eq:recursion-f} that
\begin{align*}
\mathbf{D}(k,\ell) 
&\le \mathbf{D}(k-1,\ell-1) + \frac{\eps}{\ell} \left( \max_{m \le \Delta^R} \{\mathbf{D}(k-1,m)\} + 1 \right) \\
&\le \left( 1+2\eps H(\ell-1) \right) \Delta^R + \frac{2\eps}{\ell} \cdot \Delta^R \\
&= \left( 1+2\eps H(\ell) \right) \Delta^R.
\end{align*}
This proves the lemma. 
\end{proof}

\cref{thm:cond-to-SI} then immediately follows from~\cref{lem:D-bound} since $\ell \le \Delta^R$ and $\eps \le 1/(8R\ln\Delta)$.

\subsection{$(R,\eps)$-influence decay on graphs from SSM and TID on trees}
\label{subsec:tree2girth}

Suppose that the underlying graph $G = (V,E)$ has girth at least $\girth$.
Let $K = \floor{\girth/2}-1$. We will consider the setting $K > R$. By our choice of $K$, if we pin the colors of all vertices in $S(u,K)$, the induced spin system on $B(u, K)$ is identical to that of a tree with associated boundary condition on $S(u,K)$. This fact will allow us to deduce $(R,\eps)$-influence decay on large-girth graphs from strong spatial mixing and total influence decay on trees.

We establish $(R,\eps)$-influence decay (\cref{cond:inf-decay}) using the following lemma to upper bound the sum of the distance $R$ influences;~\cref{lem:sum-infl} holds for any spin system (i.e.\ does not require assumptions about girth, SSM, or TID).
\begin{lemma}\label{lem:sum-infl}
For any pinning $(\Lambda,\tau) \in \zcT$, any vertex $u \in V \setminus \Lambda$, and any two colors $\mfb,\mfc \in \zcQ^\tau_u$, 
we have for all integers $K > R \ge 1$ that
\begin{align*}
&\sum_{v \in \zcS(u,R) \setminus \Lambda} \left\| \mu^{\tau,u\gets\mfb}_v - \mu^{\tau,u\gets\mfc}_v \right\|_{\TV} \\
\le{}&  \left\| \mu^{\tau,u\gets\mfb}_{\zcS(u,K) \setminus \Lambda} - \mu^{\tau,u\gets\mfc}_{\zcS(u,K) \setminus \Lambda} \right\|_{\TV} 
\cdot |\zcS(u,R) \setminus \Lambda| 
+ \max_{\sigma} \left\{ \sum_{v \in \zcS(u,R) \setminus \Lambda} \left\| \mu^{\tau,\sigma,u\gets\mfb}_v - \mu^{\tau,\sigma,u\gets\mfc}_v \right\|_{\TV} \right\},
\end{align*}
where the maximum is over all feasible partial colorings on $\zcS(u,K) \setminus \Lambda$.
\end{lemma}

\begin{proof}
For simplicity write $S_R = \zcS(u,R) \setminus \Lambda$ and $S_K = \zcS(u,K) \setminus \Lambda$. 
Let $\mathcal{C}$ be an optimal coupling between $\mu^{\tau,u\gets\mfb}_{S_K}$ and $\mu^{\tau,u\gets\mfc}_{S_K}$ for TV-distance. 
For every $v \in S_R$, by the triangle inequality we have
\[
\left\| \mu^{\tau,u\gets\mfb}_v - \mu^{\tau,u\gets\mfc}_v \right\|_{\TV} 
\le \sum_{(\sigma,\eta)} \Pr_\mathcal{C}(\sigma,\eta) 
\left\| \mu^{\tau,\sigma,u\gets\mfb}_v - \mu^{\tau,\eta,u\gets\mfc}_v \right\|_{\TV},
\]
where $\sigma$ and $\eta$ are both feasible partial colorings of vertices in $S_K$.
Summing over $v$, we obtain
\begin{align*}
\sum_{v \in S_R} \left\| \mu^{\tau,u\gets\mfb}_v - \mu^{\tau,u\gets\mfc}_v \right\|_{\TV} 
&\le \sum_{(\sigma,\eta)} \Pr_\mathcal{C}(\sigma,\eta) 
\sum_{v \in S_R} \left\| \mu^{\tau,\sigma,u\gets\mfb}_v - \mu^{\tau,\eta,u\gets\mfc}_v \right\|_{\TV} \\
&\le \Pr_\mathcal{C}(\sigma \neq \eta) \cdot |S_R| 
+ \Pr_\mathcal{C}(\sigma = \eta) \cdot \max_{\sigma} \left\{ \sum_{v \in S_R} \left\| \mu^{\tau,\sigma,u\gets\mfb}_v - \mu^{\tau,\sigma,u\gets\mfc}_v \right\|_{\TV} \right\}.
\end{align*}
The lemma then follows from that $\Pr_\mathcal{C}(\sigma \neq \eta) = \left\| \mu^{\tau,u\gets\mfb}_{S_K} - \mu^{\tau,u\gets\mfc}_{S_K} \right\|_{\TV}$ since $\mathcal{C}$ is an optimal coupling of the two distributions. 
\end{proof}

We then show that the influence of the center $u$ on its distance $K$ neighborhood, $S(u,K)$, is small assuming strong spatial mixing on trees and a tree topology of $B(u,K)$.

\begin{lemma}\label{lem:center-to-K}
Consider a family of spin systems that is closed under taking subgraphs and pinnings. 
Assume SSM holds on all trees in this family with uniform constants $C_\SM > 0$ and $\delta>0$. 
Let $K \ge \ceil{(\ln C_\SM)/\delta}$ be an integer. 
Then for all graphs of girth at least $2K+2$ in this family we have for any choice of pinning $(\Lambda,\tau) \in \zcT$, vertex $u \in V \setminus \Lambda$, and colors $\mfb,\mfc \in \zcQ^\tau_u$ as before,
\[
\left\| \mu^{\tau,u\gets\mfb}_{\zcS(u,K) \setminus \Lambda} - \mu^{\tau,u\gets\mfc}_{\zcS(u,K) \setminus \Lambda} \right\|_{\TV}
\le 2C_\SM(1-\delta)^{K}.
\]
\end{lemma}

\begin{proof}
To simplify notation, let $\nu = \mu^\tau_{\{u\} \cup \zcS(u,K) \setminus \Lambda}$ be the joint distribution on the center vertex $u$ and free vertices from $S(u,K) \setminus \Lambda$. 
So we have
\begin{align*}
\left\| \mu^{\tau,u\gets\mfb}_{\zcS(u,K) \setminus \Lambda} - \mu^{\tau,u\gets\mfc}_{\zcS(u,K) \setminus \Lambda} \right\|_{\TV} 
&= \left\| \nu(\cdot \mid u\gets\mfb) - \nu(\cdot \mid u\gets\mfc) \right\|_{\TV} \\
&= \frac{1}{2} \sum_{\sigma} \nu(\sigma \mid u\gets\mfc) \left| \frac{\nu(\sigma \mid u\gets\mfb)}{\nu(\sigma \mid u\gets\mfc)} - 1 \right| \\
&\le \frac{1}{2} \max_{\sigma} \left\{ \left| \frac{\nu(\sigma \mid u\gets\mfb)}{\nu(\sigma \mid u\gets\mfc)} - 1 \right| \right\},
\end{align*}
where the summation and maximization are over all feasible partial colorings $\sigma$ of free vertices in $S(u,K) \setminus \Lambda$. 
Now for any such feasible partial coloring $\sigma$, we have that
\begin{align*}
\left| \frac{\nu(\sigma \mid u\gets\mfb)}{\nu(\sigma \mid u\gets\mfc)} - 1 \right| 
&= \left| \frac{\nu(u\gets\mfb \mid \sigma)}{\nu(u\gets\mfb)} \frac{\nu(u\gets\mfc)}{\nu(u\gets\mfc \mid \sigma)} - 1 \right| \\
&\le 4 \max \left\{ \left| \frac{\nu(u\gets\mfb \mid \sigma)}{\nu(u\gets\mfb)} - 1 \right|,\, \left| \frac{\nu(u\gets\mfc)}{\nu(u\gets\mfc \mid \sigma)} - 1 \right| \right\} \\
&\le 4 \max_{\mfa \in \zcQ^\tau_u} \max_{\sigma,\eta} \left\{ \left| \frac{\nu(u\gets\mfa \mid \sigma)}{\nu(u\gets\mfa \mid \eta)} - 1 \right| \right\},
\end{align*}
where the first inequality follows from the fact that $|xy-1| = |(x-1)+(y-1)+(x-1)(y-1)| \le 4\max\{|x-1|,|y-1|\}$ for $0\le x,y \le 2$ (which holds since $K\ge (\ln C_\SM)/\delta$), 
and the second inequality follows from the law of total probability. 

Now, observe that $\nu_u(\cdot \mid \sigma) = \mu^{\tau,\sigma}_u$ is equivalent to the distribution of the root $u$ on the local tree $B(u,K)$ with boundary condition on $S(u,K)$. 
Thus, SSM on trees implies that
\begin{align*}
\left\| \mu^{\tau,u\gets\mfb}_{\zcS(u,K) \setminus \Lambda} - \mu^{\tau,u\gets\mfc}_{\zcS(u,K) \setminus \Lambda} \right\|_{\TV} 
\le 2 \max_{\mfa \in \zcQ^\tau_u} \max_{\sigma,\eta} \left\{ \left| \frac{\nu(u\gets\mfa \mid \sigma)}{\nu(u\gets\mfa \mid \eta)} - 1 \right| \right\}
\le 2C_\SM(1-\delta)^{K},
\end{align*}
as claimed.
\end{proof}

We are now ready to prove \cref{thm:tree-to-cond}.

\begin{proof}[Proof of \cref{thm:tree-to-cond}]

Combining \cref{lem:sum-infl,lem:center-to-K}, we deduce that
\begin{equation}\label{eq:cond-final}
\sum_{v \in \zcS(u,R) \setminus \Lambda} \left\| \mu^{\tau,u\gets\mfb}_v - \mu^{\tau,u\gets\mfc}_v \right\|_{\TV} 
\le 2C_{\SM}(1-\delta)^K \Delta^R + C_{\INFL}(1-\delta)^R.
\end{equation}
If we take
\begin{equation*}
R \gtrsim \frac{1}{\delta} \left( \log\log \Delta + \log C_{\INFL} + \log \frac{1}{\delta} \right)
\quad\text{and}\quad
g \ge 2K+2 \gtrsim \frac{1}{\delta} \left( R\log \Delta + \log C_{\SM} \right),
\end{equation*}
then we get that the right-hand side of \cref{eq:cond-final} is at most $1/(8R\ln\Delta)$, as desired.
\end{proof}

\section{Tighter bounds for contraction}\label{sec:colorings-weighted}
In this section, we prove~\cref{t:strong-ssm}, showing that strong spatial mixing holds with exponentially decaying rate with essentially tight bounds on $q$. The proof of~\cref{t:strong-ssm} follows the template set out in~\cref{sec:colorings-simple} to prove~\cref{t:weak-ssm}. The two main differences are the utilization of a \textit{vertex weighting} in the analysis and the associated choice of a weighted matrix norm.

Fix a vertex $v$ in a tree $T$ of maximum degree $\Delta$ under some pinning; this induces an instance of list colorings $(T,\mcL)$ where every vertex $v$ is associated with a color list $\zcQ_v$ of size $q_v = |\zcQ_v|$. 
Consider the recursion in the subtree $T_v$ rooted at $v$. 
The weighted norm is defined as follows:
for a given collection of weights $\w = (w, w_1, \ldots, w_{\Delta_v}) \succ 0$ and matrix $J \in \RR^{q \times \Delta_v q},$ define norm 
\begin{equation}\label{d:wnorm}
    \| J\|_{\w} := \sup_{\x \in \RR^{\Delta_v q}} \left\{ \frac{w \| J \x\|_2}{\max_{i \in [\Delta_v]} \left\{ w_i \|\x_i\|_2\right\}} \right\},
\end{equation}
where each $\x_i \in \RR^q$. 
Here one can think of $w$ as the weight for vertex $v$ and $w_i$ as the weight for child $i$ for each $i \in [\Delta_v]$. 

This vertex weighted norm will allow us to take advantage of discrepancies between the marginal distributions of the children of a given vertex in a way that~\cref{t:weak-ssm} could not. This will allow us to obtain the full strength of~\cref{t:strong-ssm}. For example, the following example highlights a certain ``tightness'' of the earlier analysis in~\cref{sec:colorings-simple} that must be overcome to obtain strong spatial mixing with fewer colors.

\begin{ex}\label{ex:bad}
Consider the $d$-ary tree rooted at $v$ and 
suppose $q = d +\gamma$ where $\gamma$ is some fixed constant independent of $d$. 
Let $\mfc \in [q]$ be a fixed color. 
Suppose that $\p_i(\mfc) = \frac{1}{d+1}$ for all $i\in [d]$. 
Meanwhile, every $\p_i$ is uniformly distributed on $\gamma = O(1)$ colors other than $\mfc$, so that $\p_i(\mfb) = (1-\frac{1}{d+1})\frac{1}{\gamma}$ for exactly $\gamma$ choices of $\mfb \in [q] \setminus \{\mfc\}$.
Furthermore, assume that every color $\mfb \in [q] \setminus \{\mfc\}$ is available for roughly $\frac{\gamma d}{q-1}$ children of $v$; this can be made precise under careful combinatorial constructions. 
Then our bound in \cref{l:tech1} is essentially tight in the large $d$ limit; namely, taking $d \to \infty$, our approach in \cref{sec:colorings-simple} can get at the best
\[
\left\| \left( (J_i g^{\phi})(\y) \right)_{i \in [d]} \right\|_{\star\star}^2 
\lesssim \frac{1}{e \left( 1-\frac{1}{\gamma} \right)^\gamma},
\]
which is greater than one for any fixed constant $\gamma$.
\end{ex}

We now give our main result of this section, which bounds the weighted norm of the Jacobian of our message recursion. 
\begin{prop}[Jacobian bound]\label{p:weight-jacobian}
Consider $\phi$ in~\cref{eq:phi-color}. 
If $q \ge \Delta+3$, then there exists a collection of weights $\{w_v\}_{v \in V}$, a collection of marginal-bounding parameters $\{\xi_v\}_{v \in V}$, and 
$\delta > 0$ such that for any non-root vertex $v$, for any $\p = (\p_1, \ldots, \p_{\Delta_v})$ with $\p_i \in \mcD(q,\frac{\xi_i}{q_i-1+\xi_i})$ (free children) or $\p_i = \ind_{\mfc}$ for some $\mfc \in [q]$ (pinned children), and for $\y = \phi(\p)$, we have that $g(\p) \in \mcD(q,\frac{\xi_v}{q_v-1+\xi_v})$ and
$$\left\| \left( (J_i g^{\phi})(\y) \right)_{i \in [\Delta_v]} \right\|_{\w} \le 1-\delta,$$
where $\w = (w, w_1, \ldots, w_{\Delta_v})$ consists of the weights of $v$ and all its $\Delta_v$ children. 
\end{prop}

\begin{rem}
The choice of $\frac{\xi_i}{q_i-1+\xi_i}$ for the upper bound on entries of $\p_i$ for each free child is given by \cref{l:marginal-strong}. 
In particular, $g(\p) \in \mcD(q,\frac{\xi_v}{q_v-1+\xi_v})$ follows from \cref{l:marginal-strong}.
\end{rem}

\cref{p:weight-jacobian} will enable us to establish the strong Jacobian condition (\cref{cond:jacobian-norm}) for $q$-colorings on trees in the regime of~\cref{t:strong-ssm}; we formally verify~\cref{t:strong-ssm} using the above Jacobian bound in~\cref{sec:main-result-proofs}.

\subsection{Jacobian $L^2$ norm bound}

In~\cref{sec:colorings-simple}, we used the simple marginal bound that if $q \ge \Delta + \gamma$ then $\p(\mfc) \le 1/\gamma$ for every vertex $v$. Here, we will leverage vertex-specific marginal bounds by carefully keeping track of the number of free children $d_i$ and available colors $q_i$ for each child $i$ of $v$. We will be able to use these in combination with the stronger marginal bounds of~\cref{l:marginal-strong} that arise from a \textit{two-step recursion} to show that we get an on-average gain, even when an individual vertex can exhibit worst-case behavior. The weights $\w$ encode this information and enable us to propagate the gains that we accrue.

As in \cref{sec:colorings-simple}, it suffices to consider the submatrix of the Jacobian with non-zero entries, which has dimensions $q \times dq$ where $d = d_v$ is the number of free children and $q = q_v = |\zcQ_v|$ is the number of available colors. 
This can be understood as an instance of list colorings $(T,\mcL)$ resulted from pinnings, where every vertex $v$ has $d_v$ children and color list $\zcQ_v$; in fact, all of our results hold for general list colorings as well.

Recall from~\cref{c:l2-j-bound} that 
for any $\p = (\p_1, \ldots, \p_d)$ with $\p_i \in \mcD(\mcL_v)$ and for $\y = \phi(\p),$ we have that
$$\left\| \left( (J_i g^{\phi})(\y) \right)_{i \in [d]} \right\|_2^2 \le \underbrace{\max_{\mfc \in \mcL_v} \left\{ \frac{1}{(1 - g_\mfc(\p))^2} \right\}}_{(1)} \cdot \underbrace{\max_{\mfc \in \mcL_v} \left\{ g_\mfc(\p) \sum_{i \in [d]} \p_i(\mfc) \right\}}_{(2)}.$$
We can upper bound (1) using the marginal bound of~\cref{l:marginal-strong}. To handle (2), we employ the following \textit{amortized marginal bound} that uses the stronger two-step marginal bounds of~\cref{l:marginal-strong} to keep of vertex-specific discrepancies between degree and available colors.
\begin{lemma}\label{l:am-marginal}
Consider a list coloring instance $(T,\mcL)$ where $q_v \ge d_v + \gamma$ for all $v \in T$ for some $\gamma \ge 2$. For each $v \in V(T)$ let 
\begin{equation}\label{eq:xi}
\xi_v := \left(1 + \frac{1}{\gamma - 1}\right)^{\frac{\gamma d_v}{q_v -1}}.
\end{equation}
Suppose that the root $r$ has $d = d_{r}$ children $v_1, \ldots, v_d$ and $q = q_{r} = |\zcQ_{r}|$ available colors. 
For any $\mfc \in \zcQ_{r}$ and any $\p = (\p_1, \ldots, \p_d)$ with $\p_i \in \mcD(\mcL_r,\frac{\xi_{v_i}}{q_{v_i}-1+\xi_{v_i}})$, 
we have that 
$$g_\mfc(\p) \sum_{i \in [d]} \p_i(\mfc) \le \frac{1}{e q} \exp \left(\frac{1}{q} \sum_{i \in [d]} \left(\frac{q_{v_i} - 1}{\xi_{v_i}} +1 \right) \ln \left(1 + \frac{\xi_{v_i}}{q_{v_i} - 1} \right) \right).$$
\end{lemma}

\begin{rem}
We remark that for $\p_i \in \mcD(\mcL_{r})$ constructed from $\p_{v_i} \in \mcD(\mcL_{v_i})$ as in~\cref{d:tree-recurse-coloring}, it is a subdistribution satisfying the marginal bound in \cref{l:marginal-strong} and hence belong to $\mcD(\mcL_r,\frac{\xi_{v_i}}{q_{v_i}-1+\xi_{v_i}})$. Here to prove \cref{p:weight-jacobian} we need to consider arbitrary $\p_i \in \mcD(\mcL_r,\frac{\xi_{v_i}}{q_{v_i}-1+\xi_{v_i}})$.
\end{rem}

We defer the proof to~\cref{ss:proof-am}; it arises from combining the tree recursion (\cref{d:tree-recurse-coloring}) with the associated marginal bounds (\cref{l:marginal-strong}). This enables us to bound the $L^2$ norm of the Jacobian as follows:

\begin{lemma}[$L^2$ bound]\label{l:l2-strong}
Consider a list coloring instance $(T, \mcL)$ where $q_v \ge d_v + \gamma$ for all $v \in T$ for some $\gamma \ge 2$. Let $\xi_v$ be defined as in~\cref{eq:xi} for each $v \in V(T)$. Furthermore, for each vertex $v$ define 
\begin{equation}\label{eq:weight}
    \zeta_v := \left(\frac{q_v - 1}{\xi_v} + 1\right) \ln \left(1 + \frac{\xi_v}{q_v - 1} \right) - 1.
\end{equation}
Then, $\zeta_v \in (0,1)$ for all $v \in V(T)$. 

Further, suppose that the root $r$ has $d = d_{r}$ children $v_1, \ldots, v_d$ and $q = q_{r} = |\zcQ_{r}|$ available colors, and let $\xi = \xi_{r}$. 
Then for any $\p = (\p_1, \ldots, \p_d)$ with $\p_i \in \mcD(\mcL_r,\frac{\xi_{v_i}}{q_{v_i}-1+\xi_{v_i}})$ and for $\y = \phi(\p)$, we have that 
$$\left\| \left( (J_i g^{\phi})(\y) \right)_{i \in [d]} \right\|_2^2 \le \frac{1}{q} \exp \left( \frac{2\xi}{q - 1} - \frac{\gamma}{q}\right) \exp\left( \frac{1}{q} \sum_{i \in [d]} \zeta_{v_i}\right).$$
\end{lemma}

\begin{proof}
Observe that $1 \le \xi_v \le 4$ for all $v$ and thus $(q_v - 1)/\xi_v \ge \frac12$ for all $v$ (our assumption requires $q_v \ge 3$). This makes $0 < \zeta_v < 1$ since $1 < (x+1) \ln(1 + 1/x) < 2$ for all $x \ge 1/2$; see also \cref{fact:entropy-like-ineq}. 

We apply~\cref{c:l2-j-bound} to get the desired bound in conjunction with~\cref{l:am-marginal,l:marginal-strong}.
\begin{align*}
\left\| \left( (J_i g^{\phi})(\y) \right)_{i \in [d]} \right\|_2^2 
&\le \max_{\mfc \in \mcL_r} \left\{ \frac{1}{(1 - g_\mfc(\p))^2} \right\} \cdot \max_{\mfc \in \mcL_r} \left\{ g_\mfc(\p) \sum_{i \in [d]} \p_i(\mfc) \right\} \\
&\le \left(1 + \frac{\xi}{q - 1} \right)^2 
\cdot \frac{1}{eq} \exp \left(\frac{1}{q} \sum_{i \in [d]} (\zeta_{v_i} + 1) \right) \tag{\cref{l:marginal-strong,l:am-marginal}} \\
&\le \exp\left( \frac{2\xi}{q - 1} \right) 
\cdot \frac{1}{q} \exp\left(\frac{d}{q} - 1\right) \exp \left(\frac{1}{q} \sum_{i \in [d]} \zeta_{v_i} \right) \\
&\le \frac{1}{q} \exp\left( \frac{2\xi}{q - 1} -\frac{\gamma}{q} \right) \exp \left(\frac{1}{q} \sum_{i \in [d]} \zeta_{v_i} \right). \qedhere
\end{align*}
\end{proof}

\subsection{Jacobian weighted norm bound}\label{sec:jacobian-weighted-subsec}

We begin by observing that the weighted norm $\|\cdot\|_{\w}$ can be upper bounded by a function of $\w$ and the $L^2$ norm. 
Therefore, the above bound on the $L^2$ norm yields an associated bound on the weighted norm $\|\cdot\|_{\w}$.
\begin{lemma}\label{l:generic-to-l2}
Let $\w = (w, w_1, \ldots, w_d) \succ 0$, and take $J \in \RR^{q \times dq}$. Then,
$$\|J\|_{\w}^2 \le \|J\|_2^2 \cdot w^2 \sum_{i \in [d]} \frac{1}{w_i^2}.$$
\end{lemma}
\begin{proof}
For any $\x \in \RR^{dq}$, we have
\begin{align*}
\|\x\|_2^2 = \sum_{i \in [d]} \|\x_i\|_2^2 
= \sum_{i \in [d]} \frac{1}{w_i^2} \cdot w_i^2 \|\x_i\|_2^2
\le \left( \sum_{i \in [d]} \frac{1}{w_i^2} \right) \max_{i \in [d]} \left\{ w_i^2 \|\x_i\|_2^2 \right\}.
\end{align*}
Therefore, we obtain that
\begin{align*}
\|J\|_{\w}^2 
= \sup_{\x \neq 0} \left\{ \frac{w^2 \|J \x\|_2^2}{\max_{i \in [d]} \left\{ w_i^2 \|\x_i\|_2^2 \right\} }  \right\}
= \sup_{\x \neq 0} \left\{ \frac{\|J \x\|_2^2}{\|\x\|_2^2} \cdot \frac{w^2 \|\x\|_2^2}{\max_{i \in [d]} \left\{ w_i^2 \|\x_i\|_2^2 \right\} } \right\}
\le \|J\|_2^2 \cdot w^2 \sum_{i \in [d]} \frac{1}{w_i^2},
\end{align*}
as claimed.
\end{proof}

Our choice of $\w$ comes from the vertex-specific marginal bounds of~\cref{l:marginal-strong} that allow us to overcome some of the sticking points in the proof of~\cref{t:weak-ssm}. 
We will choose weight vector $\w$ so that for $v \in V(T)$
\begin{equation}\label{eq:choose-w}
w_v = (1 - \zeta_v)^{-1/2},
\end{equation}
for $\zeta_v$ defined in~\cref{eq:weight}.
With respect to this choice of weights,~\cref{l:generic-to-l2} specializes to the following relationship between $\|J\|_{\w}$ and $\|J\|_2.$
\begin{cor}\label{l:lwtol2}
Let $\w = (w, w_1, \ldots, w_d) \succ 0$ with $w_i = (1 - \zeta_i)^{-1/2}$. Then $\|\cdot\|_{\w}$ is a well-defined norm and satisfies for any $J \in \RR^{q \times dq}$ that
$$\|J\|_{\w}^2 \le \frac{\sum_{i \in [d]} (1-\zeta_i)}{1-\zeta} \cdot \|J\|_2^2.$$
\end{cor}

We also need the following technical result.
\begin{claim}\label{claim:entropy-like-ineq}
We have that $$\frac{1}{1-\zeta} \le e^{\frac{\xi}{2(q-1)}}.$$
\end{claim}
\begin{proof}
Note that 
$1 - \zeta = 2 - \left(\frac{q - 1}{\xi} + 1\right) \ln\left(1 + \frac{\xi}{q - 1} \right).$ It thus suffices to show that 
\begin{align}\label{eq:1-zeta}
2 - \left(\frac{q - 1}{\xi} + 1\right) \ln\left(1 + \frac{\xi}{q - 1} \right) \ge e^{-\frac{\xi}{2(q-1)}}.
\end{align}
We may assume $q \ge 4$.
Since $\gamma \ge 2$, we have that
$$\frac{\xi}{q-1} = \frac{\left(1 + \frac{1}{\gamma - 1}\right)^{\gamma \cdot \frac{d}{q-1}}}{q-1} \le \frac{4}{q-1} < \frac{3}{2}.$$
\cref{eq:1-zeta} then follows from an application of \cref{fact:entropy-like-ineq} in \cref{sec:technical2}; we omit its proof here since it is technical and can also be easily verified numerically.
\end{proof}

We are ready to prove \cref{p:weight-jacobian} by combining \cref{l:l2-strong,l:lwtol2,claim:entropy-like-ineq}. 
\begin{proof}[Proof of~\cref{p:weight-jacobian}]
We choose weight vector $\w = (w_v)_{v \in V}$ where 
$w_v = (1 - \zeta_v)^{-1/2}$ for $\zeta_v$ defined as in~\cref{eq:weight}. 
Applying~\cref{l:lwtol2} in conjunction with~\cref{l:l2-strong,claim:entropy-like-ineq} gives that 
\begin{align*}
\left\|\left( (J_i g^{\phi})(\y) \right)_{i \in [d]}\right\|_{\w}^2 
&\le \frac{\sum_{i \in [d]} (1-\zeta_i)}{1-\zeta}\cdot \left\|\left( (J_i g^{\phi})(\y) \right)_{i \in [d]} \right\|_2^2 \\
&\le \exp\left( \frac{\xi}{2(q-1)} \right) 
\cdot \exp\left( \frac{2\xi}{q - 1} - \frac{\gamma}{q} \right) 
\cdot \left( \frac{d}{q} - \frac{1}{q} \sum_{i \in [d]} \zeta_i \right) \exp \left( \frac{1}{q}\sum_{i \in [d]} \zeta_i \right)
\end{align*}
where $\xi = \xi_{v}$ as defined in~\cref{eq:xi} and $d=d_v$, $q = q_{v}.$

Consider the function $f(y) := (\frac{d}{q} - y) e^{y}$ at $y_{\zeta} = \frac{1}{q}\sum_{i \in [d]} \zeta_i \in (0,d/q)$. 
Note that $f'(y)= (-1+\frac{d}{q} - y) e^{y} < 0$ for all $0 \le y \le d/q$ since $d < q$. Thus, $f(y)$ is monotone decreasing on the interval $[0,d/q]$ and we have the bound
\[
\left( \frac{d}{q} - \frac{1}{q} \sum_{i \in [d]} \zeta_i \right) \exp \left( \frac{1}{q}\sum_{i \in [d]} \zeta_i \right)
\le f(0) = \frac{d}{q} \le 1 - \frac{\gamma}{q} \le \exp\left( - \frac{\gamma}{q} \right).
\]

Combining the above two inequalities and noting that $d \le q - \gamma$ gives that
\begin{align*}
\left\|\left( (J_i g^{\phi})(\y) \right)_{i \in [d]}\right\|_{\w}^2 
\le \exp \left(\frac{5\xi}{2(q-1)} - \frac{2\gamma}{q}\right).
 \end{align*}
We pick $\gamma = 4$; note that $d_v \le \Delta_v = \zcDeg(v) -1 \le \Delta-1$ for all non-root vertices so letting $\gamma=4$ is compatible with our assumption of $q \ge \Delta + 3$. From~\cref{eq:xi} we have 
$$ \xi \le \left(1 + \frac{1}{\gamma - 1} \right)^{\gamma \cdot \frac{q-\gamma}{q-1}} 
\le \left( \frac{4}{3} \right)^{4 \cdot \frac{q-4}{q-1}}
= \left( \frac{4}{3} \right)^4 \cdot \left( \frac{4}{3} \right)^{-\frac{12}{q-1}}. $$ 
Substituting shows that the above norm is $< 1$; more precisely
\[
\frac{5\xi}{2} \cdot \frac{q}{q-1}
\le \frac{5}{2} \cdot \left( \frac{4}{3} \right)^4 \cdot \left( \frac{4}{3} \right)^{-\frac{12}{q-1}} e^{\frac{1}{q-1}} 
= 8 \cdot \frac{80}{81} \cdot \exp\left( - \frac{12\ln(4/3) - 1}{q-1} \right) 
< 8 = 2\gamma,
\]
for all $q \ge 2$.
\end{proof}

\subsection{Amortized marginal bounds}\label{ss:proof-am}
We complete this section by establishing the amortized marginal bound of~\cref{l:am-marginal} which is used to prove~\cref{p:weight-jacobian}.
\begin{proof}[Proof of~\cref{l:am-marginal}]
Note that by~\cref{d:tree-recurse-coloring}, we have that
\begin{align*}
g_\mfc(\p) \sum_{i \in [d]} \p_i(\mfc) &= \frac{\prod_{i \in [d]} (1 - \p_i(\mfc)) \sum_{i \in [d]} \p_i(\mfc) }{\sum_{\mfb \in \mcL_r} \prod_{i \in [d]}(1 - \p_i(\mfb))}.
\end{align*}
We first upper bound the numerator using the AM-GM inequality:
\begin{align*}
\prod_{i \in [d]} (1 - \p_i(\mfc)) \sum_{i \in [d]} \p_i(\mfc)
&\le \left( \frac{\sum_{i \in [d]} (1 - \p_i(\mfc)) + \sum_{i \in [d]} \p_i(\mfc)}{d+1} \right)^{d+1}
= \left(1-\frac{1}{d+1}\right)^{d+1} \le \frac{1}{e}.
\end{align*}
We then give a lower bound on the denominator:
\begin{align*}
\sum_{\mfb \in \mcL_r} \prod_{i \in [d]} (1-\p_i(\mfb))
&\ge q \left( \prod_{\mfb \in \mcL_r} \prod_{i \in [d]} (1-\p_i(\mfb)) \right)^{\frac{1}{q}} \tag{AM-GM} \\
&= q \left(\prod_{i \in [d]} \prod_{\mfb \in \mcL_r}  (1-\p_i(\mfb)) \right)^{\frac{1}{q}} \\
&\ge q\left( \prod_{i \in [d]} \left( 1 - \frac{\xi_{v_i}}{q_{v_i}-1+\xi_{v_i}} \right)^{\frac{q_{v_i}-1+\xi_{v_i}}{\xi_{v_i}}} \right)^{\frac{1}{q}}, 
\end{align*}
where the last inequality follows from the assumption that $\p_i \in \mcD(\mcL_r,\frac{\xi_{v_i}}{q_{v_i}-1+\xi_{v_i}})$ for all $i$ and also \cref{lem:product-lb} in \cref{sec:technical2}.
By combining these bounds we can deduce the desired upper bound as follows:
\begin{align*}
g_\mfc(\p) \sum_{i \in [d]} \p_i(\mfc)     
&\le \frac{1}{eq} \left( \prod_{i \in [d]} \left( 1 - \frac{\xi_{v_i}}{q_{v_i}-1 + \xi_{v_i}} \right)^{\frac{q_{v_i}-1}{\xi_{v_i}} + 1} \right)^{-\frac{1}{q}} \\
&= \frac{1}{eq} \exp \left( \frac{1}{q}\sum_{i \in [d]} \left(\frac{q_{v_i} - 1}{\xi_{v_i}} +1 \right) \ln \left(1 + \frac{\xi_{v_i}}{q_{v_i} - 1} \right) \right). \qedhere
\end{align*}
\end{proof}

\section{Weak spatial mixing on trees for antiferromagnetic Potts}\label{sec:antiferro-wsm}
In this section, we prove \cref{thm:wsm-potts}, namely weak spatial mixing for the antiferromagnetic Potts model on arbitrary trees almost up to the conjectured phase transition threshold. Notably, our proof works for any tree, not just the infinite $d$-ary tree, and also does not require any lower bounds on $\Delta$. 

We first introduce the \emph{$q$-state antiferromagnetic Potts model}. 
Let $q \ge 2$ be an integer. Let $\beta \in [0,1]$ be a parameter of the model called the \emph{inverse temperature}. For a graph $G=(V,E)$, the Gibbs distribution $\mu$ of the antiferromagnetic Potts model on $G$ is expressed as
\begin{align*}
    \mu(\sigma) \propto \beta^{m(\sigma)}, \qquad \forall \sigma:V \to [q],
\end{align*}
where $m(\sigma)$ denotes the number of \emph{monochromatic} edges under $\sigma$, i.e.\ edges $e = \{u,v\} \in E$ such that $\sigma(u) = \sigma(v)$. When $0 \leq \beta \leq 1$, the system is \emph{antiferromagnetic}, since it places greater mass on configurations $\sigma$ with fewer monochromatic edges; informally, neighboring vertices ``prefer'' to have \emph{different} colors assigned to them. The limiting case $\beta = 0$ corresponds to the uniform distribution over \emph{proper $q$-colorings} of $G$ that we studied extensively earlier in this work.

\subsection{Tree recursion and marginal upper bounds}

We will leverage upper bounds on the marginal probabilities of trees that arise from a \textit{tree recursion} analogous to~\cref{d:tree-recurse-coloring} (see e.g.\ Lemma 11 of~\cite{YZ15} for a proof).

\begin{obs}[Antiferromagnetic Potts tree recursion] \label{d:tree-recurse-potts}
Consider the Gibbs distribution of the $q$-spin antiferromagnetic Potts model with parameter $0 \leq \beta \leq 1$ on a tree $T=(V,E)$ rooted at $r$, where $r$ has children $v_1, \ldots, v_{\Delta_r}$. Then for any color $\mfc \in [q]$, we have that 
$$\p_r(\mfc) = g_{\mfc}(\p_{v_1}, \ldots, \p_{v_{\Delta_r}}) = \frac{\prod_{i \in [\Delta_r]} ( 1 - (1 - \beta)\p_{v_i}(\mfc))}{\sum_{\mfb \in [q]} \prod_{i \in [\Delta_r]} (1 - (1 - \beta)\p_{v_i}(\mfb))},$$
where $\p_v(\mfc)$ denotes the marginal probability that vertex $v$ has color $\mfc.$
\end{obs}

As in the setting of proper $q$-colorings, we will repeatedly use the following strong ``two-level'' marginal bound that arises from the above tree recursion.
\begin{lemma}[Two-Level Marginal Upper Bounds]\label{lem:potts-marg-2level-new}
Consider the Gibbs distribution of the $q$-spin antiferromagnetic Potts model with parameter $0 \leq \beta \leq 1$ on a tree $T=(V,E)$ with maximum degree $\Delta$. Let $\tau : \Lambda \to [q]$ be a pinning of a subset of vertices $\Lambda \subseteq V$, and let $r \in V \setminus \Lambda$. Suppose $r$ has $\Delta_r$ total neighbors, none of which are pinned under $\tau$. Let $q \ge (1 - \beta)(\Delta + \gamma) + 1$ for $\gamma \ge 1$. Then for every color $\mfc \in [q]$, we have the marginal upper bound
\begin{align*}
    \frac{\p_{r}^{\tau}(\mfc)}{1 - \p_r^{\tau}(\mfc)} \leq  \frac{1}{q-1} \left(1 + \frac{1-\beta}{(1-\beta)(\gamma - 1)  + 1}\right)^{\frac{((1-\beta)\gamma + 1)\Delta_r}{q-1}}
\end{align*}
\end{lemma}
\begin{proof}
Let $\gamma' = (1-\beta)\gamma + 1$.
We recall the following ``one-level'' bound that generalizes the simple upper bound of~\cref{l:marginal} from the setting of proper $q$-colorings.
\begin{claim}[Lemma 13 in~\cite{YZ15}]\label{l:simple-potts-ub}
    $\p_r^{\tau}(\mfc) \le \frac{1}{\gamma'} = \frac{1}{q-(1-\beta)\Delta_r}$
\end{claim}
Let $d = \Delta_r$.
    We apply~\cref{d:tree-recurse-potts} and compute that
\begin{align*}
\frac{\p_{r}(\mfc)}{1 - \p_{r}(\mfc)} &= \frac{\prod_{i \in [d]}(1 - (1-\beta)\p_i(\mfc))}{\sum_{\mfc' \neq \mfc \in [q]} \prod_{i \in [d]} (1 - (1-\beta)\p_i(\mfc'))} \le \frac{1}{\sum_{\mfc' \neq \mfc \in [q]} \prod_{i \in [d]} (1 - (1-\beta)\p_i(\mfc'))} \\
&\le \frac{1}{q-1} \cdot \left(\prod_{\mfc' \neq \mfc \in [q]} \prod_{i \in [d]} (1 - (1-\beta)\p_i(\mfc'))\right)^{-\frac{1}{q-1}} \tag{AM-GM} \\
&\le \frac{1}{q-1} \cdot \left(\prod_{i \in [d]} \left(1 - \frac{1-\beta}{\gamma'}\right)\right)^{-\frac{\gamma'}{q-1}}  \tag{\cref{l:simple-potts-ub,lem:product-lb}} \\
&= \frac{1}{q-1} \left(1 + \frac{1-\beta}{\gamma' - (1-\beta)}\right)^{\frac{\gamma' d}{q-1}}.
\end{align*}
Substituting $\gamma' = (1-\beta)\gamma + 1$ then gives the desired result.
\end{proof}

We now combine the marginal upper bound of~\cref{lem:potts-marg-2level-new} with an adaptation of the analysis from~\cref{p:weight-jacobian} to obtain weak spatial mixing. Specifically, rearranging the above expression gives (with $\gamma' = (1-\beta)\gamma+1$):
\begin{equation}\label{e:true_ub}
\p_r(\mfc) \le \ub_{\beta, q}(\Delta_r) := \frac{1}{1 + (q-1)\left(1 - \frac{1-\beta}{\gamma'}\right)^{\frac{\gamma' \Delta_r}{q-1}}}.
\end{equation}

\color{black}

Specifically, we will use 
potential function $\phi : \R \to \R$ similar to the choice~\cref{eq:phi-color} for $q$-colorings, defined via its derivative 
\begin{equation}\label{eq:phi-potts}
\Phi(p) = \phi'(p) = \frac{1}{\sqrt{p}(1 - (1-\beta)p)}.
\end{equation}
We will then construct a weight function $\w = \w_{\beta, q}: [\Delta] \to \RR_+$ with respect to which the tree recursion of~\cref{d:tree-recurse-potts} satisfies the \textit{weak Jacobian condition} of~\cref{cond:jacobian-norm}.

In a similar fashion to the setting of $q$-colorings (see~\cref{o:jacob-phi}), the Jacobian $(J g^{\phi})$ of \textit{message recursion} $g^{\phi}$ for the antiferromagnetic Potts model is given as follows:

\begin{obs}\label{o:jacob-potts}
Let $g$ denote the tree recursion of ~\cref{d:tree-recurse-potts} and let $\phi$ denote the potential function with derivative given in~\cref{eq:phi-potts}. Then, we have that
\begin{align*}
    (J_{i}g^{\phi})(\bm{m}) &= \diag(\Phi(g(\bm{p}))) \cdot (J_i g)(\bm{p}) \cdot \diag(\Phi(\bm{p}_{i}))^{-1} \\
    &= -(1-\beta)\diag(1 - (1-\beta)g(\bm{p}))^{-1} \cdot \wrapp{I - \sqrt{g(\bm{p})}\sqrt{g(\bm{p})}^{\top}} \cdot \diag\wrapp{\sqrt{g(\bm{p}) \odot \bm{p}_{i}}}.
    \end{align*}
\end{obs}

\subsection{Weak spatial mixing}
We begin our proof of weak spatial mixing by observing an analogous $L^2$ bound on the Jacobian of the message recursion for the antiferromagnetic Potts model to that shown for $q$-colorings in~\cref{l:l2-strong}.

\begin{lemma}\label{prop:potts-jacobian-operator-norm}
We have that $\norm{\left((J_ig^{\phi})(\bm{m})\right)_{i \in [\Delta_r]}}_{2}^{2}$ is upper bounded by
\begin{align*}
\frac{\theta}{q}
\exp \wrapp{\frac{2 \theta B}{1 - \theta B} + \frac{\theta\Delta_{r}}{q} - 1} 
\exp\wrapp{\frac{\theta}{q} \sum_{i=1}^{\Delta_{r}} s\left(\theta \ub_{\beta,q}(\Delta_{v_i})\right)},
\end{align*}
where $\theta = 1-\beta$, $B = \ub_{\beta,q}(\Delta_{r})$ for $\ub_{\beta, q}$ given in~\cref{e:true_ub}, and $s:(0,1) \to \R$ is defined by
\begin{equation}\label{eq:s-func}
s(x) = \frac{1}{x} \ln \frac{1}{1-x} - 1.
\end{equation}
\color{black}
\end{lemma}
The proof of this upper bound is similar to~\cref{l:l2-strong} and thus we defer it to the end of this section.

It remains to define a weight function with respect to which we hope our Jacobian is contractive. We chart a course inspired by~\cref{sec:jacobian-weighted-subsec}, choosing weights $w_v$ very similarly to~\cref{eq:choose-w}. For $\theta = 1-\beta$ and $v \in V(T)$, let 
\begin{equation}\label{eq:weight-potts}
\zeta_v := s\left(\theta\ub_{\beta, q}(\Delta_v)\right), \quad w_v := (1 - \zeta_v)^{-1/2}.
\end{equation}
We begin by verifying that this weight function is well-defined.
\begin{claim}
We have that $\zeta_v \in (0,1)$ for all $v \in V(T)$ if $q = (1-\beta)(\Delta_r + \gamma) + 1$ for $q \ge 4$ and $\gamma \ge 4$.
\end{claim}
\begin{proof}
Consider an arbitrary vertex $r$ in the tree. 
We expand out the definition of $\ub_{\beta, q}(\Delta_r)$ from~\cref{e:true_ub} and find that 
\begin{align}\label{eq:b-potts-ub-new}
\ub_{\beta, q}(\Delta_r) &= \frac{1}{1 + (q-1)\left(1 - \frac{\theta}{\theta \gamma + 1}\right)^{\frac{(\theta \gamma + 1)\Delta_r}{q-1}}} \\
&\le  \frac{1}{1 + (q-1)e^{-\frac{\theta(\theta \gamma + 1) \Delta_r}{(\theta \gamma + 1 - \theta)(q-1)}}} \tag{$1 -x \ge \exp\left(-\frac{x}{1-x}\right)$} \\
&= \frac{1}{1 + (q-1)e^{-\frac{(\theta \gamma + 1) \Delta_r}{(\theta \gamma + 1 - \theta)( \Delta_r + \gamma)}}} \tag{$q = (1 -\beta)(\Delta_r + \gamma) + 1$} \\
&= \frac{1}{1 + (q-1)e^{-\frac43 \frac{\Delta_r}{( \Delta_r + 4)}}} \tag{decreasing in $\gamma$, $\gamma \ge 4$} \\
&\le \frac{1}{1 + (q-1)e^{-\frac43}} \tag{increasing in $\Delta_r$, $\Delta_r \ge 2$} \\
&\le \frac{1}{1+ 3 e^{-4/3}} < \frac{3}{5} \tag{decreasing in $q$, $q \ge 4$}
\end{align}
We then deduce that  $\zeta_r = s\left( \theta\ub_{\beta, q}(\Delta_r)\right) \in (0,1)$ by~\cref{fact:entropy-like-ineq}. 
\end{proof}

The main technical lemma gives contraction of the message recursion~\cref{d:tree-recurse-potts} with respect to the weights~\cref{eq:weight-potts}. This will allow us to conclude in~\cref{sec:condition} that the tree recursion for antiferromagnetic Potts model satisfies the weak Jacobian norm condition and thus prove~\cref{thm:wsm-potts}; a formal verification is in~\cref{sec:main-result-proofs}.

\begin{prop}\label{p:jacobian-potts}
Consider $\phi$ as in~\cref{eq:phi-potts} and tree $T$ with maximum degree $\Delta$ such that $q \ge (1 - \beta)(\Delta + \gamma) + 1$ for  $\gamma \ge 5$. Suppose the root $r$ has children $v_1, \ldots, v_{\Delta_r}$ and let $\w = (w, w_1, \ldots, w_{\Delta_r}) \succ 0$ with $w_i$ defined in~\cref{eq:weight-potts}. 
Then there exists $\delta > 0$ such that for any $\p = (\p_1, \ldots, \p_{\Delta_r})$ with $\p_i \in \mcD(q,\ub_{\beta,q}(\Delta_{v_i}))$ and for $\bm{m} = \phi(\p)$, we have that $g(\p) \in \mcD(q,\ub_{\beta,q}(\Delta_r))$ and
$$\norm{\left((J_ig^{\phi})(\bm{m})\right)_{i \in [\Delta_r]}}_{\w}  \le 1- \delta.$$
\end{prop}
\begin{proof}
Combining~\cref{prop:potts-jacobian-operator-norm} with~\cref{l:lwtol2} gives that
\begin{align*}
\norm{\left((J_ig^{\phi})(\bm{m})\right)_{i \in [\Delta_r]}}_{\w}^2 
&\le \frac{ \sum_{i=1}^{\Delta_{r}} (1-\zeta_{v_i}) }{1 - \zeta_r} \norm{\left((J_ig^{\phi})(\bm{m})\right)_{i \in [\Delta]}}_{2}^2 \\
&\le \frac{1}{1 - \zeta_r} \cdot 
\exp \wrapp{\frac{2 \theta B}{1 - \theta B} + \frac{\theta\Delta_{r}}{q} - 1} 
\cdot \left( \frac{\theta \Delta_r}{q} - \frac{\theta}{q} \sum_{i=1}^{\Delta_{r}} \zeta_{v_i} \right)
\exp\wrapp{\frac{\theta}{q} \sum_{i=1}^{\Delta_{r}} \zeta_{v_i}},
\end{align*}
where $\theta = 1-\beta$ and $B = \ub_{\beta,q}(\Delta_{r})$. 

By \cref{fact:entropy-like-ineq} in \cref{sec:technical2}, we have that
\[
\frac{1}{1 - \zeta_r} = \frac{1}{1 - s\left(\theta B\right)} \le \exp\left( \frac{\theta B}{2(1-\theta B)} \right).
\]
Note that to apply~\cref{fact:entropy-like-ineq}, we used that $q \ge \theta(\Delta + \gamma) + 1$ for $\gamma \ge 4$ and thus $\theta B \le B \le 3/5$ as shown in~\cref{eq:b-potts-ub-new}.

Using the fact that the function $f(y) := (\theta \Delta_r/q - y)e^y$ is monotone decreasing on the interval $[0,\theta \Delta_r/q]$ since $q \ge \theta \Delta_r$, we obtain that
\[
\left( \frac{\theta \Delta_r}{q} - \frac{\theta}{q} \sum_{i=1}^{\Delta_{r}} \zeta_{v_i} \right)
\exp\wrapp{\frac{\theta}{q} \sum_{i=1}^{\Delta_{r}} \zeta_{v_i}}
\le f(0)
= \frac{\theta \Delta_r}{q} 
\le \exp\left( \frac{\theta\Delta_r}{q} - 1 \right),
\]
where the last inequality follows from $x = 1+(x-1) \le e^{x-1}$ for all $x \in \R$. 

Combining everything above, we deduce that
\begin{align*}
\norm{\left((J_ig^{\phi})(\bm{m})\right)_{i \in [\Delta_r]}}_{\w}^2 
\le \exp \wrapp{\frac{5 \theta B}{2(1 - \theta B)} + \frac{2\theta\Delta_{r}}{q} - 2}.
\end{align*}
We will show that the argument of the $\exp(\cdot)$ is less than $0$. 
We first deduce from \cref{eq:b-potts-ub-new} that
$$\frac{\theta B}{1 - \theta B} \le \frac{\frac{\theta}{1+(q-1)e^{-4/3}}}{1-\frac{\theta}{1+(q-1)e^{-4/3}}} = \frac{\theta}{(q-1)e^{-4/3} + 1-\theta} \le \frac{e^{4/3}}{\Delta + \gamma}$$
Meanwhile, we have from $q \ge \theta(\Delta+\gamma)+1$ that
\[
1 - \frac{\theta\Delta_{r}}{q} 
\ge 1 - \frac{\theta\Delta}{\theta(\Delta+\gamma)} 
= \frac{\gamma}{\Delta+\gamma}.
\]
Combining the above two inequalities, we obtain
\begin{align*}
\frac{5 \theta B}{2(1 - \theta B)} + \frac{2\theta\Delta_{r}}{q} - 2
\le \left( \frac{5e^{4/3}}{2} - 2 \gamma \right) \cdot \frac{1}{\Delta+\gamma} < 0,
\end{align*}
when $\gamma \ge 5 > 5e^{4/3}/4$. 
Therefore, we obtain the desired contraction for the Jacobian whenever $q \ge (1 - \beta)(\Delta + 5) + 1$, as claimed.
\end{proof}

\subsection{$L^2$ norm upper bound}
We now complete the proof of~\cref{prop:potts-jacobian-operator-norm}, giving an $L^2$ bound on the Jacobian of our message recursion. 
\begin{proof}[Proof of \cref{prop:potts-jacobian-operator-norm}]
Given the form of \cref{o:jacob-potts}, we have that for all $\x \in \RR^{\Delta_r q}$,
\begin{align*}
    & \norm{\sum_{i=1}^{\Delta_{r}} (J_{i}g^{\phi})(\bm{m}) \cdot \bm{x}_{i}}_{2}^{2} \\
    \leq{}& \theta
    \cdot \norm{\diag(1 - \theta g(\bm{p}))^{-1}}_{2}^{2} 
    \cdot \norm{I - \sqrt{g(\bm{p})}\sqrt{g(\bm{p})}^{\top}}_2^2 
    \cdot \theta \norm{\sum_{i=1}^{\Delta_{r}} \diag\wrapp{\sqrt{g(\bm{p}) \cdot \bm{p}_{i}}} \cdot \bm{x}_{i}}_{2}^{2}. 
\end{align*}
Note that
\[
\norm{I - \sqrt{g(\bm{p})}\sqrt{g(\bm{p})}^{\top}}_2^2 = 1
\]
since it is a projection matrix. 
Also, we deduce from \cref{e:true_ub} that 
\begin{align*}
\norm{\diag(1 - \theta g(\bm{p}))^{-1}}_{2}^{2} 
&= \max_{\mfc \in [q]} \wrapc{\frac{1}{(1 - \theta g_{\mfc}(\bm{p}))^2}} 
\le \frac{1}{(1-\theta B)^2} 
\le \exp\left( \frac{2\theta B}{1-\theta B} \right),
\end{align*}
where $B = \ub_{\beta,q}(\Delta_{r})$.
Finally, it follows from \cref{lem:norm-concat-diag} that
\begin{align*}
\theta \norm{\sum_{i=1}^{\Delta_{r}} \diag\wrapp{\sqrt{g(\bm{p}) \cdot \bm{p}_{i}}} \cdot \bm{x}_{i}}_{2}^{2} 
\le \max_{\mfc \in [q]} \wrapc{\sum_{i=1}^{\Delta_{r}} g_{\mfc}(\bm{p}) \cdot \theta \bm{p}_{i}(\mfc)} \cdot \norm{\bm{x}}_{2}^{2}.
\end{align*}
It remains to bound $\sum_{i=1}^{\Delta_{r}} g_{\mfc}(\bm{p}) \cdot \theta \bm{p}_{i}(\mfc)$. 
For fixed $\mfc \in [q]$, we have that
\begin{align*}
\sum_{i=1}^{\Delta_{r}} g_{\mfc}(\bm{p}) \cdot \theta\bm{p}_{i}(\mfc) =   \frac{\prod_{i=1}^{\Delta_{r}} (1 - \theta\bm{p}_{i}(\mfc)) \sum_{i=1}^{\Delta_{r}} \theta\bm{p}_{i}(\mfc)}{\sum_{\mfb \in [q]} \prod_{i=1}^{\Delta_{r}} (1 - \theta\bm{p}_{i}(\mfb))}.
\end{align*}
We upper bound the numerator and lower bound the denominator separately.
By the AM-GM inequality, we have $\prod_{i=1}^{\Delta_{r}} (1 - \theta\bm{p}_{i}(\mfc)) \leq \wrapp{1 - \frac{\theta}{\Delta_{r}} \sum_{i=1}^{\Delta_{r}} \bm{p}_{i}(\mfc)}^{\Delta_{r}}$. Consequently, the numerator is upper bounded by 
$$\wrapp{1 - \frac{1}{\Delta_{r}} \sum_{i=1}^{\Delta_{r}} \theta\bm{p}_{i}(\mfc)}^{\Delta_{r}}\sum_{i=1}^{\Delta_{r}} \theta \bm{p}_{i}(\mfc) 
\le \exp\left(-\sum_{i=1}^{\Delta_{r}} \theta\bm{p}_{i}(\mfc)\right) \left( \sum_{i=1}^{\Delta_{r}} \theta\bm{p}_{i}(\mfc)\right) \le \frac{1}{e},$$
since $x e^{-x} \le 1/e$ for $x \ge 0$. We then proceed to lower bound the denominator.
\begin{align*}
    \sum_{\mfb \in [q]} \prod_{i=1}^{\Delta_{r}} (1 - \theta \bm{p}_{i}(\mfb)) 
    &\geq q \prod_{i=1}^{\Delta_{r}} \prod_{\mfb \in [q]} (1 - \theta \bm{p}_{i}(\mfb))^{1/q} \tag{AM-GM}\\
    &\geq q \prod_{i=1}^{\Delta_{r}} \wrapp{1 - \theta \ub_{\beta,q}(\Delta_{v_i})}^{\frac{1}{q \cdot \ub_{\beta,q}(\Delta_{v_i})}} \tag{\cref{lem:product-lb}} \\
    &= q \exp\wrapp{-\frac{\theta}{q} \sum_{i=1}^{\Delta_{r}} s\left(\theta\ub_{\beta,q}(\Delta_{v_i}) \right)} \cdot \exp\wrapp{-\frac{\theta \Delta_{r}}{q}} \tag{\cref{eq:s-func}}.
\end{align*}
Combining these two inequalities, it follows that
\begin{align*}
    \max_{\mfc \in [q]} \wrapc{\sum_{i=1}^{\Delta_{r}} g_{\mfc}(\bm{p}) \cdot \theta \bm{p}_{i}(\mfc)} 
    &\leq \frac{1}{q} \cdot \exp\wrapp{\frac{\theta\Delta_{r}}{q}-1} \cdot \exp\wrapp{\frac{\theta}{q} \sum_{i=1}^{\Delta_{r}} s\left(\theta\ub_{\beta,q}(\Delta_{v_i})\right)},
\end{align*}
concluding the proof.
\end{proof}

\section{From contraction to correlation decay properties}
\label{sec:condition}

Earlier, we established contraction of the Jacobian of the tree recursions associated to $q$-colorings and the antiferromagnetic Potts models on trees. In this section, we give a black-box condition, the \textit{Jacobian norm condition}, that allows us to derive a variety of useful \textit{structural properties} of a Gibbs distribution from contraction of an associated Jacobian. We begin by formally introducing this unifying condition in~\cref{sec:jacobian-norm-cond}. In~\cref{a:contraction-ssm} we deduce strong (resp.\ weak) spatial mixing from the strong (resp.\ weak) Jacobian norm condition and in~\cref{a:contraction-si}, we deduce spectral independence from the strong Jacobian norm condition. We specialize these reductions to complete the proofs of~\cref{t:weak-ssm},~\cref{t:strong-ssm}, and~\cref{thm:wsm-potts} in~\cref{sec:main-result-proofs}.

\subsection{Jacobian norm condition}\label{sec:jacobian-norm-cond}

Motivated by~\cref{p:weak},~\cref{p:weight-jacobian}, and~\cref{p:jacobian-potts}, we formulate below a single unifying condition on the Jacobian of the tree recursion that is sufficient to imply all of the structural results we desire, including weak/strong spatial mixing, total influence decay, spectral independence. In~\cref{sec:main-result-proofs}, we will verify that this condition holds in the setting of each of our main results.

In our arguments, having control over the marginal distributions of vertices was critical. We formulate the notion of a \textit{marginal bounding function} to emphasize that our bounds only depended on \emph{limited local data}, namely the degrees of vertices.
\begin{defn}[Marginal Bounding Function]\label{def:marginal-bound-func}
Fix a positive integer $\Delta \in \N$. For a $q$-spin system $(A,h)$, we say $\lb = \lb_{A,h} : [\Delta] \times [\Delta] \to [0,1]$ (resp.\ $\ub = \ub_{A,h} : [\Delta] \times [\Delta] \to [0,1]$) is a \emph{marginal lower (resp.\ upper) bounding function} if for every tree $T=(V,E)$ of maximum degree $\Delta$ with associated Gibbs measure $\mu$, every pinning $\tau : \Lambda \to [q]$ where $\Lambda \subseteq V$, and every $r \in V \setminus \Lambda$ with $\Delta_{r} \leq \Delta$ total children and $d_{r} \leq \Delta_{r}$ children not pinned under $\tau$, we have the marginal bound $\mu_{r}^{\tau}(\mfc) \geq \lb(d_{r},\Delta_{r})$ (resp.\ $\mu_{r}^{\tau}(\mfc) \leq \ub(d_{r},\Delta_{r})$) for every $\mfc \in [q]$. For convenience, we write $\lb^{*} \defeq \inf_{d,d' \leq \Delta} \lb(d,d')$, and $\ub^{*} \defeq \sup_{d,d' \leq \Delta} \ub(d,d')$.
\end{defn}
We have formulated this definition only for trees because we focus primarily on the setting of trees in this section. We note that all of the bounding functions we have used in this work extend much more generally (e.g.\ to all triangle-free graphs).

\begin{cond}[Jacobian Norm Condition]\label{cond:jacobian-norm}
Fix a positive integer $\Delta \in \N$ and a real number $0 < \delta < 1$. Let $\ub : [\Delta] \times [\Delta] \to [0,1]$ be a marginal upper bounding function (as in~\cref{def:marginal-bound-func}), $w : [\Delta] \times [\Delta] \to \R_{\geq0}$ be a weight function, and $\phi:[0,1] \to \R_{\geq0} \cup \{\infty\}$ be a potential function which is strictly increasing on $[0,1]$ and concave on $[0, \ub^{*}]$. We say the tree recursion $g$ satisfies the \emph{strong Jacobian condition with respect to $(\phi,\ub,w,\Delta)$ with rate $1 - \delta$} if for every $d_{r} \leq \Delta_{r} \leq \Delta$, every $d_{i} \leq \Delta_{i} \leq \Delta - 1$ where $i \in [\Delta_{r}]$, 
\begin{align}\label{eq:jacobian-norm}
    \norm{(Jg^{\phi})(\bm{m}_{1},\dots,\bm{m}_{\Delta_{r}})}_{\bm{w}} = \sup_{\bm{x} \neq \bm{0}} \wrapc{\frac{w(d_{r},\Delta_{r}) \cdot \norm{\sum_{i=1}^{d_{r}} (J_{i}g^{\phi})(\bm{m}_{1},\dots,\bm{m}_{\Delta_{r}}) \cdot \bm{x}_{i}}_{2}}{\max_{i\in[d_{r}]}\wrapc{w(d_{i},\Delta_{i}) \cdot \norm{\bm{x}_{i}}_{2}}}} \leq 1 - \delta,
\end{align}
for all $\bm{m}_{i} = \phi(\bm{p}_{i})$, where $\bm{w} = (w(d_{r},\Delta_{r}), w(d_{1},\Delta_{1}),\dots,) \in \R_{\geq0}^{d_{r} + 1}$, and the $\bm{p}_{i}$ satisfy the following restrictions:
\begin{enumerate}
    \item For every $i\in[d_{r}]$, $\bm{p}_{i} \in \mcD(q,\ub(d_{i},\Delta_{i}))$.
    \item For every $i \in [\Delta_{r}] \setminus [d_{r}]$, $\bm{p}_{i}$ is equal to $\ind_{\mfc_{i}}$ for some $\mfc_{i} \in [q]$.
    \item $g(\bm{p}_{1},\dots,\bm{p}_{\Delta_{r}}) \in \mcD(q,\ub(d_{r},\Delta_{r}))$.
\end{enumerate}
We say $g$ satisfies the \emph{weak Jacobian condition with respect to $(\phi,\ub,w,\Delta)$ with rate $1 - \delta$} if the above only holds when $d_{r} = \Delta_{r}$.
\end{cond}
By relatively standard arguments in the literature, this condition will imply weak/strong spatial mixing, influence bounds, and spectral independence on trees. We will state and prove these implications formally in \cref{thm:ssm-from-contraction,thm:si-from-contraction}.

We emphasize that we make no requirement that the subdistributions $\p_i$ satisfy nontrivial marginal lower bounds. This is critical, since we compose these subdistributions with nonlinear functions $\phi$ and $\phi^{-1}$. 

For intuition on the distinction between the strong and weak versions of~\cref{cond:jacobian-norm}, note that the second restriction on the $\bm{p}_{i}$ along with $d_{r} \leq \Delta_{r}$ reflect the fact that for strong spatial mixing, one must handle the possibility of pinned vertices which are arbitrarily close to the root. On the other hand, the fact that we only optimize over $d_{r}$ many test vectors $\bm{x}_{1},\dots,\bm{x}_{d_{r}} \in \R^{q}$ (corresponding to the \emph{unpinned} children), as opposed to $\Delta_{r}$ many, reflects the fact that the pinned children must agree under different boundary conditions $\tau,\tau'$ (this is despite there being $\Delta_{r}$ many input vectors to the recursion $g^{\phi}$).

Our main theorems will follow from the technical work in earlier sections via suitable choices of parameters $(\phi, \ub, w, \Delta)$. As a reference, we summarize our parameter choices to prove the main results in~\cref{table:jacobian-cond-params}, which we make rigorous in~\cref{sec:main-result-proofs}.

\begin{table}[t] 
\centering
\begin{tabular}{|c|c|c|c|c|c|c|}
\hline
Result & Condition & Contraction & $\Phi = \phi'$ & $\ub$ & $\w$ \\
 \hline
\cref{t:weak-ssm} & $q \ge \Delta + 3\sqrt{\Delta}$ & \cref{p:weak} & $\frac{1}{\sqrt{x}(1-x)}$ & \cref{l:marginal} & $\ind$ \\
\hline 
\cref{t:strong-ssm}  & $q \ge \Delta + 3$ & \cref{p:weight-jacobian} & $\frac{1}{\sqrt{x}(1-x)}$ & \cref{l:marginal-strong} & \cref{eq:choose-w} \\
\hline 
\cref{thm:wsm-potts}  & $q \ge (1- \beta)(\Delta + 5) + 1$ & \cref{p:jacobian-potts} & $\frac{1}{\sqrt{x}(1-(1 - \beta)x)}$ & \cref{e:true_ub} & \cref{eq:weight-potts}  \\
\hline 
\end{tabular}
\caption{Parameter choices with which~\cref{cond:jacobian-norm} holds in our main results}
\label{table:jacobian-cond-params}
\end{table}

\subsection{Strong and weak spatial mixing via contraction}\label{a:contraction-ssm}
In this section, we show that \cref{cond:jacobian-norm} implies weak/strong spatial mixing of the Gibbs distribution. 
\begin{thm}\label{thm:ssm-from-contraction}
Let $\Delta \in \N$ be a positive integer, and $0 < \delta < 1$ be a real. Suppose the tree recursion $g$ satisfies the strong (resp.\ weak) version of \cref{cond:jacobian-norm} with rate $1 - \delta$ with respect to $(\phi,\ub,w,\Delta)$.
Then for every tree $T=(V,E)$ of maximum degree $\Delta$, the corresponding Gibbs measure $\mu$ exhibits strong (resp.\ weak) spatial mixing with rate $1 - \delta$ and constant
\begin{equation}\label{eq:csm}
    C_{\SM} \defeq \frac{\sqrt{q}}{1-\delta} \cdot \frac{w_{\max} \cdot \Phi(\lb^{*})}{w_{\min} \cdot \Phi(\ub^{*})},
\end{equation}
where $w_{\max} \defeq \max_{d,d' \leq \Delta} w(d,d')$, $w_{\min} \defeq \min_{d,d' \leq \Delta} w(d,d')$, $\Phi = \phi'$ is the derivative of the potential, and $\ub, \lb$ are marginal upper and lower bounding functions (respectively) as in~\cref{def:marginal-bound-func}.
\end{thm}

The key step towards proving this theorem is to show that bounds on the Jacobian norm imply that each individual application of the tree recursion reduces the discrepancy between the marginal distributions induced by two different boundary conditions. Once we establish this, we can iterate this one-step contractive inequality.
\begin{prop}[One-step contraction]\label{p:weight-contract}
Suppose we are in the setting of \cref{thm:ssm-from-contraction}. Let $T=(V,E)$ be a tree, $\tau,\tau' : \Lambda \to [q]$ be two pinnings on $\Lambda \subseteq V$, and $r \in V \setminus \Lambda$ be a fixed root vertex with $\Delta_{r}$ total children $u_{1},\dots,u_{\Delta_{r}}$. Suppose $\dist_T(r, \partial_{\tau, \tau'}) \ge 2$. Let $\bm{p}_{i}$ (resp.\ $\bm{p}_{i}'$) be the marginal distribution of $u_{i}$ with respect to the subtree $T_{u_{i}}$ conditioned on $\tau$ (resp.\ $\tau'$), and let $\bm{m}_{i} = \phi(\bm{p}_{i})$ entrywise (resp.\ $\bm{m}_{i}' = \phi(\bm{p}_{i}')$). If the strong version of \cref{cond:jacobian-norm} holds, and $u_{1},\dots,u_{d_{r}}$ are unpinned for $d_{r} \leq \Delta_{r}$, then we have the one-step contractive inequality
\begin{align*}
    w(d_{r},\Delta_{r}) \cdot \norm{g^{\phi}(\bm{m}_{1},\dots,\bm{m}_{\Delta_{r}}) - g^{\phi}(\bm{m}_{1}',\dots,\bm{m}_{\Delta_{r}}')}_{2} \leq (1 - \delta) \cdot \max_{i \in [d_{r}]} \wrapc{w(d_{i},\Delta_{i}) \cdot \norm{\bm{m}_{i} - \bm{m}_{i}'}_{2}}.
\end{align*}
Similarly, this inequality holds assuming $d_{r} = \Delta_{r}$ and the weak version of \cref{cond:jacobian-norm}.
\end{prop}
\begin{proof}
For convenience, write $w_{r} = w(d_{r},\Delta_{r})$, $w_{i} = w(d_{i},\Delta_{i})$, $b_{r} = \ub(d_{r},\Delta_{r})$, and $\ub_{i} = \ub(d_{i},\Delta_{i})$. By the Mean Value Theorem, if $\bm{m}(t) = t \cdot \bm{m} + (1 - t) \cdot \bm{m}'$ denotes the linear interpolation, then
\begin{align*}
    g^{\phi}(\bm{m}) - g^{\phi}(\bm{m}') = \int_{0}^{1} (Jg^{\phi})(\bm{m}(t)) 
    \cdot (\bm{m} - \bm{m}') \,dt.
\end{align*}
It follows that
\begin{align*}
    w_{r} \cdot \norm{g^{\phi}(\bm{m}) - g^{\phi}(\bm{m}')}_{2} &= w_{r} \cdot \norm{\int_{0}^{1} (Jg^{\phi})(\bm{m}(t)) \cdot (\bm{m} - \bm{m}') \,dt}_{2} \\
    &\leq \int_{0}^{1} w_{r} \cdot \norm{(Jg^{\phi})(\bm{m}(t)) \cdot (\bm{m} - \bm{m}')}_{2} \,dt \\
    &\leq \int_{0}^{1} \norm{(Jg^{\phi})(\bm{m}(t))}_{\bm{w}} \cdot \max_{i\in[d_{r}]}\wrapc{w_{i} \cdot \norm{\bm{m}_{i} - \bm{m}_{i}'}_{2}} \,dt \\
    &\leq \sup_{0 \leq t \leq 1} \wrapc{\norm{(Jg^{\phi})(\bm{m}(t))}_{\bm{w}}} \cdot \max_{i\in[d_{r}]}\wrapc{w_{i} \cdot \norm{\bm{m}_{i} - \bm{m}_{i}'}_{2}} \\
    &\leq (1 - \delta) \cdot \max_{i\in[d_{r}]}\wrapc{w_{i} \cdot \norm{\bm{m}_{i} - \bm{m}_{i}'}_{2}} \tag{$\ast$}.
\end{align*}
All that remains is to justify $(\ast)$. By assumption, it suffices to show that for every $0 \leq t \leq 1$, the vector $\phi^{-1}(\bm{m}_{i}(t))$ lies in $\mcD(q, \ub_{i})$. This is nontrivial since we do not impose linearity on $\phi$, so we cannot say that $\phi^{-1}(\bm{m}_{i}(t)) = t \cdot \bm{p}_{i} + (1 - t) \cdot \bm{p}_{i}'$. However, since our norm bound holds even for subdistributions, concavity and monotonicity of $\phi$ on $[0,\ub_{i}]$ are enough. In particular, $\phi$ is strictly increasing and concave on $[0,\ub^{*}]$, $\phi^{-1}$ is convex on $\phi([0,\ub^{*}])$, so
\begin{align*}
    \phi^{-1}(t \cdot \bm{m}_{i} + (1 - t) \cdot \bm{m}_{i}') \leq t \cdot \phi^{-1}(\bm{m}_{i}) + (1 - t) \cdot \phi^{-1}(\bm{m}_{i}') = t \cdot \bm{p}_{i} + (1 - t) \cdot \bm{p}_{i}' \leq \ub_{i}
\end{align*}
holds entrywise. The same argument also shows that $\phi^{-1}(\bm{m}_{i}(t))$ remains a subdistribution, and so we indeed have $\phi^{-1}(\bm{m}_{i}(t)) \in \mcD(q, \ub_{i})$ for all $i\in[d_{r}]$. Thus $\norm{(Jg^{\phi})(\bm{m}(t))}_{\bm{w}} \leq 1 - \delta$ is valid.
\end{proof}

We now prove~\cref{thm:ssm-from-contraction}, recalling that for a vertex $u \in V(T)$, we let $S(u, \ell) = \{v \in V(T) : \dist_T(u, v) = \ell\}$ be the set of vertices in $T$ at graph distance exactly $\ell$ from $u$.
\begin{proof}[Proof of \cref{thm:ssm-from-contraction}]
We show that the strong version of \cref{cond:jacobian-norm} implies strong spatial mixing; essentially the same argument shows that the weak version of \cref{cond:jacobian-norm} implies weak spatial mixing.

Let $T = (V, E)$ be a tree with maximum degree $\Delta$ and let $\tau,\tau' : \Lambda \to [q]$ be distinct pinnings on $\Lambda \subseteq V$. Fix a root vertex $r \in V \setminus \Lambda$. For convenience, write $w_{v} = w(d_{v},\Delta_{v})$ for every vertex. Let $\ell = \dist_T(r,\partial_{\tau,\tau'})$, where recall $\partial_{\tau,\tau'} = \{u \in \Lambda : \tau(u) \neq \tau'(u)\}$ is the collection of vertices on which $\tau,\tau'$ differ. We first prove that
\begin{align}\label{eq:decay-message}
    w_{r} \cdot \norm{\phi(\bm{p}_{r}) - \phi(\bm{p}_{r}')}_{2} \leq (1 - \delta)^{\ell} \cdot \max_{v \in \partial_{\tau,\tau'}} \wrapc{w_{v} \cdot \norm{\phi(\bm{p}_{v}) - \phi(\bm{p}_{v}')}_{2}}
\end{align}
by repeatedly applying \cref{p:weight-contract}, and then convert this inequality back to a total variation bound. We note that the appropriate modification for weak spatial mixing is to set $\ell = d(r,\Lambda)$, and stop iterating \cref{p:weight-contract} immediately when one reaches any vertex of $\Lambda$.

To prove \cref{eq:decay-message}, we apply \cref{p:weight-contract} iteratively. Writing $\bm{m}_{v} = \phi(\bm{p}_{v})$, observe that
\begin{align*}
    w_{r} \cdot \norm{\bm{m}_{r} - \bm{m}_{r}'}_{2} &= w_{r} \cdot \norm{g^{\phi}(\bm{m}_{1},\dots,\bm{m}_{\Delta_{r}}) - g^{\phi}(\bm{m}_{1}',\dots,\bm{m}_{\Delta_{r}}')}_{2} \\
    &\leq (1 - \delta) \cdot \max_{u \in S(r, 1)}\wrapc{w_{u} \cdot \norm{\bm{m}_{u} - \bm{m}_{u}'}_{2}} \tag{\cref{p:weight-contract}} \\
    &\leq \dotsb \tag{Induction} \\
    &\leq (1 - \delta)^{\ell-1} \cdot \max_{v \in S(r, \ell-1)} \wrapc{w_{v} \cdot \norm{\bm{m}_{v} - \bm{m}_{v}'}_{2}}.
\end{align*}
We now convert \cref{eq:decay-message} back to total variation distance to conclude the theorem. This is where we incur the constant factor loss, and where we must use the marginal lower bounding function $\lb$. If $v$ is any vertex, by the Mean Value Theorem,
\begin{align*}
    \phi(\bm{p}_{v}) - \phi(\bm{p}_{v}') &= \int_{0}^{1} \diag(\Phi(\bm{p}_{v}(t))) \cdot (\bm{p}_{v} - \bm{p}_{v}') \,dt.
\end{align*}
Furthermore, note that the entries of $\bm{p}_{v},\bm{p}_{v}'$ are all lower bounded by $\lb^{*}$ and upper bounded by $\ub^{*}$. Applying this to any vertex in $S(r, \ell) \cap \partial_{\tau,\tau'}$, we get upper bound
\begin{align*}
    \norm{\phi(\bm{p}_{v}) - \phi(\bm{p}_{v}')}_{2} 
    &\leq \sup_{0 \leq t \leq 1} \eigval_{\max}\wrapp{\diag(\Phi(\bm{p}_{v}(t)))} \cdot \norm{\bm{p}_{v} - \bm{p}_{v}'}_{2} \\
    &\leq 2 \cdot \sup_{\lb^{*} \leq p \leq \ub^{*}}\{\Phi(p)\} \cdot \norm{\bm{p}_{v} - \bm{p}_{v}'}_{\TV}.
\end{align*}
Similarly, applying this to $v = r$, we get the lower bound
\begin{align*}
    \norm{\phi(\bm{p}_{r}) - \phi(\bm{p}_{r}')}_{2} 
    &\geq \inf_{0 \leq t \leq 1} \eigval_{\min}\wrapp{\diag(\Phi(\bm{p}_{r}(t)))} \cdot \norm{\bm{p}_{r} - \bm{p}_{r}'}_{2} \\ 
    &\geq \frac{2}{\sqrt{q}} \cdot \inf_{\lb^{*} \leq p \leq \ub^{*}}\{\Phi(p)\} \cdot \norm{\bm{p}_{r} - \bm{p}_{r}'}_{\TV}.
\end{align*}
Note that $\sup_{\lb^{*} \leq p \leq \ub^{*}} \Phi(p) = \Phi(\lb^{*})$, since $\phi$ being concave on $[0,\ub^{*}]$ implies $\Phi$ is decreasing on $[0,\ub^{*}]$. By the same reasoning, $\inf_{\lb^{*} \leq p \leq \ub^{*}} \Phi(p) = \Phi(\ub^{*})$. Combining these inequalities, we see that
\begin{align*}
    \norm{\bm{p}_{r} - \bm{p}_{r}'}_{\TV} &\leq \frac{\sqrt{q}}{2} \cdot \frac{1}{w_{r} \cdot \Phi(\ub^{*})} \cdot \norm{\phi(\bm{p}_{r}) - \phi(\bm{p}_{r}')}_{2} \\
    &\leq \frac{\sqrt{q}}{2} \cdot \frac{1}{w_{r} \cdot \Phi(\ub^{*})} \cdot (1 - \delta)^{\ell-1} \cdot \max_{v \in S(r, \ell-1)} \wrapc{w_{v} \cdot \norm{\phi(\bm{p}_{v}) - \phi(\bm{p}_{v}')}_{2}} \\
    &\leq \sqrt{q} \cdot \frac{w_{\max} \cdot \Phi(\lb^{*})}{w_{\min} \cdot \Phi(\ub^{*})} \cdot (1 - \delta)^{\ell-1} \cdot \max_{v \in S(r, \ell-1)} \norm{\bm{p}_{v} - \bm{p}_{v}'}_{\TV} \\
    &\leq C_{\SM} \cdot (1-\delta)^{\ell}. \qedhere
\end{align*}
\end{proof}

\subsection{Total influence decay and spectral independence via contraction}\label{a:contraction-si}

In this section, we show that \cref{cond:jacobian-norm} implies strong quantitative bounds on influences and spectral independence for the Gibbs distribution. 
\begin{thm}\label{thm:si-from-contraction}
Let $\Delta \in \N$ be a positive integer, and $0 < \delta < 1$ be a real. Suppose the tree recursion $g$ satisfies the strong version of \cref{cond:jacobian-norm} with rate $1 - \delta$ with respect to $(\phi,\ub,w,\Delta)$. Additionally, let $\lb$ be a marginal lower bounding function. Then for every tree $T=(V,E)$ of maximum degree $\Delta$ with corresponding Gibbs measure $\mu$, and every pinning $\tau:\Lambda \to [q]$ on $\Lambda \subseteq V$, the following statements hold:
\begin{enumerate}
    \item\label{item:contraction-to-specind} \textbf{Spectral independence:} The conditional Gibbs measure $\mu^{\tau}$ satisfies $\eigval_{\max}(\infl_{\mu^{\tau}}) \leq \frac{C_{\SI}}{\delta}$, where
    \begin{align*}
        C_{\SI} \defeq \frac{w_{\max} \cdot \wrapp{\ub^{*}}^{3/2} \cdot \Phi(\lb^{*})}{w_{\min} \cdot \wrapp{\lb^{*}}^{3/2} \cdot \Phi(\ub^{*})},
    \end{align*}
    $w_{\max} \defeq \max_{d,d' \leq \Delta} w(d,d')$, $w_{\min} \defeq \min_{d,d' \leq \Delta} w(d,d')$, and $\Phi = \phi'$ is the derivative of the potential.
    \item\label{item:contraction-to-influence-decay} \textbf{Total influence decay:} For every vertex $r \in V \setminus \Lambda$, and every $\ell \in \N$, we have the influence bound
    \begin{align*}
        \max_{\mfb, \mfc \in [q]} \sum_{v \in S(r, \ell)} \norm{\mu_{v}^{\tau, r \gets \mfb} - \mu_{v}^{\tau, r \gets \mfc}}_{\TV} \leq C_{\INFL} \cdot (1 - \delta)^{\ell}
    \end{align*}
    where
    \begin{align*}
        C_{\INFL} \defeq \sqrt{q} \cdot C_{\SI} = \sqrt{q}\cdot \frac{w_{\max} \cdot \ub^{*} \cdot \Phi(\lb^{*})}{w_{\min} \cdot \lb^{*} \cdot \Phi(\ub^{*})}.
    \end{align*}
\end{enumerate}
\end{thm}

The key intermediate result we prove using the Jacobian norm bound is the following. It establishes a decay of total influence, but measured in an $L^{2}$ sense, which is more amenable to applying \cref{cond:jacobian-norm}.
\begin{prop}\label{prop:level-norm-decay}
Under the conditions of \cref{thm:si-from-contraction}, for every $\ell \in \N$ and every $\bm{x} \in \R^{[q] \times S(r, \ell)}$,
\begin{align*}
    \frac{\norm{\sum_{v \in S(r, \ell)} \infl_{\mu^{\tau}}^{r \to v} \bm{x}_{v}}_{2}}{\max_{v \in S(r, \ell)} \norm{\bm{x}_{v}}_{2}} \leq C_{\SI} \cdot (1-\delta)^{\ell}.
\end{align*}
\end{prop}
We first use this proposition to deduce the theorem.
\begin{proof}[Proof of \cref{thm:si-from-contraction}: \cref{item:contraction-to-specind}]
To bound $\eigval_{\max}(\infl_{\mu^{\tau}})$, we introduce an appropriate vector norm $\norm{\cdot}$ on $(\R_{\geq0}^{q})^{n}$ and bound the induced matrix norm of $\infl_{\mu^\tau}$. This suffices since one always has the inequality $\eigval_{\max}(\infl_{\mu^\tau}) \leq \norm{\infl_{\mu^\tau}}$ for any matrix norm $\norm{\cdot}$ induced by a vector norm.

Our choice of vector norm is a hybrid ``$L^{\infty}$ of $L^{2}$'' norm. More precisely, for a vector $\bm{x} \in (\R_{\geq0}^{q})^{n}$ written as a concatenation of vectors $\bm{x}_{1},\dots,\bm{x}_{n} \in \R^{q}$, recall from~\cref{sec:colorings-simple} that 
\begin{align*}
    \norm{\bm{x}}_\star = \max_{i=1,\dots,n} \norm{x_{i}}_{2}.
\end{align*}
For a matrix $M \in \R^{nq \times nq}$, we write
\begin{align*}
    \norm{M} \defeq \sup_{\bm{x} \neq \bm{0}} \frac{\norm{M\bm{x}}}{\norm{\bm{x}}_\star}
\end{align*}
for the matrix norm induced by the vector norm $\norm{\cdot}$; note the similarity with the one used in \cref{cond:jacobian-norm} for the Jacobian. Intuitively, the $L^{\infty}$ captures the fact that it is often more tractable to bound absolute row sums of the influence matrix, as was employed in prior works. The $L^{2}$ captures the fact that we are working with a (weighted) $L^{2}$ norm for the Jacobian of the tree recursion.

Applying this matrix norm to the influence matrix $\infl_{\mu^{\tau}}$, we have
\begin{align*}
    \norm{\infl_{\mu^{\tau}}} &= \sup_{\bm{x} \neq \bm{0}} \frac{\max_{r \in V} \norm{\sum_{v \in V} \infl_{\mu^{\tau}}^{r \to v} \bm{x}_{v}}_{2}}{\max_{r \in V} \norm{\bm{x}_{r}}_{2}} \\
    &\leq \sup_{\bm{x} \neq \bm{0}} \frac{\max_{r \in V} \sum_{\ell=1}^{\infty} \norm{\sum_{v \in S(r, \ell)} \infl_{\mu^{\tau}}^{r \to v} \bm{x}_{v}}_{2}}{\max_{r \in V} \norm{\bm{x}_{r}}_{2}} \\
    &\leq \sup_{\bm{x} \neq \bm{0}} \frac{\max_{r \in V} \sum_{\ell=1}^{\infty} C_{\SI}(1-\delta)^{\ell} \cdot \max_{v \in S(r, \ell)}\norm{\bm{x}_{v}}_{2}}{\max_{r \in V} \norm{\bm{x}_{r}}_{2}} \tag{\cref{prop:level-norm-decay}} \\
    &\leq C_{\SI}\sum_{\ell=1}^{\infty} (1-\delta)^{\ell} \cdot \sup_{\bm{x} \neq \bm{0}} \underset{\le 1}{\underbrace{\frac{\max_{r \in V} \max_{\ell \in \N^+} \max_{v \in S(r, \ell)} \norm{\bm{x}_{v}}_{2}}{\max_{r \in V} \norm{\bm{x}_{r}}_{2}}}} \\
    &\le \frac{C_{\SI}}{\delta}.
\end{align*}
Thus, we obtain the same upper bound for $\eigval_{\max}(\infl_{\mu^\tau})$ as wanted.
\end{proof}
\begin{proof}[Proof of \cref{thm:si-from-contraction}: \cref{item:contraction-to-influence-decay}]
The main work that has to be done is converting between the correct matrix norms. Observe that for every $\mfb,\mfc \in [q]$ and every $v \in S(r, \ell)$, we have that
\begin{align*}
    \norm{\mu_{v}^{\tau, r \gets \mfb} - \mu_{v}^{\tau, r \gets \mfc}}_{\TV} &\le \norm{\mu_{v}^{\tau, r \gets \mfb} - \mu_{v}^{\tau}}_{\TV} + \norm{\mu_{v}^{\tau, r \gets \mfc} - \mu_{v}^{\tau}}_{\TV} \\
    &= \frac{1}{2} \sum_{\mfa \in [q]} \abs{\infl_{\mu^{\tau}}^{r \to v}(\mfb, \mfa)} + \frac{1}{2} \sum_{\mfa \in [q]} \abs{\infl_{\mu^{\tau}}^{r \to v}(\mfc, \mfa)}.
\end{align*}
It follows that
\begin{align*}
    \max_{\mfb,\mfc\in[q]} \sum_{v \in S(r, \ell)} \norm{\mu_{v}^{\tau, r \gets \mfb} - \mu_{v}^{\tau, r \gets \mfc}}_{\TV} &\leq \max_{\mfc \in [q]} \sum_{v \in S(r, \ell)} \sum_{\mfa \in [q]} \abs{\infl_{\mu^{\tau}}^{r \to v}(\mfc,\mfa)} \\
    &= \sup_{\bm{x} \neq \bm{0}} \frac{\norm{\sum_{v \in S(r, \ell)} \infl_{\mu^{\tau}}^{r \to v} \bm{x}_{v}}_{\infty}}{\max_{v \in S(r, \ell)} \norm{\bm{x}_{v}}_{\infty}} \\
    &\leq \sqrt{q} \cdot \sup_{\bm{x} \neq \bm{0}} \frac{\norm{\sum_{v \in S(r, \ell)} \infl_{\mu^{\tau}}^{r \to v} \bm{x}_{v}}_{2}}{\max_{v \in S(r, \ell)} \norm{\bm{x}_{v}}_{2}} \\
    &\leq \sqrt{q} \cdot C_{\SI}, \tag{\cref{prop:level-norm-decay}}
\end{align*}
as claimed.
\end{proof}
All that remains is to prove \cref{prop:level-norm-decay}. The proof of this rests on the following lemma, which relates influence matrices to Jacobians of the (modified) tree recursion.
\begin{lemma}[Influence Matrices and Jacobians]\label{lem:infl-jacobian}
Let $\mu$ be the Gibbs distribution of a $q$-spin system on a tree $T=(V,E)$ rooted at $r \in V$. Let $u_{1},\dots,u_{d}$ denote the children of $r$, with marginal distributions $\bm{p}_{1},\dots,\bm{p}_{d}$ with respect to their corresponding subtrees, respectively. Then for every $i \in [d]$ and every vertex $v$ in the subtree of $T$ rooted at $u_{i}$, we have the identity
\begin{align*}
    \infl^{r \to v} = (J_{i}g^{\log})(\log \bm{p}) \cdot \infl^{u_{i} \to v}.
\end{align*}
In particular, if $\phi : \R \to \R$ is any other monotone potential function with derivative $\Phi = \phi'$, then writing $\bm{m} = \phi(\bm{p})$ entrywise, we have
\begin{align*}
    \infl^{r \to v} = \diag(g(\bm{p}) \odot \Phi(g(\bm{p})))^{-1} \cdot (J_{i}g^{\phi})(\bm{m}) \cdot \diag(\bm{p}_{i} \odot \Phi(\bm{p}_{i})) \cdot \infl^{u_{i} \to v}.
\end{align*}
\end{lemma}
While we are not aware of this lemma being stated at this level of generality in the literature, its proof involves standard techniques. Indeed, it is well-known that influences have an analytic interpretation as derivatives. We give a proof at the end of the section.

With this lemma in hand, we are now ready to prove \cref{prop:level-norm-decay}
\begin{proof}[Proof of \cref{prop:level-norm-decay}]
We first factorize our influence matrices $\infl_{\mu^{\tau}}^{r \to v}$ as products of Jacobians along edges of the tree. We then finally use our assumption, namely \cref{cond:jacobian-norm}.

To implement this, fix $r \in V$, $\ell \in \N$, and for each $v \in S(r, \ell)$, we write $r = u_{0},\dots,u_{\ell} = v$ for the unique path from $r$ to $v$. We now view the tree $T$ as rooted at $r$. For each $v \in V$, we then define $D_{r,v}$ as the diagonal matrix $D_{r,v} \defeq \diag(\bm{p}_{v} \cdot \Phi(\bm{p}_{v}))$, where $\bm{p}_{v} \in \R_{\geq0}^{q}$ is the marginal distribution for $v$ \emph{with respect to the subtree $T_{v}$} (conditioned on $\tau$); note that $D_{r,v}$ does have a (mild) dependence on the choice of root $r$, since this choice can affect the subtree $T_{v}$. Finally, we also write $\bm{m}_{v} = \phi(\bm{p}_{v})$, and for each vertex $v$ with child $u$, we abbreviate $J_{u}g^{\phi}$ for the Jacobian $(J_{u}g^{\phi})(\bm{m}_{w} : w \in S(v, 1))$ of $g^{\phi}$ computed in $T_{v}$. Then
\begin{align*}
    \norm{\sum_{v \in S(r, \ell)} \infl_{\mu^{\tau}}^{r \to v} \bm{x}_{v}}_{2} 
    &= \norm{\sum_{r=u_{0},\dots,u_{\ell}=v} \prod_{i=1}^{\ell} \left(D_{u_{i-1}}^{-1} \cdot J_{u_{i}}g^{\phi} \cdot D_{u_{i}}\right) \infl_{\mu^{\tau}}^{v \to v} \bm{x}_{v}}_{2} \tag{\cref{lem:infl-jacobian} and Induction} \\
    &= \norm{D_{r,r}^{-1} \sum_{r=u_{0},\dots,u_{\ell}=v} \prod_{i=1}^{\ell} J_{u_{i}}g^{\phi} \cdot D_{r,v} \infl_{\mu^{\tau}}^{v \to v}\bm{x}_{v}}_{2} \tag{Telescoping} \\
    &\leq \norm{D_{r,r}^{-1}}_{2} \cdot \norm{\sum_{r=u_{0},\dots,u_{\ell}=v} \prod_{i=1}^{\ell} J_{u_{i}}g^{\phi} \cdot D_{r,v}\infl_{\mu^{\tau}}^{v \to v}\bm{x}_{v}}_{2}.
\end{align*}
We now inductively ``peel off'' one layer of Jacobians at a time and apply contraction. Observe that
\begin{align*}
    &w_{r} \cdot \norm{\sum_{r=u_{0},\dots,u_{\ell}=v} \prod_{i=1}^{\ell} J_{u_{i}}g^{\phi} \cdot D_{r,v}\infl_{\mu^{\tau}}^{v \to v}\bm{x}_{v}}_{2} \\
    &= w_{r} \cdot \norm{\sum_{u_{1} \in S(r, 1)} J_{u_{1}}g^{\phi} \cdot \sum_{L_{u_{1}}(1) \ni u_{2},\dots,u_{\ell} = v} \prod_{i=2}^{\ell} J_{u_{i}}g^{\phi} \cdot D_{r,v}\infl_{\mu^{\tau}}^{v \to v}\bm{x}_{v}}_{2} \\
    &\leq (1 - \delta) \cdot \max_{u_{1} \in S(r, 1)} \wrapc{w_{u_{1}} \cdot \norm{\sum_{L_{u_{1}}(1) \ni u_{2},\dots,u_{\ell} = v} \prod_{i=2}^{\ell} J_{u_{i}}g^{\phi} \cdot D_{r,v}\infl_{\mu^{\tau}}^{v \to v}\bm{x}_{v}}_{2}} \tag{Jacobian Norm Bound} \\
    &\leq \dotsb \tag{Induction} \\
    &\leq (1-\delta)^{\ell} \cdot \max_{v \in S(r, \ell)} \wrapc{w_{v} \cdot \norm{D_{r,v}\infl_{\mu^{\tau}}^{v \to v}\bm{x}_{v}}_{2}} \\
    &\leq (1 - \delta)^{\ell} \cdot \max_{v \in S(r, \ell)} \wrapc{w_{v} \cdot \norm{D_{r,v}}_{2} \cdot \norm{\infl_{\mu^{\tau}}^{v \to v}}_{2}} \cdot \max_{v \in S(r, \ell)}\wrapc{\norm{\bm{x}_{v}}_{2}}.
\end{align*}
Noting that
\begin{align*}
    \infl_{\mu^{\tau}}^{v \to v} &= I - \allone\bm{p}_{v}^{\top} = \diag\wrapp{\sqrt{\bm{p}_{v}}}^{-1} \cdot \underset{\text{Projection}}{\underbrace{\wrapp{I - \sqrt{\bm{p}_{v}}\sqrt{\bm{p}_{v}}^{\top}}}} \cdot \diag\wrapp{\sqrt{\bm{p}_{v}}},
\end{align*}
we have $\norm{\infl_{\mu^{\tau}}^{v \to v}}_{2} \leq \sqrt{\frac{\ub^{*}}{\lb^{*}}}$.
Since the entries of each $D_{r,v}$ are upper bounded by $\ub^{*} \cdot \Phi(\lb^{*})$ and lower bounded by $\lb^{*} \cdot \Phi(\ub^{*})$, we have
\begin{align*}
    \max_{r,v \in V} \wrapc{w_{r}^{-1}\norm{D_{r,r}^{-1}}_{2} \cdot w_{v}\norm{D_{r,v}}_{2} \cdot \norm{\infl_{\mu^{\tau}}^{v \to v}}_{2}} \leq \frac{w_{\max} \cdot \wrapp{\ub^{*}}^{3/2} \cdot \Phi(\lb^{*})}{w_{\min} \cdot \wrapp{\lb^{*}}^{3/2} \cdot \Phi(\ub^{*})},
\end{align*}
so the desired claim follows.
\end{proof}

\begin{proof}[Proof of \cref{lem:infl-jacobian}]
Note that the second identity follows from the first because the Chain Rule and the Inverse Function Theorem together imply that
\begin{align*}
    (J_{i}g^{\phi})(\bm{m}) = \diag(\Phi(g(\bm{p}))) \cdot (J_{i}g)(\bm{p}) \cdot \diag(\Phi(\bm{p}_{i}))^{-1}.
\end{align*}
Hence, it suffices to establish the first identity. We do this entrywise via an analytic method which is \emph{agnostic} to the specific form of the tree recursion $g$; all that matters is that $g$ \emph{computes} the marginals of $r$ given the marginals of the $u_{1},\dots,u_{d}$ (in their respective subtrees).

Consider the multivariate partition function $Z_{\mu}(\bm{\lambda}) = \sum_{\sigma : V \to [q]} \mu(\sigma) \prod_{v \in V} \lambda_{v, \sigma(v)}$, where we have associated to each vertex-color pair $v,\mfc$ a variable $\lambda_{v,\mfc}$; these $\lambda_{v,\mfc}$ are often interpreted as \emph{external fields}. Now, fix two colors $\mfb,\mfc \in [q]$. Let
\begin{align*}
    P_{r,\mfc}(\bm{t}) \defeq \frac{\partial}{\partial t_{r,\mfc}} \log Z_{\mu}(\exp(\bm{t})) = \frac{\sum_{\sigma : \sigma(r) = \mfc} \mu(\sigma) \prod_{v \in V} \exp(t_{v,\sigma(v)})}{\sum_{\sigma} \mu(\sigma) \prod_{v \in V} \exp(t_{v,\sigma(v)})}.
\end{align*}
Note that $P_{r,\mfc}(\bm{0}) = \mu_{r}(\mfc)$, and by the same token, we have that
\begin{align*}
    \infl^{r \to v}(\mfc, \mfb) = \frac{\partial}{\partial t_{v,\mfb}} \log P_{r,\mfc}(\bm{0}).
\end{align*}
Now, by definition, the recursion $g^{\log}$ satisfies
\begin{align*}
    \log P_{r,\mfc}(\bm{t}) = g_{\mfc}^{\log}(\log P_{u,\mfa}(\bm{t}) : u \sim r, \mfa \in [q]),
\end{align*}
where each $P_{u,\mfa}$ is computed with respect to its subtree $T_{u}$. Differentiating this recursion and applying the Chain Rule, we obtain
\begin{align*}
    &\frac{\partial}{\partial t_{v,\mfb}} \log P_{r,\mfc}(\bm{0}) \\
    &= \sum_{j=1}^{d} \sum_{\mfa \in [q]} (J_{j}g^{\log})_{\mfc, \mfa}(\log \bm{p}) \cdot \frac{\partial}{\partial t_{v,\mfb}} \log P_{u_{j},\mfa}(\bm{t}) \\
    &= \sum_{\mfa \in [q]} (J_{i}g^{\log})_{\mfc, \mfa}(\log \bm{p}) \cdot \frac{\partial}{\partial t_{v,\mfb}} \log P_{u_{i},\mfa}(\bm{t}) \tag{$P_{u_{j},\mfa}$ doesn't depend on $t_{v,\mfb}$ for $j \neq i$} \\
    &= \sum_{\mfa \in [q]} (J_{i}g^{\log})_{\mfc, \mfa}(\log \bm{p}) \cdot \infl^{u_{i} \to v}(\mfa, \mfb)
\end{align*}
as desired. 
\end{proof}

\subsection{Proofs of main results}\label{sec:main-result-proofs} 

We will now give the formal proofs of~\cref{t:weak-ssm,t:strong-ssm,thm:SSM+SSI,thm:wsm-potts} via the same recipe, namely verifying that for suitable choice of parameters $(\phi, \ub, w, \Delta)$, the Jacobian of the appropriate tree recursion has the strong (or weak) Jacobian norm condition. The desired results will then immediately follow by some combination of~\cref{thm:si-from-contraction,thm:ssm-from-contraction}.

As an illustration, we begin by verifying~\cref{t:weak-ssm}, namely that SSM holds with exponential decay rate for $q$-colorings on maximum degree $\Delta$ trees, whenever $q \ge \Delta + 3\sqrt{\Delta}.$

\begin{proof}[Proof of~\cref{t:weak-ssm}]
Consider $q, \Delta \ge 2$ such that $q = \Delta + \gamma \ge \Delta + 3\sqrt{\Delta}$ and a tree $T$ with maximum degree $\Delta$. We will show that the Jacobian of the tree recursion~\cref{d:tree-recurse-coloring} satisfies the strong Jacobian norm condition of~\cref{cond:jacobian-norm} with respect to $(\phi, \ub, w, \Delta)$ with rate $1 - \delta$ for some $\delta > 0$. As noted in~\cref{table:jacobian-cond-params}, we make the following parameter choices: 
\begin{itemize}
    \item Let $\phi =  2 \arctanh(\sqrt{x})$ as in~\cref{p:weak}, so that $\phi$ is strictly increasing on $[0, 1]$ and concave on $[0, 1/\gamma]$ by~\cref{fact:lu-yin-properties}
    \item Take $\ub = 1/\gamma$ for $\gamma = q - \Delta \ge 3\sqrt{\Delta}$, noting that by~\cref{l:marginal}, $\p_v \preceq 1/\gamma$ for any unpinned vertex $v \in V(T)$.
    \item Let $w = 1$ (i.e.\ $\|J\|_{\w} = \|J\|_{\star \star}$ as defined in~\cref{d:norm-star})
\end{itemize}
\cref{p:weak} then implies that there exists $\delta = \delta(q, \Delta) > 0$ such that the tree recursion satisfies the strong Jacobian norm condition with respect to $(\phi, \ub, w, \Delta)$ with rate $1 - \delta$. Consequently by~\cref{thm:ssm-from-contraction}, there exists some $C_{\SM} > 0$ such that strong spatial mixing holds with rate $1 - \delta$ and constant $C_{\SM}$. 
\end{proof}

With some slight complications, the proof of~\cref{t:strong-ssm} is analogous.

\begin{proof}[Proof of~\cref{t:strong-ssm}]
Let $T$ be an arbitrary tree of maximum degree $\Delta$ rooted at $\rho \in V$. Fix $q \ge \Delta + 3$. We make the following parameter choices with respect to which we will verify~\cref{cond:jacobian-norm}.
\begin{itemize}
   \item As before $\phi =  2 \arctanh(\sqrt{x})$ as in~\cref{p:weak}, so that $\phi$ is strictly increasing on $[0, 1]$ and concave on $[0, 1/\gamma]$ by~\cref{fact:lu-yin-properties}
    \item For vertex $v \in V(T)$, we let 
    $$\ub(d_v, \Delta_v) := \ub(d_v) = 1-\frac{1}{\frac{1}{q-1}\left(1 + \frac{1}{\gamma
   -1}\right)^{\frac{d_v \gamma }{q-1}}+1},$$
    where we observe that if $v$ has $d_v$ unpinned neighbors,~\cref{l:marginal-strong} implies that $\p_v \le \ub(d_v)$.
    \item For $v \in V(T)$, we let
    $$w(d_v, \Delta_v) = (1 - \zeta_v)^{-1/2},$$
    for $\zeta_v$ defined in~\cref{eq:weight,eq:xi} that only depends on $d_v, \Delta_v$.
\end{itemize}
With respect to $(\phi, \ub, w, \Delta)$ as above,~\cref{p:weight-jacobian} implies that there exists $\delta = \delta(q, \Delta)$ such for any vertex $v \in V(T)$ that is \textit{not} the root $\rho$, the tree recursion for the subtree of $T$ rooted at $v$ satisfies the strong Jacobian norm condition. The proof of~\cref{p:weight-jacobian} implies an explicit upper bound on the Jacobian of the tree recursion of the subtree rooted at \textit{any} $w \in V(T)$ in terms of the number of colors available after a subset of vertices are pinned. In particular, at the root vertex $\rho$, if $q \ge \Delta + 3$, we can compute that $\sup_{\bm{m} = \phi(\p)}\|J g^{\phi}(\bm{m})\|_{\w} \le 17$. We then observe that the proof of~\cref{p:weight-contract} actually implies that 
$$ w_{\rho} \cdot \norm{g^{\phi}(\bm{m}) - g^{\phi}(\bm{m}')}_{2} \leq \sup_{0 \leq t \leq 1} \wrapc{\norm{(Jg^{\phi})(\bm{m}(t))}_{\bm{w}}} \cdot \max_{i\in[d_{\rho}]}\wrapc{w_{i} \cdot \norm{\bm{m}_{i} - \bm{m}_{i}'}_{2}},$$
and thus we have that 
$$ w_{\rho} \cdot \norm{g^{\phi}(\bm{m}) - g^{\phi}(\bm{m}')}_{2} \leq 17 \max_{i\in[d_{\rho}]}\wrapc{w_{i} \cdot \norm{\bm{m}_{i} - \bm{m}_{i}'}_{2}}$$
Therefore, by paying an additional multiplicative factor of $C = \frac{17}{w_{\rho}(1- \delta)}$ (i.e.\ not contracting on the first step in the proof of~\cref{thm:ssm-from-contraction}) we see that~\cref{thm:ssm-from-contraction} implies that if $q \ge \Delta + 3$, the $q$-colorings model on trees exhibits strong spatial mixing with constant $C' = C \cdot C_{\SM}$ and rate $1 - \delta.$
\end{proof}

In the course of proving~\cref{t:strong-ssm}, we showed that the tree recursion satisfies the strong Jacobian norm condition. This property also implies total influence decay and thus gives~\cref{thm:SSM+SSI} as we verify below.

\begin{proof}[Proof of~\cref{thm:SSM+SSI}]
Part (1) follows from~\cref{t:strong-ssm}. Similar to the proof of~\cref{t:strong-ssm} above, we note that for a tree $T = (V, E)$ rooted at $\rho$ of maximum degree $\Delta$, the tree recursion of the subtree rooted at any $v \neq \rho$ satisfies the strong Jacobian norm condition with respect to $(\phi, \ub, w, \Delta)$ as defined in the proof of~\cref{t:strong-ssm} immediately above with rate $1 - \delta$ for some $\delta > 0$. Since $q \ge \Delta + 3$, at the root vertex $\rho$, we have that that $\sup_{\bm{m} = \phi(\p)}\|J g^{\phi}(\bm{m})\|_{\w} \le 17$ as noted in the above proof. By considering the proof of~\cref{prop:level-norm-decay}, we observe that~\cref{prop:level-norm-decay} holds with constant $C_{\SI}' \le 17 C_{\SI} \cdot \frac{1}{1 - \delta}$ for $C_{\SI}$ defined in~\cref{thm:si-from-contraction}:~\cref{item:contraction-to-specind}. Thus, by~\cref{thm:si-from-contraction}:~\cref{item:contraction-to-influence-decay} that total influence decay holds with rate $1 - \delta$ and constant $C_{\INFL} = \sqrt{q}C_{\SI}'$, verifying part (2) of~\cref{thm:SSM+SSI}.
\end{proof}

With $q = (1+\eps) \Delta$ colors, we can get a better bound for $\delta$ which is independent of the maximum degree $\Delta$.
\begin{prop}\label{prop:(1+eps)Delta}
If $q = (1 + \eps) \Delta$ for $\eps \in (0,1)$ and $\Delta \ge 7/\eps^2$, 
then for all trees of maximum degree $\Delta$
the uniform distribution of proper $q$-colorings exhibits:
\begin{itemize}
    \item Strong spatial mixing with exponential decay rate $e^{-\eps/4}$ and constant $C_\SM = O_\eps(\poly(\Delta))$;
    \item Total influence decay with exponential decay rate $e^{-\eps/4}$ and constant $C_\INFL = O_\eps(\poly(\Delta))$.
\end{itemize}
\end{prop}
\begin{proof}
\cref{coro:(1+eps)Delta} implies that the tree recursion satisfies the strong Jacobian condition with rate $\exp\left(-\eps/4\right)$ with respect to $(\phi, \ub = 1/(\eps \Delta), w = 1, \Delta)$ for $\phi' = \Phi = \frac{1}{\sqrt{p}(1 - p)}$ (as noted in~\cref{table:jacobian-cond-params}).
By~\cref{l:marginal}, we observe that 
$$\ub^* \le \frac{1}{q - \Delta} \le \frac{1}{\eps \Delta},$$
and by~\cref{lem:marginal-lb}, we observe that 
$$\lb^* \ge \frac{1}{q} \left(1 - \frac{1}{\eps \Delta} \right)^{\Delta}.$$
Thus by~\cref{thm:ssm-from-contraction}, we have that the $q$-colorings model for $q = (1 + \eps)\Delta$ with $\eps \in (0,1)$ and $\Delta \ge 7/\eps^2$ exhibits strong spatial mixing with rate $\exp\left(-\eps/4\right)$ and constant
\begin{align*}
C_{\SM} &= \frac{\sqrt{q}}{\exp\left(-\eps/4\right)} \cdot \frac{\Phi(\lb^{*})}{\Phi(\ub^{*})} \le \frac{q^2 e^{\eps/4}}{q-1} \sqrt{\frac{1}{\eps \Delta}} \left(1 + \frac{1}{\eps \Delta  - 1} \right)^{\frac12 \Delta - 1} \le \frac{q^2 e^{\eps/4}}{q-1} \sqrt{\frac{1}{\eps \Delta}}  e^{\frac{\Delta}{2(\eps \Delta  - 1)}} \\
&\le 2q \sqrt{\frac{1}{\eps \Delta}} e^{1/\eps + \eps/4} = O_\eps(\poly(\Delta))
\end{align*}
Similarly, 
by~\cref{thm:si-from-contraction}:~\cref{item:contraction-to-influence-decay}, the strong Jacobian norm condition implies that the $q$-colorings model has total influence decay with rate $1 - \delta$ and constant 
\begin{align*}
C_{\INFL} &= \sqrt{q} \cdot \frac{\ub^* \cdot \Phi(\lb^*)}{\lb^* \cdot \Phi(\ub^*)} \\
&\le   \frac{q^{3}}{ (\Delta \eps)^{3/2} \left(q - 1\right)} \left(1 + \frac{1}{\Delta    \eps - 1}\right)^{3\Delta/2 } \le   \frac{q^{3}}{ (\Delta \eps)^{3/2} \left(q - 1\right)} \exp\left(\frac{3\Delta}{2(\Delta    \eps - 1)}\right) \\
&\le   \frac{q^{3}}{ (\Delta \eps)^{3/2} \left(q - 1\right)}e^{3/\eps} =  O_\eps(\poly(\Delta)).
\end{align*}
\end{proof}

\begin{proof}[Proof of \cref{thm:wsm-potts}]
Let $T$ be an arbitrary tree of maximum degree $\Delta$ rooted at $\rho \in V$ and  let $q \ge (1 - \beta)(\Delta + \gamma) + 1$ for some $\gamma \ge 5, \beta \in (0, 1)$.  We make the following parameter choices with respect to which we will verify~\cref{cond:jacobian-norm}.
\begin{itemize}
   \item  Let $\phi$ be defined implicitly via its derivative $\Phi(p) = \phi'(p) = \frac{1}{\sqrt{p}(1 - (1-\beta)p)}$ as in~\cref{eq:phi-potts}.
    \item For $v \in V(T)$, let 
    $\ub(d_v, \Delta_v) := \ub_{\beta, q}(\Delta_v)$ as defined in~\cref{e:true_ub}, so that by~\cref{lem:potts-marg-2level-new}, for any unpinned vertex $v$ with $\Delta_v$ children and no pinned neighbors $\p_v \le \ub(\Delta_v)$.
    \item For $v \in V(T)$, we let
    $$w(d_v, \Delta_v) = w_v = (1 - \zeta_v)^{-1/2},$$
    for $\zeta_v$ defined in~\cref{eq:weight-potts} that only depends on $\Delta_v$.
\end{itemize}
\cref{p:jacobian-potts} implies that there exists $\delta = \delta(\beta, \Delta, q) > 0$ such that the tree recursion~\cref{d:tree-recurse-potts} satisfies the \textit{weak Jacobian condition} of~\cref{cond:jacobian-norm}
with respect to $(\phi, \ub_{\beta, q}, w, \Delta)$ with rate $1- \delta$. Weak spatial mixing for the antiferromagnetic Potts model then follows by~\cref{thm:ssm-from-contraction}.
\end{proof}

\section{Conclusion and open problems}
\label{sec:conclusion}

In this paper we established SSM for $q$-colorings on all trees of maximum degree $\Delta$ whenever $q \ge \Delta+3$, almost resolving \cref{c:coloring-folklore}. Similarly, for the antiferromagnetic Potts model we proved WSM on trees when $q \ge (1-\beta)(\Delta+5) + 1$. It would be interesting to 
fully resolve \cref{c:coloring-folklore} and \cref{conjecture:antiferro-Potts-wsm}, namely reducing the number of colors to $\Delta + 1$ and achieving the full range $\beta \ge \max\{0,1-q/\Delta\}$ respectively.

We also prove spectral independence and rapid mixing of Glauber dynamics on graphs of maximum degree $\Delta$ and girth $g = \Omega_\Delta(1)$ whenever $q \ge \Delta+3$, establishing \cref{c:color-mix} for a general class of graphs with a nearly optimal bound. 
Can one prove SSM or total influence decay for such graphs too?
More generally, it is still unclear the relationship between spectral independence, SSM, and total influence decay for general spin systems.

\bibliographystyle{alpha}
\bibliography{tssm.bib}

\appendix

\section{Proofs for marginal bounds}\label{sec:pf-marg-bounds}

\begin{proof}[Proof of~\cref{l:marginal-strong}]
We apply~\cref{d:tree-recurse-coloring} and compute that
\begin{align*}
\frac{\p_{r}(\mfc)}{1 - \p_{r}(\mfc)} &= \frac{\prod_{i \in [d]}(1 - \p_i(\mfc))}{\sum_{\mfc' \neq \mfc \in \mcL_r} \prod_{i \in [d]} (1 - \p_i(\mfc'))} \le \frac{1}{\sum_{\mfc' \neq \mfc \in \mcL_r} \prod_{i \in [d]} (1 - \p_i(\mfc'))} \\
&\le \frac{1}{q-1} \cdot \left(\prod_{\mfc' \neq \mfc \in \mcL_r} \prod_{i \in [d]} (1 - \p_i(\mfc'))\right)^{-\frac{1}{q-1}} \tag{AM-GM} \\
&\le \frac{1}{q-1} \cdot \left(\prod_{i \in [d]} \left(1 - \frac{1}{\gamma}\right)\right)^{-\frac{\gamma}{q-1}} 
= \frac{1}{q-1} \left(1 + \frac{1}{\gamma - 1}\right)^{\frac{\gamma d}{q-1}}. \tag{\cref{lem:product-lb}} 
\end{align*}
\end{proof}

\section{Proofs of technical lemmata}\label{sec:technical2}

We begin by verifying the computation of the Jacobian of the tree recursion.
\begin{proof}[Proof of~\cref{o:jacob-plain}]
Let $g(\p) = (g_{\mfc}(p))_{c \in [q]}$ for tree recursion $g_{\mfc}(p)$ defined in~\cref{d:tree-recurse-coloring}.
We compute the entry-wise derivatives and see that for $\mfb \neq \mfc$,
\begin{align*}
(J_i g)(\p)_{\mfc, \mfb} &= \partial_{p_i(\mfb)} g_{\mfc}(\p)  
= \frac{\prod_{j \in [d]}(1 - \p_{j}(\mfc)) \prod_{j \in [d],\,j\neq i}(1 - \p_{j}(\mfb))  }{\left(\sum_{\mfc' \in [q]}  \prod_{j \in [d]} (1 - \p_j(\mfc')) \right)^2} 
= \frac{g_{\mfc}(\p) g_{\mfb}(\p)}{(1 - \p_i(\mfb))},
\end{align*}
and for $\mfb = \mfc$,
\begin{align*}
& (J_i g)(\p)_{\mfc, \mfb} = \partial_{p_i(\mfb)} g_{\mfc}(\p) \\
&= \frac{-\prod_{j \in [d]; j \neq i}(1 - \p_{j}(\mfc)) \left(\sum_{\mfc' \in [q]}  \prod_{j \in [d]} (1 - \p_j(\mfc')) \right) + \prod_{j \in [d],\,j\neq i}(1 - \p_{j}(\mfc)) \prod_{j\in [d]} (1 - \p_j(\mfc)) }{\left(\sum_{\mfc' \in [q]}  \prod_{j \in [d]} (1 - \p_j(\mfc')) \right)^2} \\
&= \frac{g_{\mfc}(\p)^2 - g_{\mfc}(\p)}{1 - \p_i(\mfc)}.
\end{align*}
Thus, grouping terms gives as desired
\begin{align*}
(J_i g)(\p) &= (g(\p) g(\p)^{\top} - \diag(g(\p))) \diag(1 - \p_i)^{-1} \in \RR^{q \times q}. \qedhere
\end{align*}
\end{proof}

\begin{proof}[Proof of~\cref{o:jacob-phi}]
Applying the chain rule, we find that 
\begin{align*}
(J_i g^{\phi})_{\mfc, \mfb}(\y) &= \partial_{\y_{i, \mfb}} g_{\mfc}^{\phi}(\y) = \partial_{\y_{i, \mfb}} \phi(g_{\mfc}(\phi^{-1}(\y))) = \phi'(g_{\mfc}(\phi^{-1}(\y))) (\partial_{\p_{i, \mfb}} g_{\mfc})(\p) (\phi^{-1})'(\y) \\
&= \Phi(g_{\mfc}(\p)) (J_i g)_{\mfb, \mfc}(\p) \Phi(p_{i, \mfb})^{-1},
\end{align*}
which implies that 
$$(J_i g^{\phi})(\y) = \diag(\Phi(g(\p))) (J_i g)(\p) \diag(\Phi(\p_i))^{-1}.$$
Substituting in the expression from~\cref{o:jacob-plain} gives the desired result.
\end{proof}

In this work, we repeatedly used the following sharp upper bound on the operator norm of a concatenation of diagonal matrices, which can be proved via Cauchy--Schwarz.
\begin{lemma}\label{lem:norm-concat-diag}
For positive integers $d,q \geq 1$, and diagonal matrices $D_{1},\dots,D_{d} \in \R^{q \times q}$,
\begin{align*}
    \sup_{\bm{x}_{1},\dots,\bm{x}_{d} \in \R^{q}} \frac{\norm{\sum_{i=1}^{d} D_{i} \bm{x}_{i}}_{2}^{2}}{\sum_{i=1}^{d} \norm{\bm{x}_{i}}_{2}^{2}} \leq \max_{\mfc \in [q]} \sum_{i=1}^{d} D_{i}(\mfc,\mfc)^{2}.
\end{align*}
\end{lemma}
\begin{proof}
Expanding out the numerator, we have
\begin{align*}
    \norm{\sum_{i=1}^{d} D_{i} \bm{x}_{i}}_{2}^{2} &= \sum_{\mfc \in [q]} \wrapp{\sum_{i=1}^{d} D_{i}(\mfc,\mfc) \cdot \bm{x}_{i}(\mfc)}^{2} \\
    &\leq \sum_{\mfc \in [q]} \wrapp{\sum_{i=1}^{d} D_{i}(\mfc,\mfc)^{2}}\wrapp{\sum_{i=1}^{d} \bm{x}_{i}(\mfc)^{2}} \tag{Cauchy--Schwarz} \\
    &\leq \max_{\mfc \in [q]} \wrapc{\sum_{i=1}^{d} D_{i}(\mfc,\mfc)^{2}} \cdot \sum_{\mfc \in [q]} \sum_{i=1}^{d} \bm{x}_{i}(\mfc)^{2} \\
    &= \max_{\mfc \in [q]} \wrapc{\sum_{i=1}^{d} D_{i}(\mfc,\mfc)^{2}} \cdot \norm{\bm{x}}_{2}^{2}. \qedhere
\end{align*}
\end{proof}

We will also leveraged by following ``Lagrangian'' lower bound throughout this work.
\begin{lemma}\label{lem:product-lb}
Let $\Omega$ be a finite set, and let $b$ be a positive real number satisfying $1 / \abs{\Omega} \leq b \leq 1$. Let $\nu \in \R_{\geq0}^{\Omega}$ be a vector satisfying $\nu(\omega) \leq b$ for every $\omega \in \Omega$ and $s = \sum_{\omega \in \Omega} \nu(\omega) \leq 1$. Then we have the lower bound
\begin{align*}
    \prod_{\omega \in \Omega} (1 - \nu(\omega)) \geq (1 - b)^{s/b}.
\end{align*}
\end{lemma}
\begin{proof}
We claim the minimizer must be a vector which supported on $\ceil{s/b}$ elements and has value $b$ uniformly on its support (except, perhaps a single ``overflow'' element, which has value $s - b \cdot \floor{s/b}$). If we can establish this, then the lower bound would follow just by calculation.

Fix $\nu \in \R_{\geq0}^{\Omega}$ satisfying $\nu(\omega) \leq b$ for every $\omega \in \Omega$ and $\sum_{\omega \in \Omega} \nu(\omega) = s \leq 1$. Suppose $\nu$ does not have the claimed structure. Then there exist distinct elements $x,y \in \Omega$ such that $b > \nu(x) \geq \nu(y)$. In this case, for $0 < \epsilon \leq b - \nu(x)$, define $\nu' \in \R_{\geq0}^{\Omega}$ by setting $\nu'(\omega) = \nu(\omega)$ for all $\omega \neq x,y$, and $\nu'(x) = \nu(x) + \epsilon, \nu'(y) = \nu(y) - \epsilon$. We claim that $\prod_{\omega \in \Omega} (1 - \nu'(\omega)) < \prod_{\omega \in \Omega} (1 - \nu(\omega))$. As $\nu'$ agrees with $\nu$ for all $\omega \neq x,y$, this is equivalent to
\begin{align*}
    (1 - \nu'(x))(1 - \nu'(y)) < (1 - \nu(x))(1 - \nu(y)).
\end{align*}
To see that this holds, we simply expand the left-hand side using the definition of $\nu'$:
\begin{align*}
    (1 - \nu'(x))(1 - \nu'(y)) &= (1 - \nu(x) - \epsilon)(1 - \nu(y) + \epsilon) \\
    &= (1 - \nu(x))(1 - \nu(y)) - \epsilon^{2} - \epsilon(1 - \nu(y)) + \epsilon(1 - \nu(x)) \\
    &= (1 - \nu(x))(1 - \nu(y)) - \epsilon^{2} - \epsilon(\nu(x) - \nu(y)) \\
    &\leq (1 - \nu(x))(1 - \nu(y)) - \epsilon^{2} \tag{Using $\nu(x) \geq \nu(y)$} \\
    &< (1 - \nu(x))(1 - \nu(y)) \tag{Using $\epsilon > 0$}.
\end{align*}
Hence, the lemma follows.
\end{proof}

To show contraction for $q$-colorings, we studied the potential function $\phi(x) = 2\arctanh(\sqrt{x})$, which enjoys several nice properties that aided in our analysis.
\begin{fact}\label{fact:lu-yin-properties}
Define $\phi:[0,1] \to \R \cup \{\infty\}$ by the function $\phi(x) \defeq 2\arctanh(\sqrt{x})$, which has derivative $\Phi(x) = \phi'(x) \defeq \frac{1}{\sqrt{x}(1-x)}$. Then $\phi$ is nonnegative and strictly increasing on all of $[0,1]$, with range $\R_{\geq0} \cup \{\infty\}$. Furthermore, $\phi$ is concave on $[0,1/3]$.
\end{fact}
\begin{proof}
Note $\arctanh(y)$ is nonnegative for $0 \le y \le 1$ and thus so is $\phi(x)$. Further, $\phi$ is continuous and twice differentiable in $(0, 1)$, and thus we can verify that $\phi$ is concave on $[0, 1/3]$ by noting that $\phi''(x) = \frac{3 x-1}{2 (1-x)^2 x^{3/2}}$, which is nonpositive for $x \le 1/3$; thus $\phi$ is concave on $[0, 1/3].$
\end{proof}

\begin{lemma}
If $\phi : \R \to \R$ is smooth, strictly increasing, and concave, then its inverse $\phi^{-1} : \R \to \R$ is smooth, strictly increasing, and convex.
\end{lemma}
\begin{proof}
This is an immediate consequence of the inverse function theorem.
\end{proof}

We used the entropy-type quantity $s(x)$ defined in~\cref{eq:s-func} to define our weights in giving essentially sharp bounds on strong spatial mixing for colorings on trees. In the course of establishing contraction, we leveraged several technical properties of this function.
\begin{fact}\label{fact:entropy-like-ineq}
\begin{enumerate}[(1)]
\item Define $f: \R^+ \to \R$ by
$$ f(x) \defeq \left( \frac{1}{x} + 1 \right) \ln (1+x) - 1. $$
Then for any $x \in (0,3/2)$ it holds
$$ 0 < f(x) \le 1-e^{-x/2} < 1. $$

\item Define $s:\R^+ \to \R$ by 
$$ s(x) \defeq \frac{1}{x} \ln \frac{1}{1-x} - 1. $$
Then for any $x \in (0,3/5)$ it holds
$$ 0 < s(x) \leq 1 - \exp\wrapp{-\frac{1}{2} \cdot \frac{x}{1-x}} < 1. $$
\end{enumerate}
\end{fact}
\begin{proof}
The lower bound on $f$ follows from that
\begin{align*}
f(x) = \left( \frac{1}{x} + 1 \right) \ln (1+x) - 1 
> \left( \frac{1}{x} + 1 \right) \frac{x}{1+x} - 1 = 0
\end{align*}
for all $x>0$. For the upper bound on $f$, consider the function 
$$ \psi(x) \defeq (1+x)\ln(1+x) + x e^{-x/2} -2x. $$
It suffices to show that $\psi(x) \le 0$ for all $x \in (0,3/2)$. 
We compute the derivative as
$$ \psi'(x) = \ln(1+x) + \left( 1-\frac{x}{2} \right) e^{-x/2} -1, $$
and the second derivative as
$$ \psi''(x) = \frac{1}{1+x} - \left( 1-\frac{x}{4} \right) e^{-x/2}. $$
Now note that the quadratic function $(1+x)(1-x/4)$ and the exponential function $e^{x/2}$ has exactly two intersection points, $0$ and some $x_0 \in (0,3/2)$, as the former is concave and the latter is convex. 
Hence, $\psi''(x)$ has a unique root on $(0,3/2)$ which is $x_0$. 
Then $\psi'(x)$ is decreasing on $(0,x_0)$ and increasing on $(x_0,3/2)$, implying that it has a unique root on $(0,3/2)$, denoted by $x_1$.
Similarly, $\psi(x)$ is decreasing on $(0,x_1)$ and increasing on $(x_1,3/2)$, and we conclude that
$$ \psi(x) \le \max\{ \psi(0), \psi(3/2) \} = 0. $$

The bounds on $s$ follow immediately from the observation that $s(x) = f(\frac{x}{1-x})$ and $\frac{x}{1-x} \in (0,3/2)$ whenever $x \in (0,3/5)$.
\end{proof}

\end{document}